\documentclass[preprint]{aastex63}
\usepackage{xcolor}

\usepackage{natbib}
\usepackage{amsmath}
\usepackage{enumitem}

\shortauthors{Bernardinelli et al.}

\defcitealias{Bernstein2000}{BK}

\newcommand\change[1]{#1}
\newcommand{\ie}{\textit{i.e.}}
\newcommand{\eg}{\textit{e.g.}}
\newcommand{\E}{\mathrm{E}}

\newcommand{\des}{\textit{DES}}

\newcommand{\eqq}[1]{Equation~(\ref{#1})}

\newcommand{\comment}[1]{}
\newcommand{\vecc}{\ensuremath{\mathbf{c}}}
\newcommand{\vecP}{\ensuremath{\mathbf{P}}}
\newcommand{\likeli}{\mathcal{L}}
\newcommand{\anomaly}{\mathcal{M}}
\newcommand{\dcdhdpdA}{\mathrm{d}\vecc\,\mathrm{d}H\,\mathrm{d}A\,\mathrm{d}\vecP\,}
\newcommand{\nirb}{\mathrm{NIRB}}
\newcommand{\nirf}{\mathrm{NIRF}}

\reportnum{FERMILAB-PUB-24-0939-PPD}
\reportnum{DES-2024-0873}

\begin{document}

\title{Photometry of outer Solar System objects from the Dark Energy Survey II: a joint analysis of trans-Neptunian absolute magnitudes, colors, lightcurves and dynamics}

\author[0000-0003-0743-9422]{Pedro H. Bernardinelli}
\altaffiliation{DiRAC Postdoctoral Fellow}
\affiliation{DiRAC Institute, Department of Astronomy, University of Washington, 3910 15th Ave NE, Seattle, WA, 98195, USA}
\email{phbern@uw.edu}

\author[0000-0002-8613-8259]{Gary M. Bernstein}
\affiliation{Department of Physics and Astronomy, University of Pennsylvania, Philadelphia, PA 19104, USA}
\email{garyb@upenn.edu}

\author{T.~M.~C.~Abbott}
\affiliation{Cerro Tololo Inter-American Observatory, NSF's National Optical-Infrared Astronomy Research Laboratory, Casilla 603, La Serena, Chile}

\author{M.~Aguena}
\affiliation{Laborat\'orio Interinstitucional de e-Astronomia - LIneA, Rua Gal. Jos\'e Cristino 77, Rio de Janeiro, RJ - 20921-400, Brazil}

\author[0000-0002-7069-7857]{S.~S.~Allam}
\affiliation{Fermi National Accelerator Laboratory, P. O. Box 500, Batavia, IL 60510, USA}

\author[0000-0002-8458-5047]{D.~Brooks}
\affiliation{Department of Physics \& Astronomy, University College London, Gower Street, London, WC1E 6BT, UK}

\author[0000-0003-3044-5150]{A.~Carnero~Rosell}
\affiliation{Instituto de Astrofisica de Canarias, E-38205 La Laguna, Tenerife, Spain}
\affiliation{Laborat\'orio Interinstitucional de e-Astronomia - LIneA, Rua Gal. Jos\'e Cristino 77, Rio de Janeiro, RJ - 20921-400, Brazil}
\affiliation{Universidad de La Laguna, Dpto. Astrofísica, E-38206 La Laguna, Tenerife, Spain}

\author[0000-0002-3130-0204]{J.~Carretero}
\affiliation{Institut de F\'{\i}sica d'Altes Energies (IFAE), The Barcelona Institute of Science and Technology, Campus UAB, 08193 Bellaterra (Barcelona) Spain}

\author{L.~N.~da Costa}
\affiliation{Laborat\'orio Interinstitucional de e-Astronomia - LIneA, Rua Gal. Jos\'e Cristino 77, Rio de Janeiro, RJ - 20921-400, Brazil}

\author{M.~E.~S.~Pereira}
\affiliation{Hamburger Sternwarte, Universit\"{a}t Hamburg, Gojenbergsweg 112, 21029 Hamburg, Germany}

\author[0000-0002-4213-8783]{T.~M.~Davis}
\affiliation{School of Mathematics and Physics, University of Queensland,  Brisbane, QLD 4072, Australia}

\author[0000-0001-8318-6813]{J.~De~Vicente}
\affiliation{Centro de Investigaciones Energ\'eticas, Medioambientales y Tecnol\'ogicas (CIEMAT), Madrid, Spain}

\author[0000-0002-0466-3288]{S.~Desai}
\affiliation{Department of Physics, IIT Hyderabad, Kandi, Telangana 502285, India}

\author[0000-0002-8357-7467]{H.~T.~Diehl}
\affiliation{Fermi National Accelerator Laboratory, P. O. Box 500, Batavia, IL 60510, USA}

\author{P.~Doel}
\affiliation{Department of Physics \& Astronomy, University College London, Gower Street, London, WC1E 6BT, UK}

\author{S.~Everett}
\affiliation{California Institute of Technology, 1200 East California Blvd, MC 249-17, Pasadena, CA 91125, USA}

\author[0000-0002-2367-5049]{B.~Flaugher}
\affiliation{Fermi National Accelerator Laboratory, P. O. Box 500, Batavia, IL 60510, USA}

\author[0000-0003-4079-3263]{J.~Frieman}
\affiliation{Fermi National Accelerator Laboratory, P. O. Box 500, Batavia, IL 60510, USA}
\affiliation{Kavli Institute for Cosmological Physics, University of Chicago, Chicago, IL 60637, USA}

\author[0000-0002-9370-8360]{J.~Garc\'ia-Bellido}
\affiliation{Instituto de Fisica Teorica UAM/CSIC, Universidad Autonoma de Madrid, 28049 Madrid, Spain}

\author[0000-0001-9632-0815]{E.~Gaztanaga}
\affiliation{Institut d'Estudis Espacials de Catalunya (IEEC), 08034 Barcelona, Spain}
\affiliation{Institute of Cosmology and Gravitation, University of Portsmouth, Portsmouth, PO1 3FX, UK}
\affiliation{Institute of Space Sciences (ICE, CSIC),  Campus UAB, Carrer de Can Magrans, s/n,  08193 Barcelona, Spain}

\author{R.~A.~Gruendl}
\affiliation{Center for Astrophysical Surveys, National Center for Supercomputing Applications, 1205 West Clark St., Urbana, IL 61801, USA}
\affiliation{Department of Astronomy, University of Illinois at Urbana-Champaign, 1002 W. Green Street, Urbana, IL 61801, USA}

\author[0000-0003-0825-0517]{G.~Gutierrez}
\affiliation{Fermi National Accelerator Laboratory, P. O. Box 500, Batavia, IL 60510, USA}

\author[0000-0001-6718-2978]{K.~Herner}
\affiliation{Fermi National Accelerator Laboratory, P. O. Box 500, Batavia, IL 60510, USA}

\author{S.~R.~Hinton}
\affiliation{School of Mathematics and Physics, University of Queensland,  Brisbane, QLD 4072, Australia}

\author{D.~L.~Hollowood}
\affiliation{Santa Cruz Institute for Particle Physics, Santa Cruz, CA 95064, USA}

\author[0000-0002-6550-2023]{K.~Honscheid}
\affiliation{Center for Cosmology and Astro-Particle Physics, The Ohio State University, Columbus, OH 43210, USA}
\affiliation{Department of Physics, The Ohio State University, Columbus, OH 43210, USA}

\author[0000-0001-5160-4486]{D.~J.~James}
\affiliation{Center for Astrophysics $\vert$ Harvard \& Smithsonian, 60 Garden Street, Cambridge, MA 02138, USA}

\author[0000-0003-0120-0808]{K.~Kuehn}
\affiliation{Australian Astronomical Optics, Macquarie University, North Ryde, NSW 2113, Australia}
\affiliation{Lowell Observatory, 1400 Mars Hill Rd, Flagstaff, AZ 86001, USA}

\author[0000-0002-1134-9035]{O.~Lahav}
\affiliation{Department of Physics \& Astronomy, University College London, Gower Street, London, WC1E 6BT, UK}

\author{S.~Lee}
\affiliation{Jet Propulsion Laboratory, California Institute of Technology, 4800 Oak Grove Dr., Pasadena, CA 91109, USA}

\author[0000-0003-0710-9474]{J.~L.~Marshall}
\affiliation{George P. and Cynthia Woods Mitchell Institute for Fundamental Physics and Astronomy, and Department of Physics and Astronomy, Texas A\&M University, College Station, TX 77843,  USA}

\author[0000-0001-9497-7266]{J. Mena-Fern{\'a}ndez}
\affiliation{LPSC Grenoble - 53, Avenue des Martyrs 38026 Grenoble, France}

\author[0000-0002-6610-4836]{R.~Miquel}
\affiliation{Instituci\'o Catalana de Recerca i Estudis Avan\c{c}ats, E-08010 Barcelona, Spain}
\affiliation{Institut de F\'{\i}sica d'Altes Energies (IFAE), The Barcelona Institute of Science and Technology, Campus UAB, 08193 Bellaterra (Barcelona) Spain}

\author{J.~Myles}
\affiliation{Department of Astrophysical Sciences, Princeton University, Peyton Hall, Princeton, NJ 08544, USA}

\author[0000-0002-2598-0514]{A.~A.~Plazas~Malag\'on}
\affiliation{Kavli Institute for Particle Astrophysics \& Cosmology, P. O. Box 2450, Stanford University, Stanford, CA 94305, USA}
\affiliation{SLAC National Accelerator Laboratory, Menlo Park, CA 94025, USA}

\author{S.~Samuroff}
\affiliation{Department of Physics, Northeastern University, Boston, MA 02115, USA}
\affiliation{Institut de F\'{\i}sica d'Altes Energies (IFAE), The Barcelona Institute of Science and Technology, Campus UAB, 08193 Bellaterra (Barcelona) Spain}

\author[0000-0002-9646-8198]{E.~Sanchez}
\affiliation{Centro de Investigaciones Energ\'eticas, Medioambientales y Tecnol\'ogicas (CIEMAT), Madrid, Spain}

\author{B.~Santiago}
\affiliation{Instituto de F\'\i sica, UFRGS, Caixa Postal 15051, Porto Alegre, RS - 91501-970, Brazil}
\affiliation{Laborat\'orio Interinstitucional de e-Astronomia - LIneA, Rua Gal. Jos\'e Cristino 77, Rio de Janeiro, RJ - 20921-400, Brazil}

\author[0000-0002-1831-1953]{I.~Sevilla-Noarbe}
\affiliation{Centro de Investigaciones Energ\'eticas, Medioambientales y Tecnol\'ogicas (CIEMAT), Madrid, Spain}

\author[0000-0002-3321-1432]{M.~Smith}
\affiliation{Physics Department, Lancaster University, Lancaster, LA1 4YB, UK}

\author[0000-0002-7047-9358]{E.~Suchyta}
\affiliation{Computer Science and Mathematics Division, Oak Ridge National Laboratory, Oak Ridge, TN 37831}

\author[0000-0003-1704-0781]{G.~Tarle}
\affiliation{Department of Physics, University of Michigan, Ann Arbor, MI 48109, USA}

\author[0000-0001-7211-5729]{D.~L.~Tucker}
\affiliation{Fermi National Accelerator Laboratory, P. O. Box 500, Batavia, IL 60510, USA}

\author{V.~Vikram}
\affiliation{}

\author[0000-0002-7123-8943]{A.~R.~Walker}
\affiliation{Cerro Tololo Inter-American Observatory, NSF's National Optical-Infrared Astronomy Research Laboratory, Casilla 603, La Serena, Chile}

\author{N.~Weaverdyck}
\affiliation{Department of Astronomy, University of California, Berkeley,  501 Campbell Hall, Berkeley, CA 94720, USA}
\affiliation{Lawrence Berkeley National Laboratory, 1 Cyclotron Road, Berkeley, CA 94720, USA}

\collaboration{1000}{(The \des\ Collaboration)}

\suppressAffiliations

\begin{abstract}
	\vspace{0.2in}
	To the extent that the physical properties of trans-Neptunian objects (TNOs) are determined before any migration, we can attempt to connect members of present-day TNO dynamical classes to their regions of birth by studying the classes' distribution of physical properties.  For the well-defined sample of 696 TNOs with absolute magnitudes $5.5 < H_r < 8.2$ detected in the Dark Energy Survey (DES), we characterize the relationships between their dynamical state and their physical properties---namely $H_r$, indicative of size; their colors $c=\{g-r,r-i,i-z\}$, indicative of surface composition; and their semi-amplitude $A$ of flux variation, indicative of asphericity and surface inhomogeneity.  We seek ``birth'' physical distributions $p_\beta(H_r, c, A)$ that can recreate all the distribution of these parameters in every dynamical class $d$ as a mixture $p_d=\sum_\beta f_{\beta|d} p_\beta$ with fractions $f_{\beta|d}.$
We first show that the observed colors of all 696 are consistent with arising from 2 Gaussian distributions in $griz$ space, ``near-IR bright'' (NIRB) and ``near-IR faint'' (NIRF), presumably arising from an inner and outer birth population, respectively,   quantifying the trend seen by \citet{fraser2023}.  We find that the DES TNOs favor a model in which the $H_r$ and $A$ distributions of NIRF-colored objects are independent of their current dynamical state, and likewise for NIRB objects, supporting the assignment of NIRF and NIRB as birth populations.  All objects are consistent with a common $p(H_r)$ that departs from a power law, but NIRF objects are significantly more time-variable.  Cold classical (CC) TNOs are purely NIRF while hot classical (HC), scattered, and detached TNOs are consistent with $\approx70\%$ NIRB, and there are significant variations among resonances' NIRB fractions.  The NIRB component of HC TNOs (and of some resonances) have broader inclination distributions than the NIRF, \ie\ the current dynamical state (inclination distribution) retains information about birth location. We also find potential evidence for radial stratification within the birth NIRB population, in that HC NIRBs are on average redder than detached or scattered NIRBs, and a similar effect distinguishes CC NIRF objects from dynamically excited NIRFs.  We estimate total object counts and masses of each dynamical class's members within our $H_r$ range.  These results will strongly constrain models of planetesimal formation and migration in the outer solar system.
\end{abstract}

\section{Introduction} \label{sec:intro}

The present-day population of trans-Neptunian objects (TNOs) can be grouped into broadly defined classes of dynamical behavior \citep{Gladman2008,Khain2020,gladman2021}, with only the low-inclination ($I<5\arcdeg$) \emph{cold classicals} (CC) at semi-major axes $42<a<48$~AU believed to have been formed near their current heliocentric distance.  The other dynamical classes include:
the \emph{resonant} population,  trapped in mean motion resonances with Neptune; the
\emph{scattering} population, currently undergoing significant evolution in $a,$ primarily because their lower perihelia expose them to significant dynamical influence from Neptune; the \emph{hot classical} (HC) population, at higher incinlations than the CC's in the $39<a<49$ region between the 3:2 and 2:1 mean-motion resonances with Neptune; and
the \emph {detached} population, with larger semi-major axes that are not experiencing any significant dynamical perturbations from any of the known planets.  Most scenarios for the origin of these dynamical classes have them populated by objects created in quiescent orbits at heliocentric distances $\lesssim42$~AU, and scattered into higher-excitation orbits by Neptune as it migrates outward to its current position \citep[see review by][]{morbidelli2020}.

This scenario is supported not just by dynamical modeling, but by observations of the physical properties of the TNOs.
The photometric colors of TNOs are indicative of their surface composition.  The bulk abundances of the TNOs will depend on the location of their formation within the protoplanetary disk that formed the Solar System; if surface processing is independent of heliocentric distance after any dynamical migrations begin, then surface colors will retain memory of the region of origin.

The color diversity of TNOs became clear from the first measurements of their colors by \cite{Luu1996},~\cite{Tegler1998} and~\cite{Jewitt2001}.  Subsequent surveys have identified strong correlation between dynamical state and color; in particular, the CCs are seen to be exclusively red, in contrast to a mix of redder and bluer objects among the HCs and other classes
\citep{tegler2000,doressoundiram2001,trujillo2002}. Correlations between color and inclination have been extensively revisited in the literature for the distinct dynamical populations of the trans-Neptunian region \citep[\eg][]{Peixinho2015,marsset2019,marsset2023}. There have been multiple attempts at creating a taxonomical classification for TNO colors, with two \citep{Fraser2012,Tegler2016,Wong2017}, three \citep{Pike2017b} and four \citep{Barucci2001,Barucci2005,Fulchignoni2008} color classes proposed out of small samples of optical and near-IR data, while \cite{DalleOre2013} combined color and albedo to suggest 10 distinct taxa. Recent detailed investigations, including a large sample of near-IR colors, by \citet{fraser2023}, suggests two sequences in the visible-NIR color space, with CC's exclusively resident in one space.

In parallel with the study of TNO colors, the absolute magnitude distribution, a proxy for the size distribution, of the TNO populations has also been extensively studied. This distribution is known to transition from a steeper to a shallower behavior at diameters of roughly 100 km \citep{Bernstein2004,Fraser2014,Lawler2018a,kavelaars2021,napier2024} for the distinct trans-Neptunian populations. To date, \cite{Wong2017} has been the only attempt at constraining TNO $H$ distributions as a function of color, finding $n(H)$ for $7.5\lesssim H \lesssim 9.5$ to be indistinguishable between dynamically cold and hot populations. 

Further evidence of a distinct formation history for the CCs from the dynamically excited populations is the absence of large bodies \change{\citep[$H_r\lesssim<5.5,$][]{NA22}} from the CCs, also noted early in the study of TNOs \citep{Bernstein2004,Petit2011,kavelaars2021}. Most recently, \citet{bernardinelli2023} demonstrated that the CC's have a distinct distribution of light-curve semi-amplitude $A$ from the HCs, with higher variability (suggesting less-spherical shapes) for the CC's within a fixed range of $H_r.$

In this work, we extend the work of \citet{bernardinelli2023} to examine the joint distributions $p_d(H_r, c, A)$ of absolute magnitude (indicating size), color(s) $c$ (surface composition), and light-curve amplitude (shape)  of the 696 TNOs with $5.5 < H_r < 8.2$ found in different dynamical classes $d$ by the Dark Energy Survey \citep[DES,][]{Bernardinelli2022}.  We will treat these physical properties purely as ``birthmarks'' tagging their heliocentric distance of formation, rather than investigating the physics of the formation or surface properties themselves. Since migration processes should be independent of the physical properties, we can, ideally, identify distributions $p_\beta(H_r, c, A)$ of physical properties that are characteristic of regions $\beta=\{1,2,\ldots\}$ of \emph{birth}.  Determining the coefficients $f_{\beta|d}$ such that $p_d = \sum_\beta f_{\beta|d} p_\beta$ would then indicate what fraction of each \emph{dynamical} class originated at each formation distance range.

The DES sample is the best available for such studies at present, because it is the only large sample with well-defined selection function for which multi-band, multi-epoch observations are available for every object, as well as complete determination of dynamical state (i.e. orbital parameters) and variability estimates (i.e. lightcurve amplitudes).  \citet{Bernardinelli2022} detail the detection process and selection function for TNOs in the DES---of particular note is that the selection function is largely independent of color or variability.  \citet{bernardinelli2023} produce photometric measurements $D_i$ for TNO $i$ consisting of flux measurements in the $griz$ bands and their statistical uncertainties, from which they derive a posterior probability $p(H_r, c, A | D_i)$ for the physical observables of each object conditioned on its observations. These probability distributions are realized as ``measurement swarms''---the results of a Monte Carlo sampling from the posterior probability for $H_r, c,$ and $A$ under an assumption of uniform prior on the mean fluxes of the source.

The principal goal of this paper is to determine the minimal number of distinct physical distributions $N_\beta$ required to be consistent with the full sample of DES objects in all dynamical classes $d$. The known discrepancies between the CCs' and the excited populations' distributions of color, size, and variability require that $N_\beta\ge 2.$  If each $\beta$ is indeed the birthmark of a distinct formation distance, then the $f_{\beta|d}$ will tell us the distribution of formation distances for the TNOs that ended up in each dynamical class---a powerful constraint on scenarios for the dynamical evolution of the outer solar system, and potentially a constraint on models of the physics of planetesimal formation as well.

The largest TNOs are known to have color and albedo distributions that are quite distinct from smaller objects \citep{Brown2012}, perhaps due to frost cycles, differentiation, and other processes clearly evidenced by New Horizons observations of Pluto and Charon \citep{Stern2015}. \citet{bernardinelli2023} also definitively detect lower photometric variability for larger objects.  For these reasons we will restrict this paper's analysis to a fixed, relatively narrow range of $5.5<H_r<8.2,$ to keep size dependence from being confused with any trends of physical properties with formation region.

In Section \ref{sec:colormodel}, we find the Gaussian mixture model (GMM) which best describes the intrinsic distribution of the full DES TNO sample in the space of $g-r, r-i,$ and $i-z$ colors.  A GMM represents a distribution as a sum of multivariate normal distributions.  There we find that
the color distribution of the DES TNOs as a whole can be adequately fit as the sum of two Gaussian distributions, which we call ``near-IR faint'' (NIRF) and ``near-IR bright'' (NIRB) because of their relative $i$ and $z$-band brightnesses for fixed $g-r$ visible color.  These closely resemble the two color sequences proposed from independent data by \citet{fraser2023}. \change{Our NIRB/NIRF families, like the two defined in \citet{fraser2023}, project onto visible colors to reproduce the ``red/blue'' bimodality identified in earlier work, but the inclusion of a NIR band better separates them.} Section \ref{sec:framework} presents a mathematical framework that enables the study of statistical properties (such as TNO size distribution slopes and the distribution of lightcurve amplitudes) to broadly defined ``families'' of objects. We then proceed in Section~\ref{sec:nirbnirf} to assess whether we can derive just two physical distributions $q_\nirf(H_r, c, A)$ and $q_\nirb(H_r, c, A),$ such that each dynamical class is constructed from a mixture of these two distributions. Section \ref{sec:split} assesses whether there is evidence in the DES catalog for $N_\beta>2$ distinguishable physical components among the dynamical classes, \ie\ evidence for further stratification of birth properties that has not yet been erased by dynamical mixing. Section \ref{sec:inclination} discusses the distribution of physical classes vs inclination, and derives population estimates for the distinct trans-Neptunian physical populations. Section \ref{sec:discussion} discusses our findings, \change{and places them into the context of previous results and potential physical scenarios.} We summarize our results in Section \ref{sec:summary}. The appendices give details of the statistical algorithms used in this work:  Appendix~\ref{sec:gmm} describes the optimization of the GMM in the presence of measurement errors and selection functions, \change{Appendix \ref{sec:evidence} describes our approach to computing Bayesian evidence ratios}, Appendix \ref{sec:hmc} presents our Hamiltonian Monte Carlo approach to sampling the complicated parameter space of our model, and Appendix \ref{sec:completeness} presents our strategy for deriving completeness estimates for our sample.

\section{The color distribution of the trans-Neptunian region}
\label{sec:colormodel}
We wish to model the underlying color distribution for TNOs, \ie\ before any selection or noise, implied by our TNO sample. To this end, we will posit that the underlying distribution for the colors $\mathbf{c}$ is a mixture model given by a combination of $K$ components. For a general measurement $\mathbf{x}$ in $d$ dimensions, we have that the mixture model is
\begin{equation}
	p(\mathbf{x}|\boldsymbol\theta) = \sum_{\alpha=1}^K f_\alpha p(\mathbf{x}|\boldsymbol\theta_\alpha), \label{eq:gmm}
\end{equation}
where the $\alpha^\text{th}$ component is described by a probability distribution $p(\mathbf{x}|\boldsymbol\theta_\alpha)$, and the parameters $\boldsymbol\theta = \bigcup_\alpha \{\boldsymbol\theta_\alpha, f_\alpha \} $ are subject to the constraint $\sum_\alpha f_\alpha = 1$ to ensure that $p(\mathbf{x}|\boldsymbol\theta)$ is normalized. In a \emph{Gaussian} mixture model (GMM), the probability distribution $p(\mathbf{x}|\boldsymbol\theta_\alpha)$ is a multivariate Gaussian with mean $\boldsymbol\mu_\alpha$ and covariance matrix $\boldsymbol\Sigma_\alpha$, that is,
\begin{equation}
  p(\mathbf{x} | \boldsymbol\theta_\alpha) = \mathcal{N}(\mathbf{x}|\boldsymbol\mu_\alpha, \boldsymbol\Sigma_\alpha) \equiv \frac{1}{\sqrt{(2\pi)^d \det(\Sigma_\alpha)}} \exp\left[-\frac{1}{2}(\mathbf{x} - \boldsymbol\mu_\alpha)^\top \boldsymbol\Sigma_\alpha^{-1} (\mathbf{x} - \boldsymbol\mu_\alpha)\right].
  \label{eq:multivariate}
\end{equation}
In our case the data space is $\mathbf{x}=(g-r, r-i,i-z).$  The output of the Markov chain executed in \citet{bernardinelli2023} for each TNO yields a sample of its probability density over $\mathbf{x}.$ Strictly for purposes of deriving the most likely GMM, we approximate this measurement probability as a normal distribution with mean and covariance equal to the mean and covariance matrix of the TNO's samples; once the GMM components are defined, all other analyses in this paper use the measurement-error uncertainties defined by the Markov chain samples.

Appendix~\ref{sec:gmm} describes our variant of the GMM method which optimizes the intrinsic distribution in the presence of known uncertainties and selection effects applied to each measurement.  We also discuss there the algorithm's method for selecting the number $K$ of components to allow in the model. The derivation of the color selection function is described in Appendix \ref{sec:completeness}.

\subsection{Resulting model}
Applying the GMM method to all $800+$ DES TNOs,\footnote{\change{The physical$+$dynamical analyes will be restricted to 696 of these sources that are not color outliers and lie within a specific range of $H_r.$}} and testing $K\in[1,50]$, the model robustly converges to a two-component model, independent of whether the color selection function is included in the training procedure.  Before proceeding, we wish to
determine if there are any objects whose colors are inconsistent with the model. Outliers would lead to a wider $\boldsymbol\Sigma_\alpha$ [see \eqq{eq:MstepAdaptive}]. To investigate whether there are any such outliers, we need to determine a likelihood that any particular TNO's measurements could have arisen from a member of a given component $\alpha$ of the mixture.  Our target TNO's measurement errors are defined by the samples 
$ \{\mathbf{c}_{j} \}, 1\le j \le N_{\rm samp},$ from its Markov chain.   We first define a metric
\begin{equation}
	\mathcal{L}_\alpha \equiv \frac{1}{N_{\rm samp}}\sum_j p_\alpha (\mathbf{c}_j | \mu_\alpha, \Sigma_\alpha),\label{eq:assignment}
\end{equation}
the mean probability of the color samples being drawn from the mixture component $\alpha$ in the absence of measurement errors.  We compare the source's $\mathcal{L}_\alpha$ value to the values that would be generated by objects drawn from the mixture.  To do this, we first create the color uncertainty distribution $\{\Delta \mathbf{c}_j\}\equiv \{\mathbf{c}_j -\bar{\mathbf{c}}\}$ of differences between the chain values and their mean color $\bar{\mathbf{c}}.$ Then we 
sample synthetic``truth'' colors $\tilde{\mathbf{c}}_k$ ($k=1,\ldots,N_{\rm sim})$ from $p_\alpha (\mathbf{c} | \mu_\alpha, \Sigma_\alpha,)$ and compute 
\begin{equation}
	\mathcal{L}_{k,\alpha} = \frac{1}{N_{\rm samp}} \sum_j p_\alpha(\tilde{\mathbf{c}}_k + \Delta\mathbf{c}_j | \mu_\alpha,\Sigma_\alpha)
\end{equation}
yielding a value of $\mathcal{L}$ that would be produced from a member of the mixture subject to measurement errors.
We can derive a $p$-value by measuring the fraction of the $N_{\rm sim}$ draws yielding $\mathcal{L}_{k,\alpha} \leq \mathcal{L}_\alpha$.  If indeed the measurement errors in color space were independent of the truth color of the source, then the $p$-value of objects drawn from mixture $\alpha$ would be uniformly distributed between 0 and 1.

\begin{figure}[ht!]
	\centering
	\includegraphics[width=\textwidth]{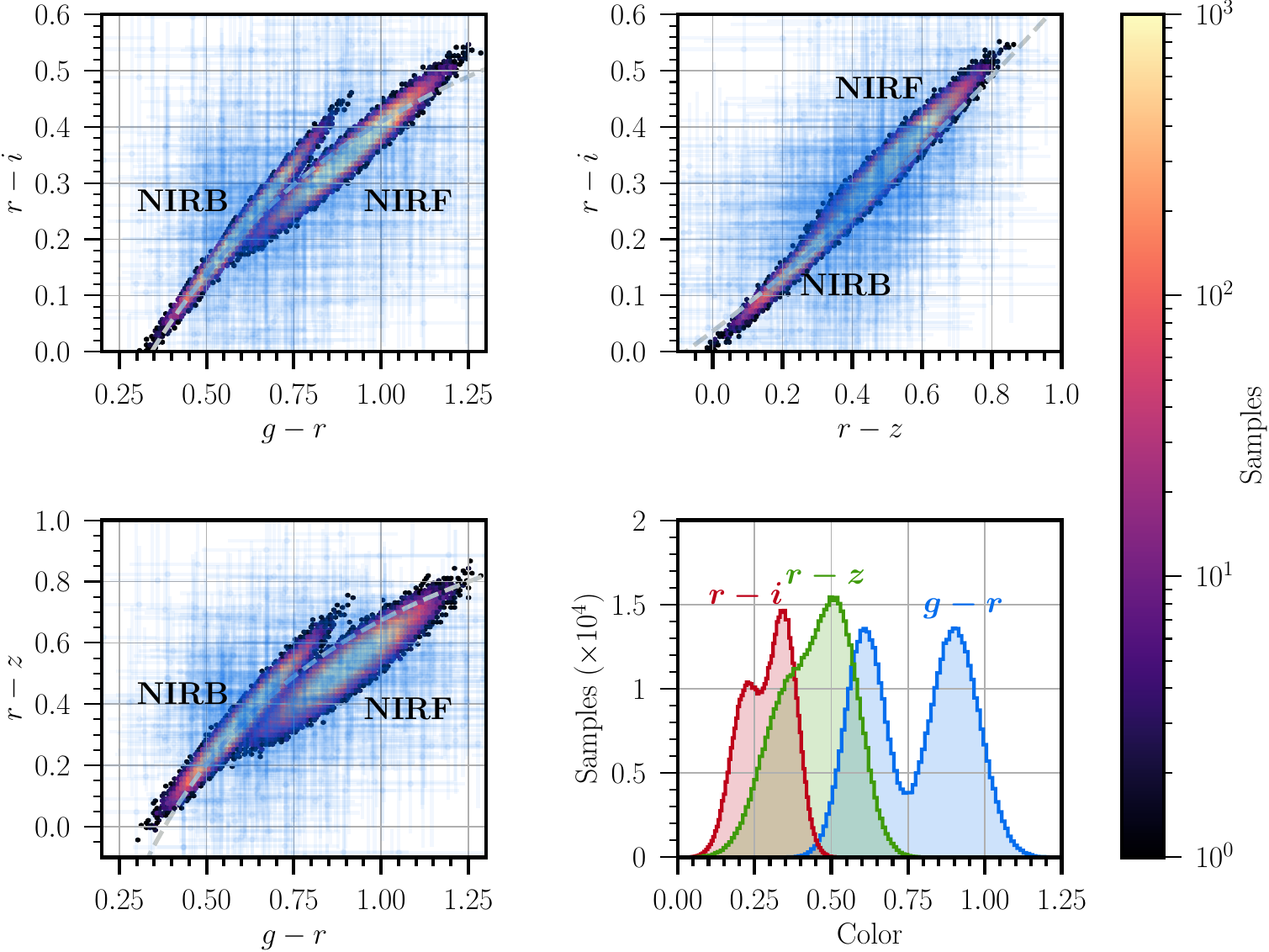}
	\caption{$g-r$, $r-i$ and $r-z$ histograms of $5 \times 10^5$ realizations of the model from Table \ref{tb:gmm}, as well as all \des\ TNO color measurements and their respective error bars (in blue). The two upper panels and the lower left panel show the two-dimensional projections of the samples, and the lower right panel the one dimensional histograms of each color. In the two-dimensional panels, the gray dashed line is the locus for sources with reflectance $\propto 1+\alpha(\lambda/550\,\text{nm})$ for varying $\alpha.$}
\label{im:model}
\end{figure}

We identify a first round of outliers as the 22 TNOs for which $p \leq 0.01$ for \emph{both} mixture components.   We then repeat the training procedure for the GMM parameters while excluding these outliers from the sample, keeping the number of components fixed, and ignoring the selection effects, which are seen to be unimportant.  This yields our best estimates of the GMM parameters describing the intrinsic color distribution, which are given in Table \ref{tb:gmm}. A visualization of this model is presented in Figure \ref{im:model}.

Note that the fractions $f_\alpha$ in the two components in the Table are given for the sample as a whole, and will depend upon how many members of each dynamical group are present in the training sample.  We derive debiased estimates of the intrinsic fractions for each dynamical class in Section \ref{sec:nirbnirf}.

\begin{deluxetable}{l|ccc}[ht!]
	\tabletypesize{\footnotesize}
	\tablecaption{Parameters for the best-fit GMM\label{tb:gmm}}
	\tablehead{\colhead{Component} &  \colhead{Mean $\mu$} &  \colhead{Covariance matrix $\Sigma$} & \colhead{Amplitude $f$}}
	\startdata
	NIRB & $\begin{pmatrix}0.609 &  0.223 & 0.355 \end{pmatrix}$ & $\begin{pmatrix} 0.00434 & 0.00337 & 0.00547 \\ 0.00337 & 0.00268 &  0.00437\\ 0.00547 &  0.00437 &  0.00727 \end{pmatrix}$  & $0.446$ \\ \\
	NIRF & $\begin{pmatrix} 0.902 &  0.347 &  0.524 \end{pmatrix}$ & $\begin{pmatrix} 0.00674 &  0.00362 &  0.00557 \\ 0.00362 &  0.00205 &  0.00321 \\ 0.00557 &  0.00321 & 0.00533 \end{pmatrix}$ & $0.554$ \\
	\enddata
\end{deluxetable}

There is a striking visual resemblance between our GMM components and the color classes proposed by \cite{fraser2023} (compare our Figure \ref{im:model} to their Figure 7). The two models were independently developed from  distinct datasets and with significantly different methodologies, strengthening the evidence that there are two distinct families of object surfaces in the trans-Neptunian region. Alluding to \cite{fraser2023}'s nomenclature, we call our components ``near-IR bright'' (NIRB), and ``near-IR faint'' (NIRF), according to whether the near-IR bands $i$ and $z$ are brighter or fainter at a fixed visible color $g-r$.

\subsection{Outliers}\label{sec:outliers}

If we assume that the $p$-values derived from $\mathcal{L}_\alpha$ have a uniform distribution, then the expected number of targets with ($p_\alpha < p_0$) for some threshold $p_0$ is $=N_\mathrm{TNO} p_0$ if all objects arose from component $\alpha.$  For our 814 objects, $p_0 = 0.001$ leads to $N_\mathrm{expected} < 1$. We have, however,  two independent components---if we assign each object $p={\rm max}(p_\alpha)$ from the two populations, then the expected number of sources with $p<p_0$ has a bound $\le N_\mathrm{TNO} p_0,$ because outliers from one component might fall in a high-probability region for the other component.   We can thus say with confidence that $\lesssim 1$ of the 814 TNOs should have $p<0.001$ if they all arise from one of the two components.

\begin{figure}[ht!]
	\centering
	\includegraphics[width=\textwidth]{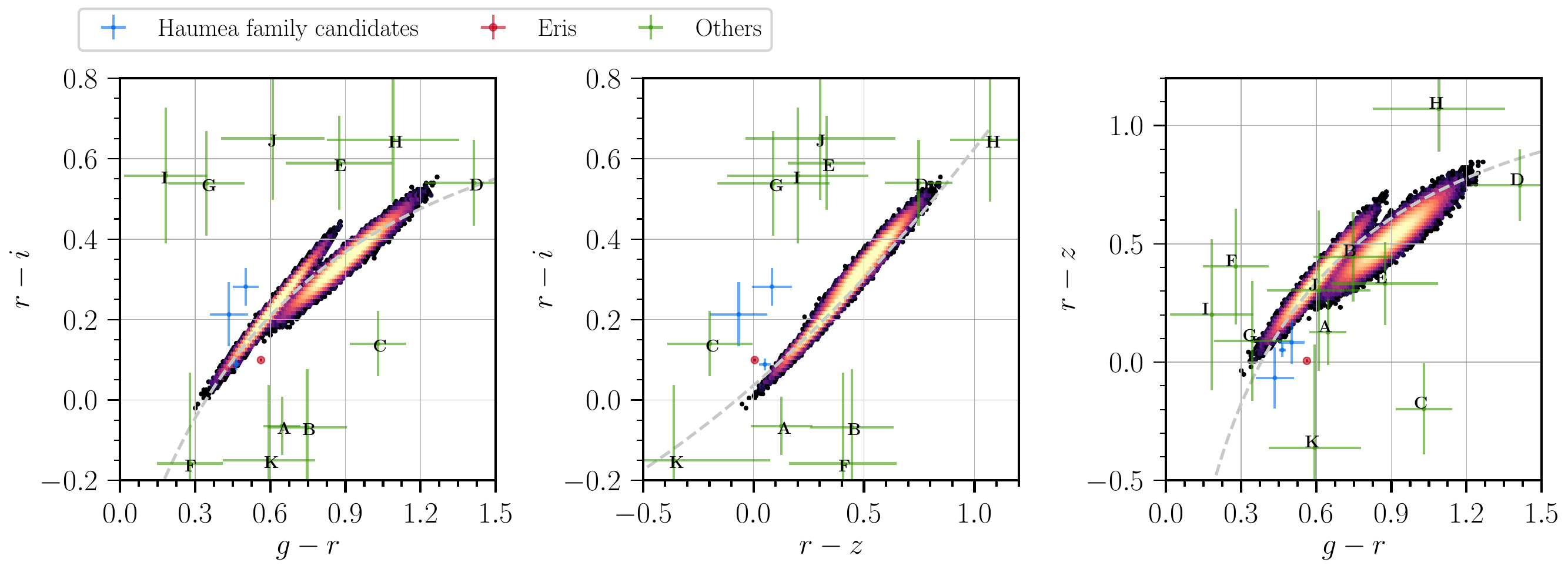}\\
	\vspace{10pt}
	\includegraphics[width=0.8\textwidth]{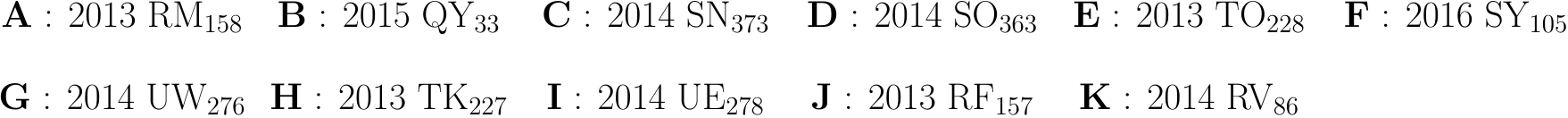}
	\caption{NIRB and NIRF realizations as in Figure \ref{im:model}, and, in green, the $g-r$, $r-i$ and $r-z$ colors and their respective uncertainties of the objects with $p_0 < 0.001$ for both components, and uncertainties $ < 0.5$ ($<0.2$) for the most (least) uncertain color. 
One of the outliers (in red) is Eris, and three of them in blue are potential Haumea family members. The other outliers are marked individually in the panels\label{im:outliers}}
\end{figure}

In fact we find that 22 TNOs (not the same set as were outliers in the first round!) have ${\rm max}(p_\alpha)<0.001,$ a statistically significant level of outliers from the GMM.
Figure~\ref{im:outliers} plots the colors of 15 outliers having at least one color measured with uncertainty $<0.2$~mag, and lists the others.

Are there any discernible characteristics for the outliers? 
Eris is one, with $p \leq 10^{-5},$ which might be expected, considering this is one of the largest dwarf planets, with different physical processes than the smaller bodies; furthermore it is the brightest object in the DES sample, leading to the smallest color uncertainties.

A group of outliers with similar colors are candidates for the Haumea collisional family: 2013 UQ$_{15}$ has been identified both dynamically \citep{Pike2019} through its colors \citep{Schwamb2019} as a member of this family. Two other objects have similar colors: the orbital elements of 2010 RF$_{64}$ place it in the region of high probability for membership in the parameter space of \cite{proudfoot2019}, as confirmed by further dynamical studies by \cite{proudfoot2024b}.  \change{2014 UQ$_{277},$ while measurably redder in $r-i$ than Haumea and our other two candidates, is possibly in} the 7:4 resonance, a potential dynamical state for escaped members of the Haumea family \citep{ragozzine2007, proudfoot_ragozzine}. These objects are of dynamical interest, and motivate further study. 

The other 18 $p<0.001$ outliers are scattered around the color space with no particular distinguishing dynamical features, except that only one of them is a CC. Further observations would determine whether these objects are real outliers, indicative of other surface classes in the trans-Neptunian region, or if these identifications are the artifacts of poor photometric measurements from the DES images.


\subsection{Probabilistic color class assignment}
We can also use the summed probability in Equation \ref{eq:assignment} for both the NIRB and NIRF components to probabilistically assign objects to each of those classes. We define the (log) probability ratio $R_{\rm NIRF} \equiv \log \mathcal{L}_\nirf - \log \mathcal{L}_\nirb$ to discriminate between these two color classes. $R_{\rm NIRF}>0$ ($R_{\rm NIRF}<0$) indicates a tendency for the object to belong to the NIRF (NIRB) component, with larger (smaller, that is, more negative) values indicating a stronger confidence in this assignment. 

\begin{figure}[ht!]
	\centering
	\includegraphics[width=\textwidth]{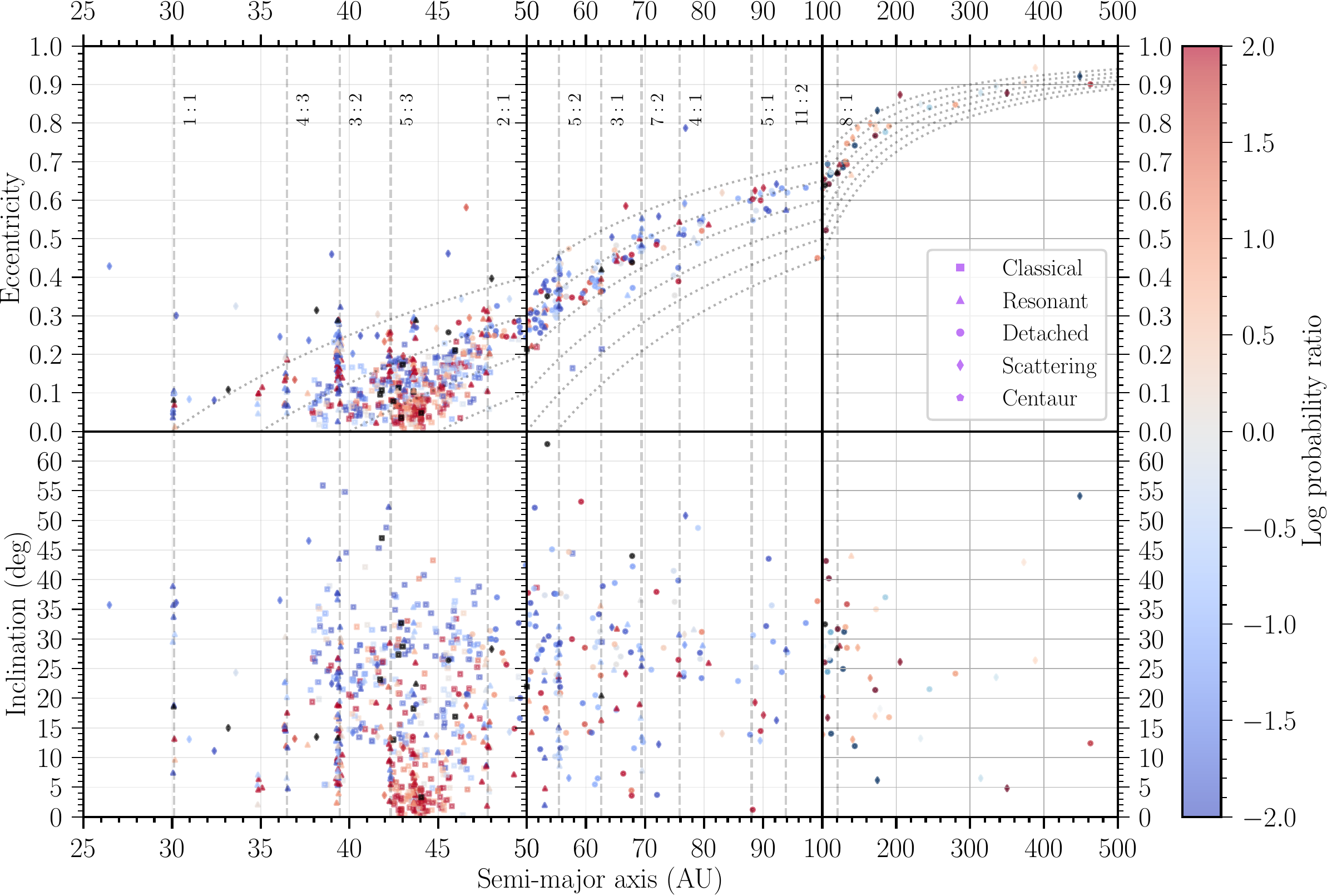}
	\caption{Semi-major axes, eccentricities and inclinations for all objects in the DES catalog. The shape of each marker indicates the dynamical classification of the object, and the color scale represents the probabilistic color assignment - objects with $|R_c| > 2$ are securely assigned to their respective classes. The outliers of Section \ref{sec:outliers} are marked in black. Also shown are the location of major Neptunian mean motion resonances (dashed lines), and lines of constant perihelion (dotted lines).}
\label{im:aeicolor}
\end{figure}

The semi-major axes, eccentricities and inclinations of all objects in the sample are presented in Figure \ref{im:aeicolor}, color-coded by their log probability ratio $R_c$. A few trends can be immediately seen from this figure: the cold Classical region ($42 \lesssim a \lesssim 48 \, \mathrm{au}$, $I < 5\degr$, $e < 0.2$) is composed almost exclusively of NIRF objects, and the majority of NIRF objects are found at lower inclinations in the other dynamical populations as well. The dynamically excited classes (hot Classicals; resonant, scattering and detached objects) appear to be NIRB dominated. Of the 10 Neptune Trojans in the DES sample, 4 are securely NIRB ($R_\nirb>2$)and 2 are securely NIRF ($R_\nirf > 2$).
This confirms the findings of \cite{Lin2019}, \cite{markwardt2023} and \cite{bolin2023} that the absence of redder objects in this population seen by \cite{Jewitt2018} was due to small sample size. 

In the next Section, we turn these qualitative impressions into quantitative measurements of the intrinsic contributions of each color family to each dynamical class, while accounting for different observational biases in each class as well as potentially distinct absolute magnitude distributions that depend on color and dynamics.

\section{Likelihoods for color/family models}\label{sec:framework}
\subsection{Mathematical framework}
Consider the TNO population to be modelled as the union of subpopulations indexed by $\beta$.  Each TNO has characteristics in the set $\{\vecc, H_r, A, \vecP\}$ where \vecc\ is an array of colors, $H_r$ is the absolute magnitude in some reference color (assumed throughout this paper to be in $r$ band), \vecP\ are the six orbital elements $\{a,e,i,\Omega,\omega,\anomaly\}$, and $A$ is the lightcurve amplitude.  We posit that the full distribution of \change{TNOs} is a Poisson draw from a distribution over all of these parameters
\begin{equation}
	N(\vecc, H, A, \vecP) \equiv \sum_\beta N_\beta p_\beta(\vecc, H, A, \vecP) . \label{eq:model}
\end{equation}
We will let $\mathcal{S}$ be a boolean variable indicating successful selection of an object in the survey.  If an object is successfully selected, our catalog gains an entry $D_i$ containing the observation data for the object.  The observational data consist of positional data that strongly constrain the orbital parameters $\vecP_i,$ plus a flux for each exposure of the target.

We have a simulator that can give the odds of detecting and selecting a TNO of given truth features
\begin{equation}
	p(\mathcal{S} | \vecc, H, A, \vecP).
\end{equation}
Notice that this selection probability depends primarily upon which exposures the source hits---determined by \vecP---and upon the apparent magnitude $m_r\approx H_r-10\log_{10}{\Delta},$ where the distance $\Delta$ to the source is another function of \vecP.  There is a weaker dependence of $p(\mathcal{S})$ on the colors and lightcurve amplitude (at a given phase): Appendix \ref{sec:completeness} describes how this function can be estimated.

We also have an equation that gives the likelihood of obtaining the data $D_i$ as a function of truth values. Remembering that the existence of $D_i$ implies there has been a successful selection $\mathcal{S}_i,$ the probability of obtaining data $D_i$ from a member of the underlying population is
\begin{equation}
	\likeli(D_i | \vecc, H, A, \vecP) =  \likeli(D_i, \mathcal{S}_i | \vecc, H, A, \vecP) =  \likeli(D_i | \mathcal{S}_i, \vecc, H, A, \vecP)  p(\mathcal{S}_i | \vecc, H, A, \vecP).
\end{equation}

The likelihood of a given catalog $D=\{D_i\}$ being observed is then given by the Poisson formula for the TNO truth values, convolved with the likelihood of the observed data in terms of the truth values:
\begin{align}
	\likeli\left(D | \{N_\beta, \boldsymbol{q}_\beta \}\right) & = \prod_i \left[ \sum_\beta N_\beta \int\dcdhdpdA \likeli(D_i | \vecc, H, A, \vecP) p_\beta(\vecc, H, A, \vecP|\boldsymbol{q}_\beta) \right] \nonumber \\
	                                     & \quad \times \exp\left[-\sum_\beta N_\beta \int\dcdhdpdA p(\mathcal{S}  | \vecc, H, A, \vecP)  p_\beta(\vecc, H, A, \vecP|\boldsymbol{q}_\beta) \right].
	\label{eq:LDN}
\end{align}
We've written this as conditioned on the subpopulation parent abundances $N_\beta,$ and other hyperparameters $\boldsymbol{q}_\beta$ of the subpopulations' $p_\beta(\vecc,H, A, \vecP)$ distributions.

The MCMC procedure used in \cite{bernardinelli2023} converts the observed $D_i,$ which are a time series of observed fluxes for the source, into $N_{\rm samp}$ samples $\hat D_{ij} = \{\vecc_{ij}, H_{ij}, A_{ij}\}$ of color, $H_r$, and LCA for object $i$
that approximate the posterior distribution of these quantities assuming a uniform prior $u(\bar f_g, \bar f_r, \bar f_i, \bar f_z, A)$ over the time-averaged, distance-adjusted mean fluxes $\bar f_b$ in each band.  Since we are working with magnitudes such as $H_r=\text{const}-2.5\log_{10}\bar f_r$ and $df/dH\propto 10^{-0.4H},$ the uniform prior in fluxes is equivalent to the prior
\begin{equation}
{\rm Pr}(\hat D) \equiv  {\rm Pr}(\vecc, H,A) \propto \prod_{b\in g,r,i,z}10^{-0.4H_b}.
\end{equation}
This means that the $\hat D_{ij}$ have been drawn from the distribution
\begin{align}
	\label{eq:sampD}
  \hat D_{ij}& = \{\vecc_{ij}, H_{ij},A_{ij}\}  \sim \frac{1}{E(D_i)} \likeli(D_i | \hat D) {\rm Pr}(\hat D), \\
  E(D_i) & \equiv \int d\hat D\, \likeli(D_i | \hat D) {\rm Pr}(\hat D).
\end{align}
The normalizing quantity $E(D_i)$ is the evidence of the observations $D_i$ under the assumed uniform-flux prior.


It is also true that the astrometric observations have determined the orbital parameters essentially perfectly for our current purposes, so that the $\vecP$ dependence of $\likeli(D_i|\vecc, H, A,\vecP)$ separates into $\delta(\vecP-\vecP_i)$.
 Given (\ref{eq:sampD}) and this delta function, the integral in the first term of the right-hand side of \eqq{eq:LDN} can be approximated using importance sampling of the $\hat D_{ij}$ as
\begin{equation}
	\int\dcdhdpdA \likeli(D_i | \vecc, H, A, \vecP ) p_\beta(\vecc, H, A, \vecP | \boldsymbol{q}_\beta)
	\approx
	\frac{E(D_i)} {N_{\rm samp}}
	\sum_j \frac{p_\beta(\vecc_{ij},H_{ij},\vecP_i, A_{ij} | \boldsymbol{q}_\beta)}{ {\rm Pr}(\vecc_{ij},H_{ij})}
\end{equation}
The factor to the left of the sum is determined by the measurement errors on the fluxes, which are independent of the subsample $\beta$ or any of its parameters $\boldsymbol{q}_\beta,$ leading to an irrelevant constant multiplicative scaling of the final likelihood in \eqq{eq:LDN}.  
The likelihood of any given model can therefore be expressed as
\begin{subequations}
\begin{align}
	\log \likeli( \{N_\beta, \boldsymbol{q}_\beta\} | D) & = \sum_i \log\left[\sum_\beta N_\beta R_{i\beta}\right] - \sum_\beta N_\beta S_\beta + \log p(\{N_\beta,\boldsymbol{q}_\beta\}) +  (\textrm{const}) \\ 
	R_{i\beta}                                    & \equiv \sum_j \left[ p_\beta(\vecc_{ij},H_{ij},\vecP_i, A_{ij} | \boldsymbol{q}_\beta)  \prod_{b\in g,r,i,z} 10^{0.4(H_{ij}+c_{br,ij})} \right] \\
	S_\beta                                       & \equiv \int\dcdhdpdA p(\mathcal{S} | \vecc, H, A, \vecP )  p_\beta(\vecc, H, A, \vecP | \boldsymbol{q}_\beta).
\label{eq:Sbeta}
\end{align}
\end{subequations}

Similar to the GMM procedure, we can reparametrize this function by defining $N \equiv \sum_\beta N_\beta$ and $f_\beta \equiv N_\beta/N$, so that $0 \leq f_\beta \leq 1$ and $\sum_\beta f_\beta = 1$. The $f_\beta$ variables are thus constrained to lie on a simplex, so there is total number of free parameters is unchanged.  The model becomes
\begin{equation}
	N(\vecc, H, A, \vecP | \{\boldsymbol{q}_\beta\} ) = N \sum_\beta f_\beta p_\beta(\vecc, H, A, \vecP | \boldsymbol{q}_\beta),
\end{equation}
where, with no loss of generality, we posit that
\begin{equation}
	\int\dcdhdpdA  p_\beta(\vecc, H, A, \vecP | \boldsymbol{q}_\beta) = 1.
\end{equation}

The likelihood becomes
\begin{equation}
	\log \likeli (\{N, f_\beta, \boldsymbol{q}_\beta \} | D) = (\textrm{const}) + \log p (\{N, f_\beta, \boldsymbol{q}_\beta \}) + \sum_i \log \left[N \sum_\beta f_\beta R_{i\beta}\right] - N \sum_\beta f_\beta S_\beta.
\label{eq:Npoisson}
\end{equation}

We can solve for $N$ analytically, under the assumption of a uniform prior distribution, as follows. The maximum likelihood is found where $\partial_N \log \mathcal{L} = 0$:
\begin{equation} \frac{\partial \log\mathcal{L}}{\partial N} = \frac{N_\mathrm{obj}}{N} - \sum_\beta f_\beta S_\beta = 0 \implies N = \frac{N_\mathrm{obj}}{\sum_\beta f_\beta S_\beta}. \label{eq:numbers}
\end{equation}
Substituting back in the likelihood, we have that
\begin{equation}
	\label{eq:loglike}
	\log \mathcal{L} (\{f_\beta,\boldsymbol{q}_\beta\} | D) =\\ (\mathrm{const}) + \log p(\{f_\beta, \boldsymbol{q}_\beta\}) + \sum_i \log \left[ \sum_\beta f_\beta R_{i\beta} \right] - N_\mathrm{obj} \log\left[ \sum_\beta f_\beta S_\beta \right].
\end{equation}

\subsection{Choice of functional forms and interpreting $f_\beta$}
To this point, we have made no strong restrictions as to what defines membership in the families $\beta$ of TNOs that comprise the mixture model defined in \eqq{eq:model}: $\beta$ could represent TNOs with some particular physical characteristics, such as colors (as in Section \ref{sec:colormodel}, where $\beta \rightarrow \alpha$), or a dynamical state $d$ of the object \citep[\eg\ as in][]{Gladman2008}, or a joint physical+dynamical distribution with subpopulations enumerated by a pair $\beta=(\alpha,d).$ 
In the subsequent analyses, we will assume that, within a family $\beta$, the orbital parameters, absolute magnitudes, colors and lightcurve amplitude functional forms are separable, that is,
\begin{equation}
	p_\beta(\vecc, H_r, A, \vecP) = p_\beta(\vecc) p_\beta(H_r) p_\beta(\vecP) p_\beta(A).
\end{equation}
This means that we can choose a functional form for each underlying distribution, and maximize the likelihood in Equation \ref{eq:loglike} to find the most likely values of the parameters $\mathbf{q}_\beta$ of the $p_\beta$ functions. Note that we require our distributions to be normalized.

We initially fix the color functions $p_\beta(\vecc)$ to be one of the Gaussians derived by the GMM, \ie\ defined in \eqq{eq:multivariate} with the parameters given in Table~\ref{tb:gmm} for $\alpha\in\{\text{NIRF,NIRB}\}.$  This naturally incorporates probabilistic color classification into our dissection of the full TNO population.
The GMM parameters will \emph{not} be maximized with the rest of the likelihood, although we will in Section~\ref{sec:split} explore cases where we subdivide each GMM component to create $>2$ possible color subpopulations.

For $p_\beta(H)$ distributions, we adopt a rolling power law (as in \citealt{Bernstein2004}), which adds curvature to a simpler power-law distribution:
\begin{equation}
	p_\beta(H_r | \theta_\beta, \theta'_\beta) \propto 10^{\theta_\beta (H_r - 7) + \theta'_\beta (H_r - 7)^2}.
\label{eq:rolling}
\end{equation}
We will restrict all of our further analyses to TNOs with $5.5<H_r<8.2,$ assume all $H$ values are in the $r$ band, and normalize $p_\beta(H)$ to unit integral over this range.  In this case, a numerical normalization is more convenient, although an analytic normalization is possible.
Some care is needed when comparing population counts or fractions $f_\beta$ that have been derived for different ranges of $H_r$.

For $p_\beta(A),$ we adopt the $\beta$-distribution models of \citet{bernardinelli2023}.
\begin{equation}
	p_\beta(A | \bar{A}_\beta, s_\beta) = A^{\bar{A}_\beta 10^{s_\beta} - 1}(1-A)^{(1-\bar{A}_\beta) 10^{s_\beta} - 1} \times \frac{\Gamma \left(10^{s_\beta}\right)}{\Gamma\left(\bar{A}_\beta 10^{s_\beta}\right) \Gamma \left((1-\bar{A}_\beta) 10^{s_\beta}\right)}, \label{eq:betadist}
\end{equation}
where $\bar{A}$ is the mean lightcurve amplitude, and $s$ is a sharpness parameter. At fixed $\bar{A}$, a larger $s$ implies a narrower distribution about $\bar A.$ 
As shown in~\cite{Bernardinelli2022}, for a fixed apparent magnitude $m_r$, the completeness function is virtually independent of $A$, so $p(A|\mathcal{S}) \approx \, \mathrm{const}$. This means that, while computing the $S_\beta$ terms, the $p_\beta(A)$ integrals always evaluate to 1, simplifying the likelihood calculation.

 The classification of TNOs into dynamical classes indexed by $d$ is incorporated by having each $\beta$ be associated with one $d$
using an indicator function
\begin{equation}
	p_\beta(\vecP) = \begin{cases} 1, & \text{if }\vecP\text{ implies the object belongs to dynamical class } d \\ 0 & \text{otherwise}  \end{cases}. \label{eq:indicator}
\end{equation}

Our analyses will define subpopulations by a combination of a physical family $\alpha$ and a dynamical family $d$, \ie\ $\beta=(\alpha,d),$ so we will be using \eqq{eq:loglike} to constrain population fractions $f_{\alpha d}$ with data from \emph{all} of our objects (in our chosen $H_r$ range) simultaneously, as well as constrain the hyperparameters $\boldsymbol{q}_\beta$ constraining size and lightcurve distributions.
This allows us to test whether distinct dynamical families share their physical parameters.

We can also define the fraction of a dynamical class $d$ that is made of a physical class $\alpha$ as
\begin{equation}
  f_{\alpha | d} \equiv \frac{f_{\alpha d}}{\sum_{\alpha^\prime} f_{\alpha^\prime d}}.
\end{equation}
The $f_{\alpha | d}$ are independent of the relative contribution of class $d$ to the overall model in Equation \ref{im:model},
so $f_{\alpha |d}$ and $f_{\alpha | d^\prime}$ can be meaningfully compared to each other.
In other words, our approach of leaving one overall $N$ for the full population and a single simplex constraint $\sum_{\alpha d} f_{\alpha d}=1$ is mathematically equivalent to having a free number density $N_d$ for each dynamical class, and distinct simplex constraint $\sum_\alpha f_{\alpha | d}=1$ for each.

Other investigators have reported ``X-to-Y'' ratio of physical properties (for example, neutral-to-red) within a dynamical class $d$.  The results of our MCMC chains for the $f$ values can be converted to such an estimate in some range $H_1<H_r<H_2$ for class $d$ as
\begin{equation}
	R = \frac{f_{Xd} \int_{H_1}^{H_2} \mathrm{d}H \, p_{X} (H)}{f_{Yd} \int_{H_1}^{H_2} \mathrm{d}H \, p_{Y} (H)}.
\end{equation}

\section{The NIRB and NIRF physical parameter distributions}
\label{sec:nirbnirf}
In this section, we will assess decompositions of the members of various dynamical classes into two physical components, one with color distribution $p_\beta(\vecc)$ following the NIRB component of the GMM, the other drawn from NIRF colors. The hyperparameters describing the $H$ and $A$ distributions for some component $\beta$ are $\boldsymbol{q}_\beta = \{ \theta_\beta, \theta^\prime_\beta, \bar A_\beta, s_\beta\}.$
Our inference will proceed as follows:
\begin{enumerate}
\item Evaluate the log-likelihood of \eqq{eq:loglike} for each dynamical class independently, yielding the posterior probability distribution $p_d(f_{\nirb|d}, \boldsymbol{q}_{\nirb,d}, \boldsymbol{q}_{\nirf,d} | D_d)$ of the NIRB fraction and the NIRB/NIRF size and variability distributions within each $d.$
\item Test the hypothesis that all $d$ share a common set of parameters, $\boldsymbol{q}_{\nirf, d} = \boldsymbol{q}_{\nirf}$ and $\boldsymbol{q}_{\nirb, d} = \boldsymbol{q}_{\nirb},$
against the hypothesis that there are distinct $\mathbf{q}$'s for each dynamical class.  If the shared-value hypothesis is favored, or at least not disfavored, then the data allow for each dynamical class to be composed of its own mixture of just two global physical populations, $\beta \in \{\nirb,\nirf\}).$
\item We can refit the full sample simultaneously to this model of shared parameters across distributions to obtain $p( \{f_{\beta,d}\}, \boldsymbol{q}_{\nirb}, \boldsymbol{q}_{\nirf}, | D).$
\item Marginalizing over the fractions $f_{\beta,d}$ yields the allowed physical hyperparameters for the NIRB and NIRF populations.
\item Marginalizing over the physical parameters yields estimates of the fractions $f_{\nirb|d}$ of each population that is in NIRB, and also of the total number of sources in our chosen $H$ range in each dynamical component.
\end{enumerate}

Our analysis uses only DES TNOs with $5.5 < H_r < 8.2,$, and we will exclude all outliers from Section \ref{sec:outliers} (of which 16 are in this size range), which leaves us with 696 objects. The left column of 
Table~\ref{tb:sample} lists the 5 distinct dynamical classes $d$ into which we divide these.  The right column gives a subdivision of the Resonant class members for investigation of their dependence vs semi-major axis.

\begin{deluxetable}{ccc|ccc}[ht!]
	\tablewidth{0pt}
        \tablecolumns{6}
	\tablecaption{Sample definitions \label{tb:sample}}
	\tabletypesize{\small}
        \tablehead{\multicolumn{3}{c}{Comprehensive subsets} & \multicolumn{3}{c}{Resonant subsets}  \\
          \colhead{Subset} & \colhead{\# objects} & \colhead{$f_\nirb$} & \colhead{Subset} & \colhead{\# objects} & \colhead{$f_\nirb$} }
	\startdata
	Cold Classical (CC) & 102 & $0.019 \pm 0.015$     & Neptunian Trojan (1:1) & 6  & $0.682 \pm 0.156 $    \\
        Hot Classical (HC) & 254  & $0.631 \pm 0.034 $   & Inner belt (5:4 \& 4:3) & 13  & $0.551 \pm 0.139$   \\
        Scattering & 33   & $0.686 \pm 0.090$                  & 3:2 & 49 & $0.559\pm0.075$  \\ 
	Resonant & 172  & \nodata                                           & Main belt (5:3 \& 7:4) & 42  & $0.279 \pm 0.075$  \\
        Detached & 135 & $0.710\pm0.045$                     &   2:1 & 18 & $0.408 \pm 0.126 $ \\
                       &       &                                                    & 5:2 & 12 & $0.792 \pm 0.110$ \\ 
	 \textbf{Total} & \textbf{696} &                            &   Distant ($a>50$ au, excluding 5:2) &  32 & $0.614 \pm 0.091$ \\ 
	\enddata
	\tablecomments{The dynamical definitions follow the scheme of \cite{Gladman2008}, and the Classicals are subdivided by their free inclinations following \cite{Huang2022} - objects with $I_\mathrm{free} \leq 5\degr$ are considered Cold Classicals, objects with higher free inclinations are Hot Classicals. A complete description of the dynamical classification procedure is presented in \cite{Bernardinelli2022}. The left column lists the 5 distinct dynamical classes which span the entire set of 696 DES TNOs with $5.5<H_r<8.2.$  The right column lists some subsets of the resonant class, which will be compared in Section \ref{sec:fractions}.  The mean and standard deviation of the fraction of the dynamical class coming from the NIRB physical class is given, assuming common $p(H)$ slopes across all dynamical classes as in Section~\ref{sec:plaw}. We do not provide $f_\nirb$ for the Resonant objects because the different resonances have significantly distinct values.}
\end{deluxetable}

\subsection{Do dynamical distributions share physical hyperparameters?}
\label{sec:plaw}
\begin{figure}[ht!]
	\centering
	\includegraphics[width=0.49\textwidth]{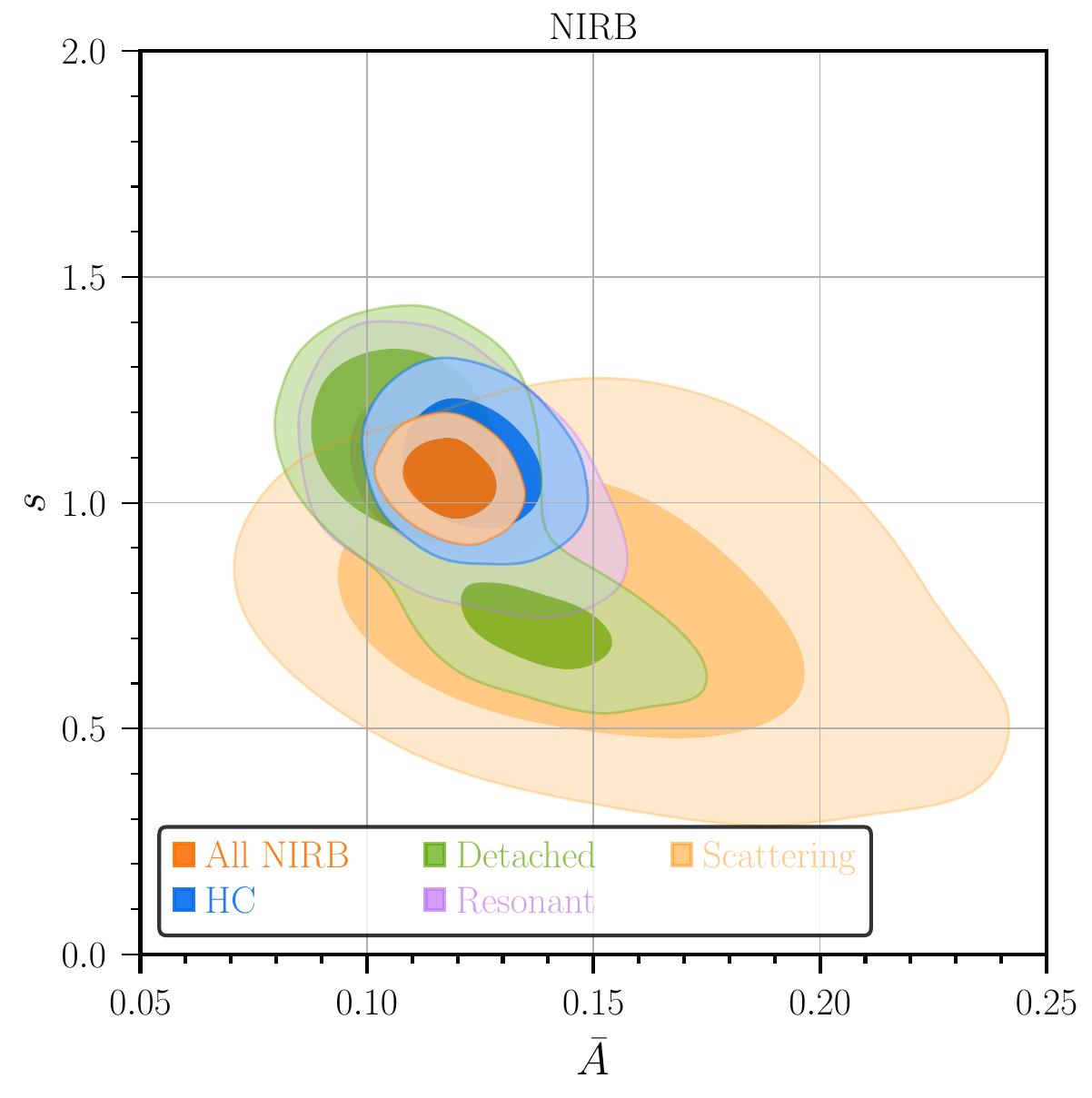}
	\includegraphics[width=0.49\textwidth]{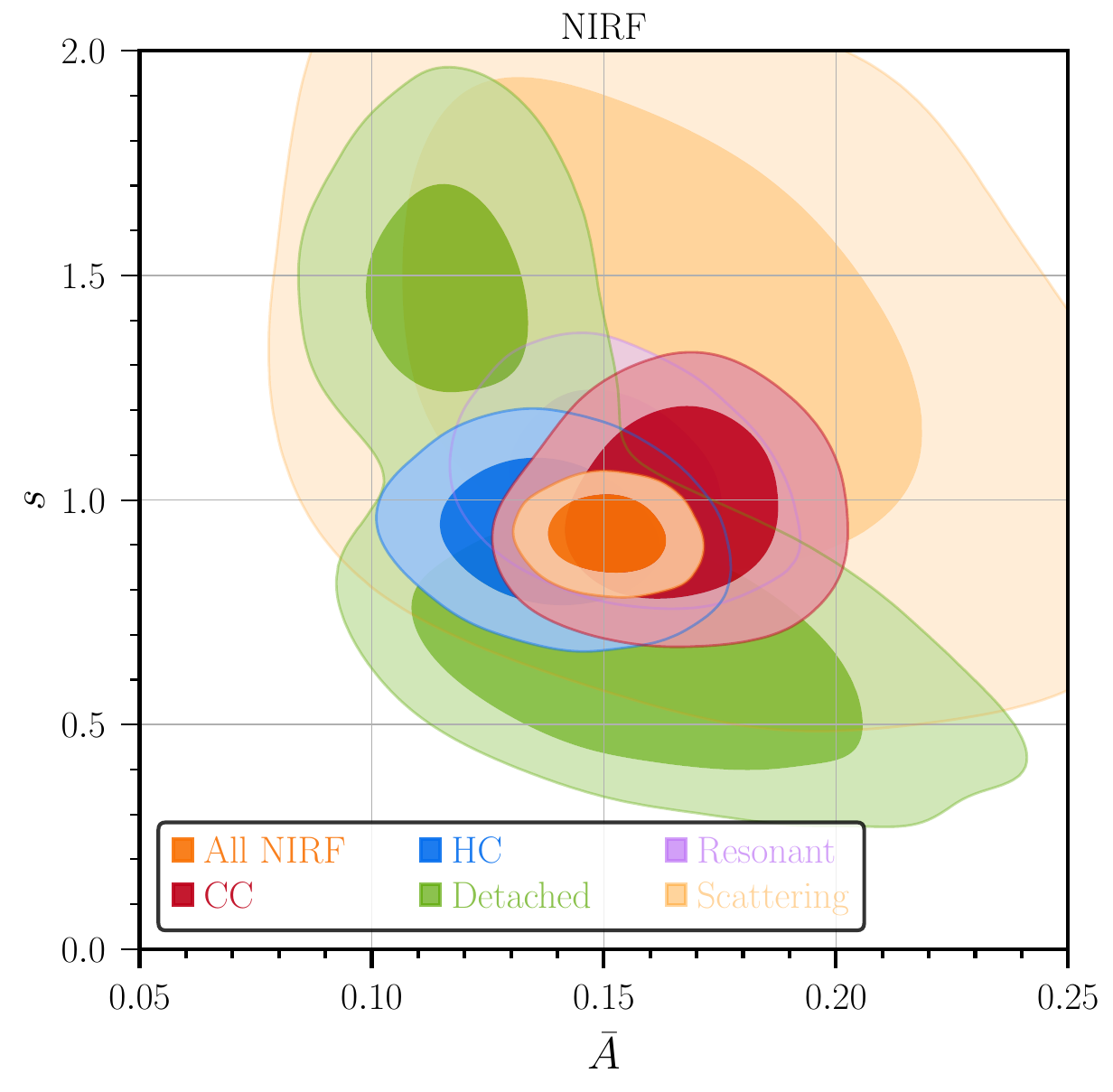}

	\caption{68\% and 95\% posterior contours for the HMC-derived samples for the NIRB (left) and NIRF (right) parameters $\bar{A}$ and $s$ parameters of the LCA distribution. Each figure contains the posterior for all disjoint sets of objects from the distinct dynamical classes, and the posterior derived from fitting all the subsets simultaneously \change{(the smallest, orange contours).} Note there is no cold-classical constraint on the NIRB component because the CC's are purely NIRF.\label{im:lcadynamics}}
\end{figure}

The left panel of Figure~\ref{im:lcadynamics} shows the posterior distributions of the NIRB $p(A)$ distribution parameters $\bar{A}_{\nirb,d}$ and $s_{\nirb,d}$, computed for each individual dynamical class $d$, and the distribution of $\bar{A}_{\nirb}$ and $s_{\nirb}$ if all classes are required to share the share common parameters.  The right panel shows the same thing for the NIRF class.
The means of these posterior distributions satisfy $\bar{A}_\nirf > \bar{A}_\nirb$ in every case; and all of the classes' 68\% credible regions overlap with the common model.  \change{The variability amplitude of the NIRF TNOs is significantly higher than that of the collected NIRB TNOs, which remains true even when CC's are removed from the NIRF population, suggesting that this is a birth characteristic of the population rather than a consequence of migration history.} 

When producing these posterior distributions for a single dynamical class $d$ in isolation, there are 9 free parameters: 2 $p(A)$ parameters $(\bar A, s),$ plus two $p(H)$ parameters $(\theta, \theta^\prime)$ for each of the NIRB and NIRF components, plus the contribution fraction $f_{\nirb|d}$.  In the case of shared distributions, there are 8 free physical parameters, plus 9 degrees of freedom among the $2\times5$ simplex-constrained fractions $f_\beta=f_{\alpha,d}$. This is too many dimensions to evaluate the likelihood over a comprehensive grid; instead, we sample the posterior distribution using a Hamiltonian Monte Carlo (HMC) chain. The HMC generates less-correlated samples with a higher acceptance rate than the simpler Metropolis-Hastings Monte Carlo approach, and readily adapts to simplex constraint(s) on the $f_\beta$ in an arbitrary number of dimensions. We present a short introduction to HMC in Appendix \ref{sec:hmc}.

For a quantitative test of whether all $d$ share the same $p(A)$ within a color class,  we can compute the log Bayesian evidence ratio $\mathcal{R}$ between hypothesis $\mathcal{H}_1$ (one shared $p(A)$ distribution for all NIRB objects, and one for all NIRF objects), and hypothesis $\mathcal{H}_2$ (each subpopulation has its own values of the $\bar A_\alpha$ and $s_\alpha$ parameters).  \change{Appendix~\ref{sec:evidence} reviews the definition of the evidence ratio and how we measure it from the MCMC chain outputs.  In the \cite{Jeffreys} scale, evidence ratios of $\mathcal{R} > 2.30$, $3.45$, $4.61$ are said to be ``strong'', ``very strong'' and ``decisive'' in favor of $\mathcal{H}_2$. As this is a logarithmic scale, negative values of $\mathcal{R}$ imply that $\mathcal{H}_1$ is preferred over $\mathcal{H}_2$, and its significance is measured on the same absolute scale.}

The resultant log evidence ratios for a shared set of $p(A)$ parameters are 5.016 and 6.143 for the NIRB and NIRF components, both showing \emph{decisive} evidence in favor of the shared-parameter models. Evidence ratios are sensitive to the priors assumed on the parameters. We have taken uniform priors for $\bar{A}_{\nirf,\nirb} \in [0.05, 0.25]$ and $s_{\nirf,\nirb} \in [0.4,1.9]$, and the evidence is decisively in favor of the shared model for any reasonable choice of prior. 

\change{We also test whether the NIRF $p(A)$ is driven by the CC population. The resulting posteriors for the non-CC NIRF distribution are near identical to the case when the CCs are included, and we have that the evidence ratio for a scenario where the non-CC NIRF population is distinct (and therefore more variable) than the NIRB TNOs is $\mathcal{R} = 4.22$, showing very strong evidence for two distinct populations.}

\begin{figure}[ht!]
	\centering
	\includegraphics[width=0.49\textwidth]{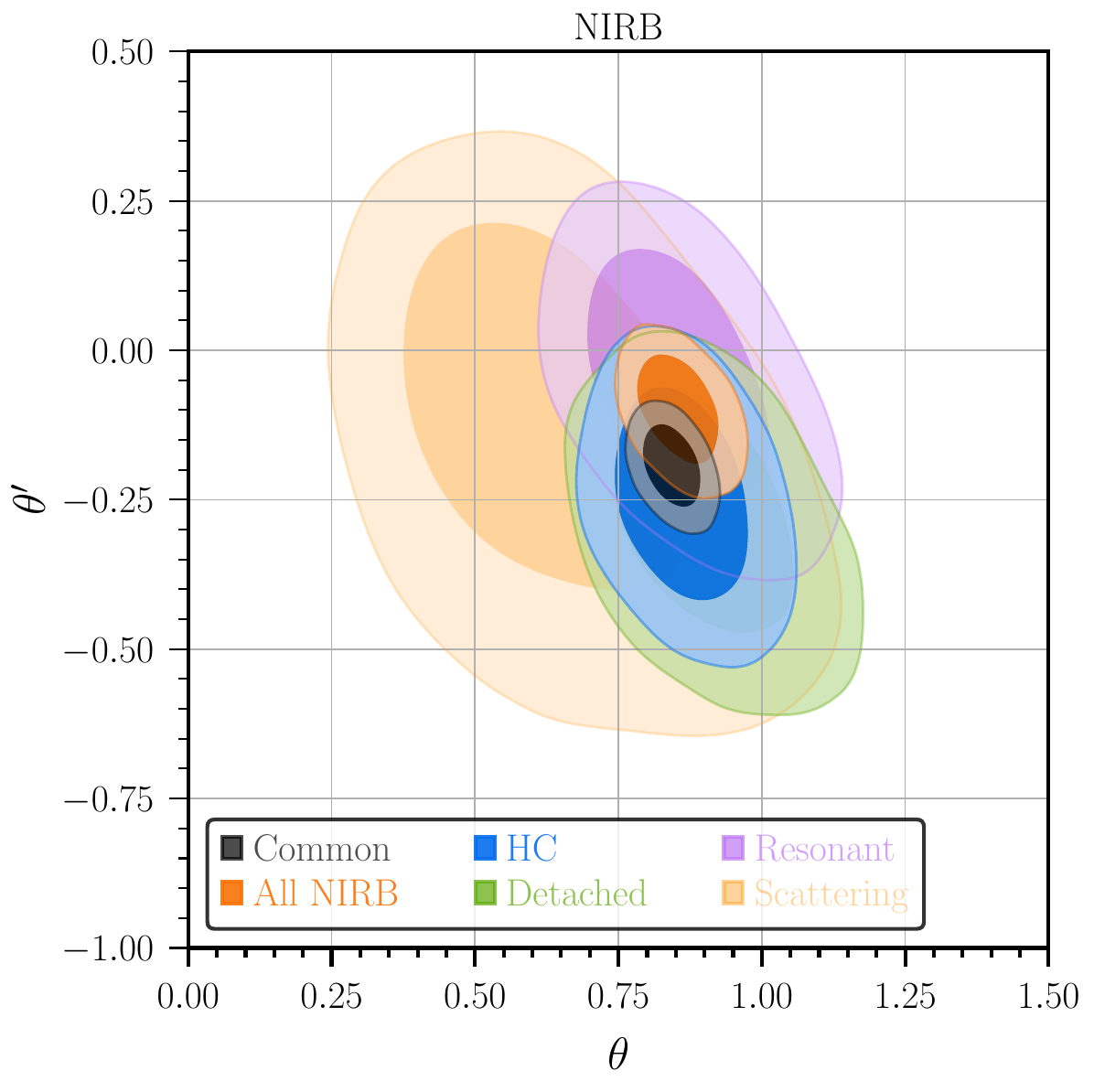}
	\includegraphics[width=0.49\textwidth]{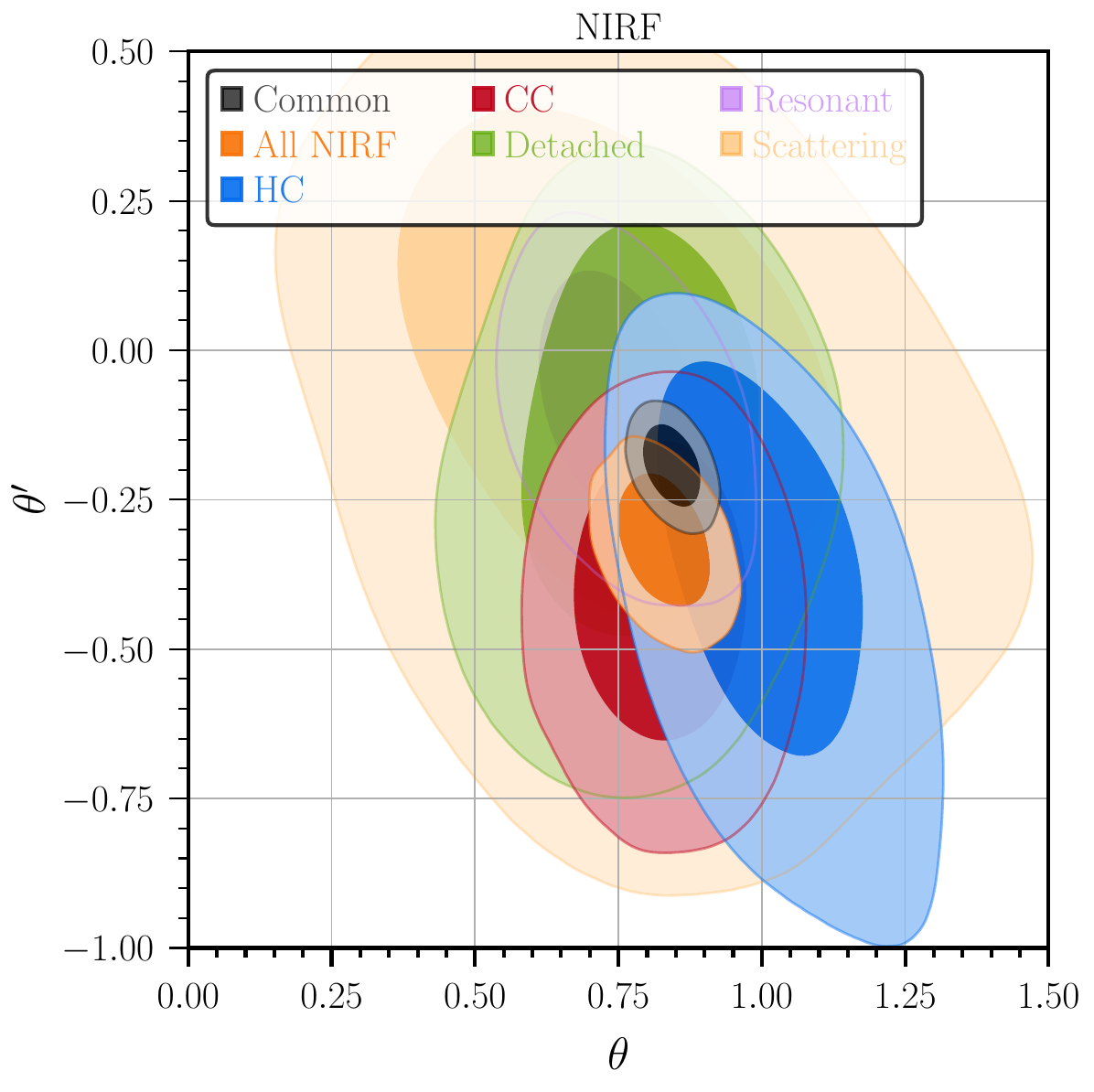}

	\caption{Similar to Figure \ref{im:lcadynamics}, this figure shows posterior contours for the NIRB (left) and NIRF (right) $(\theta,\theta')$ parameters of the rolling power law $H_r$ distribution. In addition to the disjoint sets of distinct dynamical classes, the posteriors in orange represent a single rolling power law for all members of either the NIRB or the NIRF class, and the black posterior shows a common distribution for all samples, independently of their color families.}
\label{im:rolling}
\end{figure}
We can similarly turn our attention to the parameters $\theta$ and $\theta^\prime$ of the $p(H)$ distributions.  Figures~\ref{im:rolling} plot the posterior distributions of the slope parameters for NIRB (left) and NIRF (right) components of each individual dynamical class, and for a model in which these parameters are shared across dynamical classes.  There is again visual agreement between all the classes, and we can calculate the evidence ratios between the hypotheses of common vs separate $H$ distributions per class $d$.

A common distribution is preferred for all NIRB or NIRF objects: the evidence ratio for a 4 parameter $(\theta_\nirb,\theta'_\nirb,\theta_\nirf,\theta'_\nirf)$ distribution, assuming a uniform prior that restricts $0 \leq \theta \leq 1.5$ and $-1 \leq \theta' \leq 1$, is $\mathcal{R} = 13.556$, showing decisive evidence for a common distribution instead of one $H_r$ distribution per dynamical class.
We should again take note of sensitivity of $\mathcal{R}$ to the prior, \eg\ the width $W$ of the prior on $\theta$ 
appears once per dynamical subpopulation $d$ in the numerator of the evidence ratio, while in the denominator it appears only once, which means that the evidence ratio scales as $W^{-8}.$ Hence $\mathcal{R}$ would drop by $\approx$ 6 if we were to halve the width of our prior.  A large reduction in the prior width could change the Bayes ratio from being decisively in favor of shared $p(H)$ slopes to being inconclusive.  At that point, however, most of the dynamical classes would have their posterior probabilities determined primarily by the prior, with the data being unimportant.  So it is fair to say that, to the extent that the data inform the $p(H)$ slopes and derivatives, they favor the common model.

We conclude that the DES data support a model in which each of the two GMM color components has a population with
size and variability distributions that are the same in every dynamical state, \ie\ the various dynamical classes are each composed of a mixture of shared NIRF and NIRB populations, within the selected $H$ range.

\subsection{Physical characteristics of NIRF and NIRB}
\label{sec:physical}
\begin{figure}[ht!]
	\centering
	\includegraphics[width=0.49\textwidth]{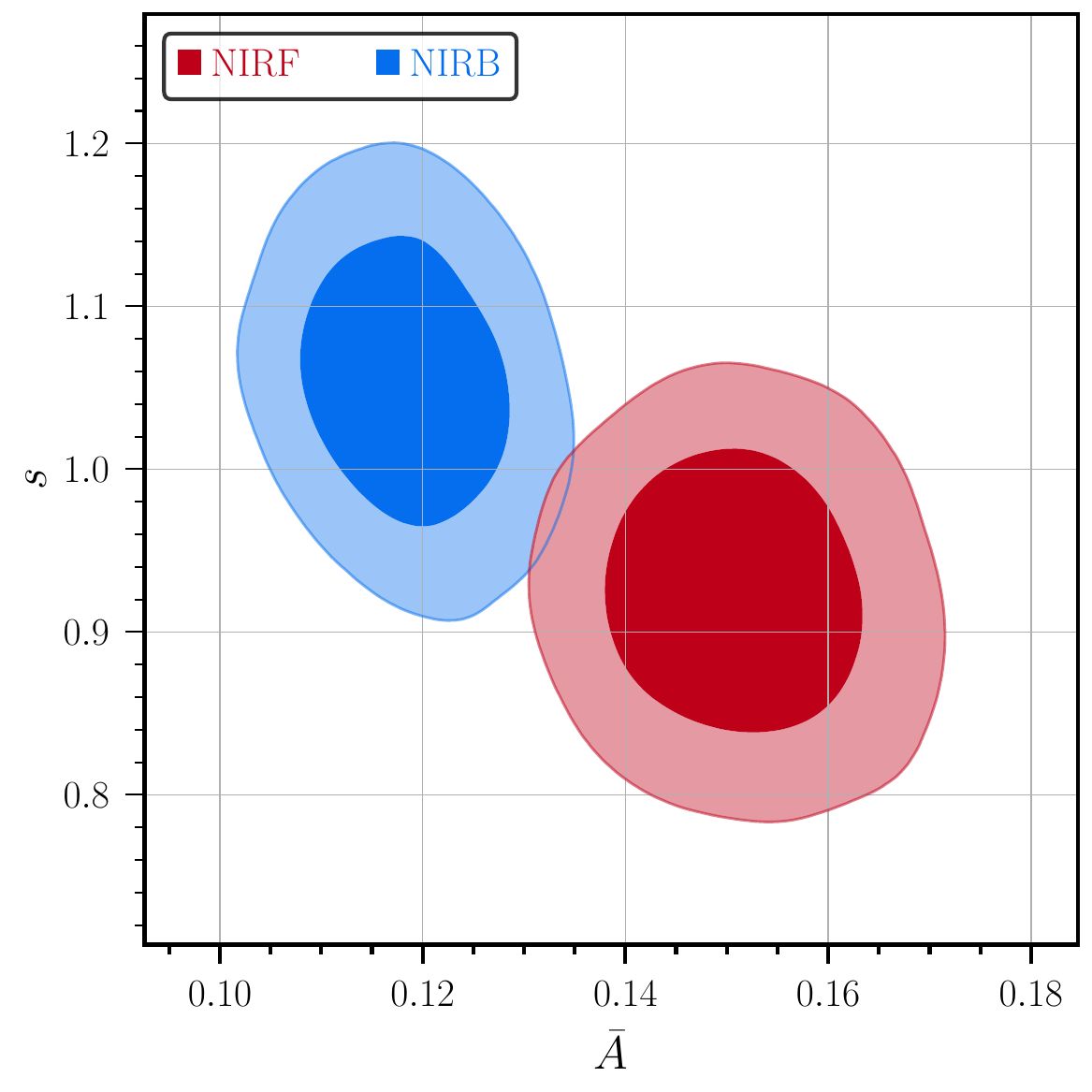}
	\includegraphics[width=0.49\textwidth]{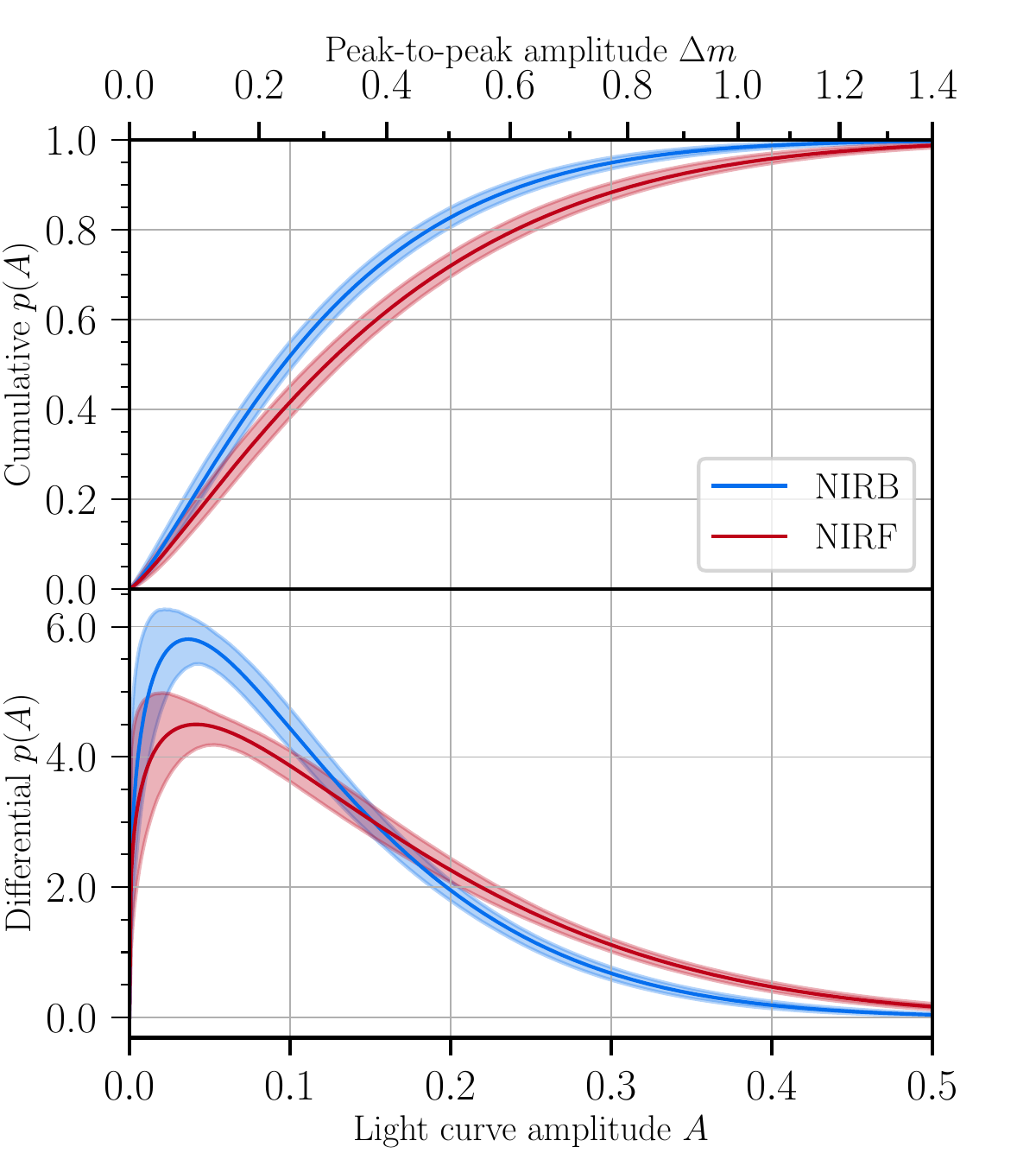}

	\caption{\emph{Left:} 68\% and 95\% posterior contours of the NIRB (blue) and NIRF (red) LCA distribution parameters. \emph{Right:} Visualization of the mean and 68\% limits of the cumulative (top) and differential (bottom) distributions for the NIRB and NIRF distributions. The horizontal axis shows both the LCA $A$ and the peak-to-peak amplitude $\Delta m \equiv 2.5 \log_{10} \left((1+A)/(1-A)\right)$.\label{im:lcadist}}
\end{figure}

We now adopt the two-component physical model and examine the $H$ and $A$ distributions implied by the 696-element sample.  The variability distributions of the NIRF and NIRB classes are distinct, in agreement with the results of \citet{bernardinelli2023}.
In this model, 
the NIRF population is more variable: $\bar{A}_\nirb = 0.118 \pm 0.007$, vs $\bar{A}_\nirf = 0.151 \pm 0.008,$ a $3\sigma$ difference; while the $s$ parameters are similar, $s_\nirb = 1.054 \pm 0.058$ vs $s_\nirf = 0.922 \pm 0.087$. We show the posterior distributions for the NIRB and NIRF parameters, and illustrate the resulting shape of the $\beta$ distribution, in Figure \ref{im:lcadist}. 

\cite{bernardinelli2023} show (using the same data) a decisive distinction between the detached and CC LCA distributions, and very strong distinction between the HC and CC distributions, setting the constraint that there are at least two distinct LCA distributions.  This is consistent with the newer results, since the CCs are almost entirely NIRF members (with higher variability) while the scattering and HC populations are NIRB-dominated.  Thus the variability amplitude can be associated either with another phyical property (color), or with dynamical state.  The data are not of sufficient power to determine whether color or dynamical state is the stronger predictor of $\bar A.$

The posteriors of the $H$ distributions of NIRF and NIRB, on the other hand, overlap substantially: we have that $\theta_\nirb = 0.856 \pm 0.047$, $\theta'_\nirb = -0.099 \pm 0.059$ and $\theta_\nirf = 0.829 \pm 0.053$, $\theta'_\nirf = -0.321 \pm 0.073$.
We can test whether a single distribution is sufficient to describe the data, that is, if the NIRF and NIRB $H_r$ distributions are distinguishable from each other assuming a rolling power law. The evidence ratio is $\mathcal{R} = 2.370$, showing strong evidence for a common distribution. In this case, $\theta = 0.844 \pm 0.033$ and $\theta' = -0.227 \pm 0.045$. This is in concurrence with the observation by \citet{petit2023} that the $H$ distribution of CC's in the OSSOS survey has the same shape as the dynamically ``hot'' population in the range $5.5<H_r<8.3,$ \change{nearly the same as we examine.} 

Furthermore, as has been well established in \cite{Bernstein2004} and reproduced in subsequent works \citep[\eg][]{napier2024}, we find that $\theta' <0$ to high significance, indicating a deficit of small objects relative to power-law extrapolations of brighter sources. 

Our range of $H_r$ is sufficient to clearly exclude a single-power-law $p(H_r),$ however it is too narrow to differentiate between different formulae for the departure from power-law.  For instance the exponentially tapered power law of \citet{kavelaars2021}, motivated by simulations of the streaming instability hypothesis \citep{youdinStreamingInstabilitiesProtoplanetary2005}, can fit our data equally well with the proper choice of parameters.  We do not quantify other $p(H_r)$ models because the data lack the power to discriminate them from the rolling power law.

\subsection{NIRF/NIRB fractions of dynamical classes}
\label{sec:fractions}
\begin{figure}[ht!]
	\centering
	\includegraphics[width=0.49\textwidth]{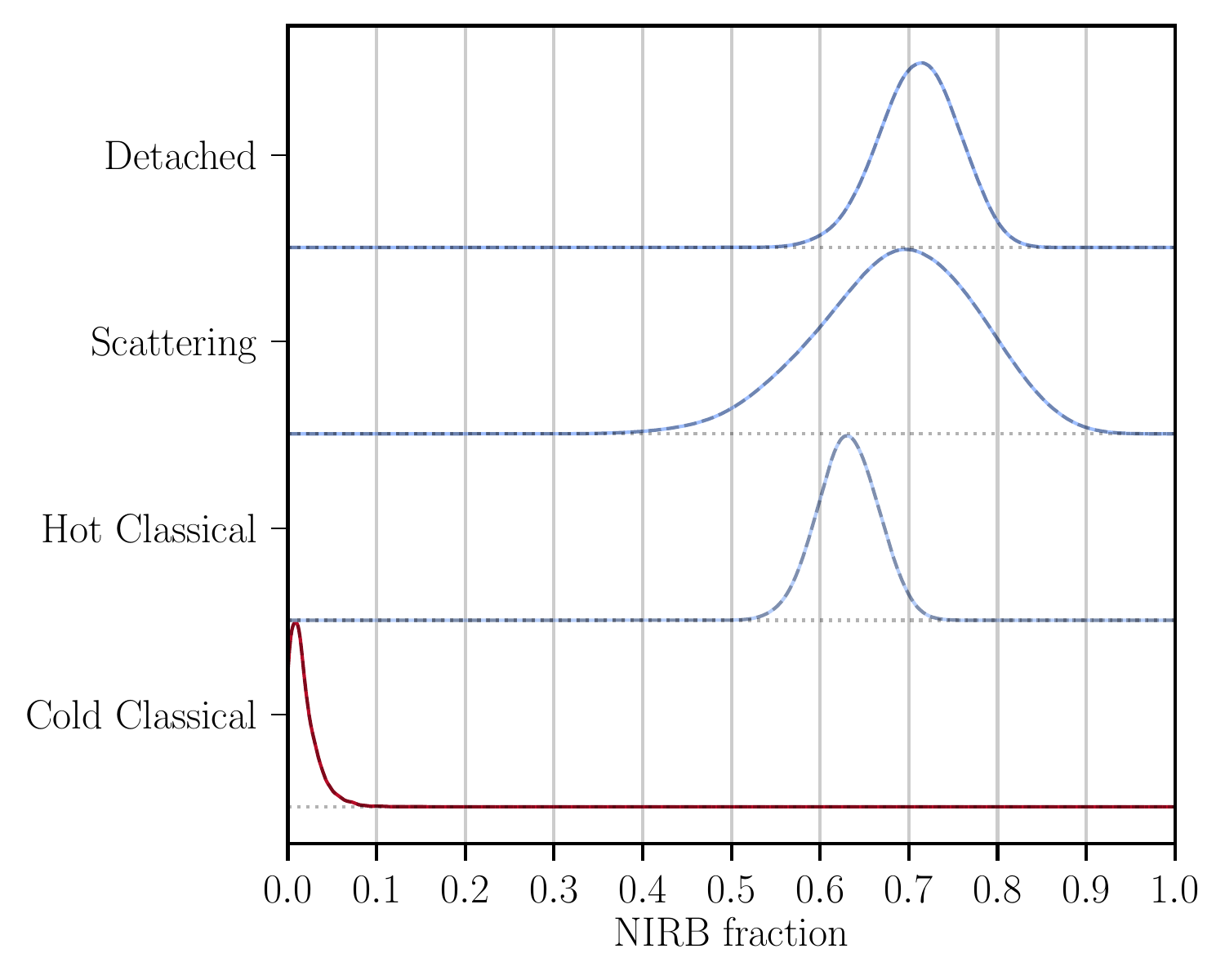}
	\includegraphics[width=0.49\textwidth]{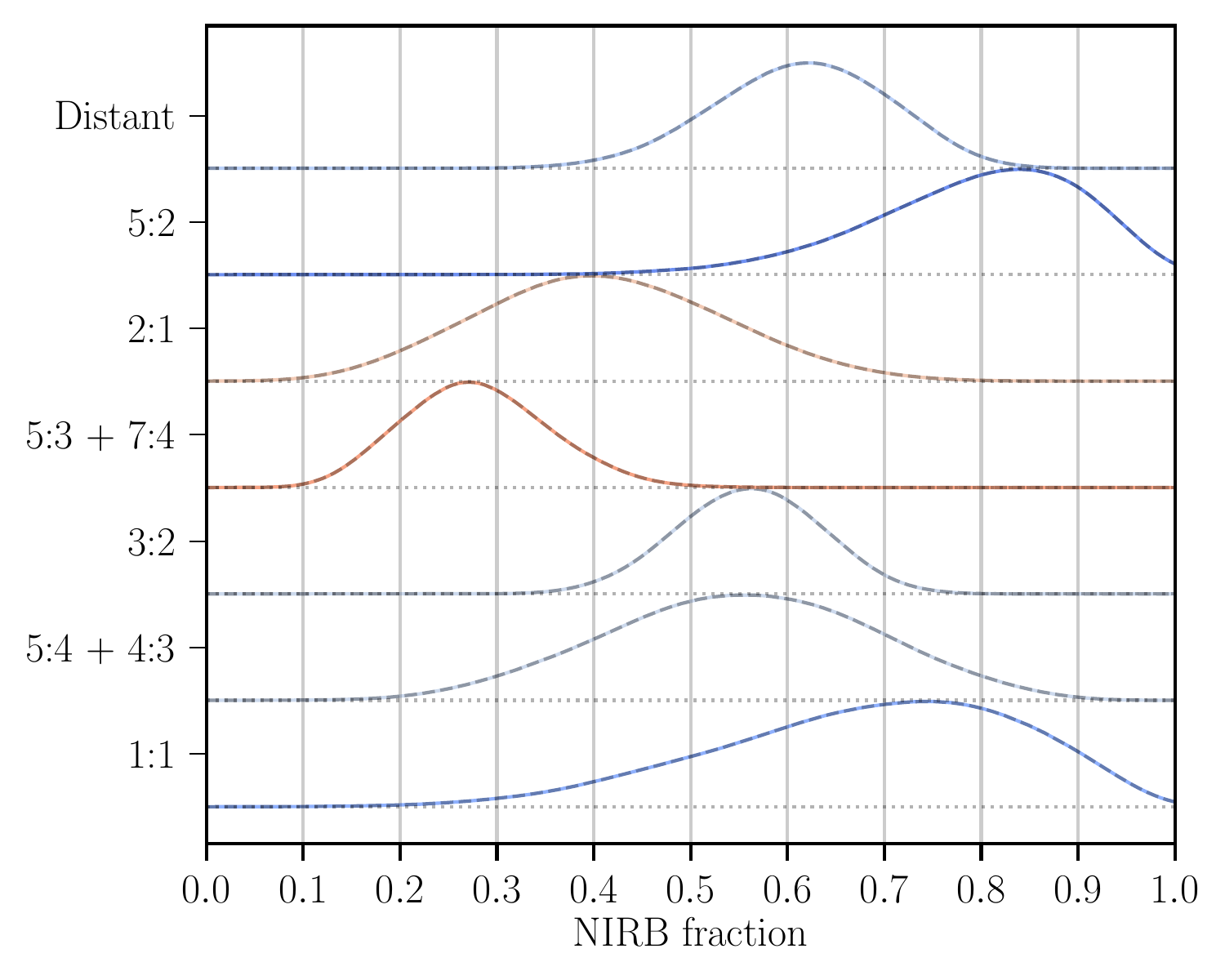}
	\caption{Posterior distributions of $f_{\nirb|d}$ after marginalizing over physical hyperparameters of the NIRF and NIRB populations. The vertical axis represents the distinct subsets of the data, and, inside a certain subset, this axis represents the relative probability of the posterior. The color of each curve relates to the peak of the distribution, going from red (NIRF-dominated) to blue (NIRB-dominated). The left panel shows the posteriors for the 4 main dynamical subsets (cold and hot Classicals, scattering and detached objects), while the right panel shows the posteriors for resonant subpopulations, stacked in order of $a$. \label{im:fclass}}
\end{figure}

We next examine the posterior probability distributions of the NIRB fraction
$f_{\nirb|d}$ for each class $d$ by marginalizing over the shared parameters $\boldsymbol{q}$.
These are plotted on the left half of Figure~\ref{im:fclass}, and the means and standard deviations are given in Table~\ref{tb:sample}.
As expected, there is substantial variation among the 5 dynamical classes, and the Bayesian evidence test completely rules out the hypothesis that all classes share a common $f_\nirb$ with  $\mathcal{R} = 62.185.$
If we accept the NIRB and NIRF classes are representing two different birth regions in the solar system, these values constrain the fraction of each class that was born in the nearer heliocentric distance range.

As expected, the CC dynamical class is consistent with 100\% NIRF composition, with the dynamically excited classes mixing the NIRF and NIRB populations.  

It is also of interest to divide the Resonant class into subclasses for different resonances with corresponding distinct values of semi-major axes $a.$ In this case the values of $\boldsymbol\theta$ are still being informed by the other dynamical classes, but each resonance subset has its own free $f_\nirb.$  These are shown on the right side of Figure~\ref{im:fclass}, and the means and standard deviations are given in Table~\ref{tb:sample}, and indicate a diversity of physical origins for distinct resonances' populations.  In particular, the resonances amidst the ``main belt'' of classical TNOs, the 5:3 and 7:4, are unique in being dominated by the NIRF population, similar to the CCs.
A common $f_\nirb$ distribution for all resonant subgroups is also ``strongly''  excluded by the Bayesian evidence ratio, at $\mathcal{R} = 2.403$.

In summary, the CC populations is essentially pure NIRF, and all of the dynamically excited populations are consistent with $f_{\rm NIRB}$ in the 0.6--0.8 range, with the notable exception that the ``main belt'' resonances (definitely) and 2:1 resonance (probably) have a larger fraction of NIRF objects.

\subsection{Inclination dependence of NIRF/NIRB fractions}

\begin{figure}[ht!]
	\centering
	\includegraphics[width=0.32\textwidth]{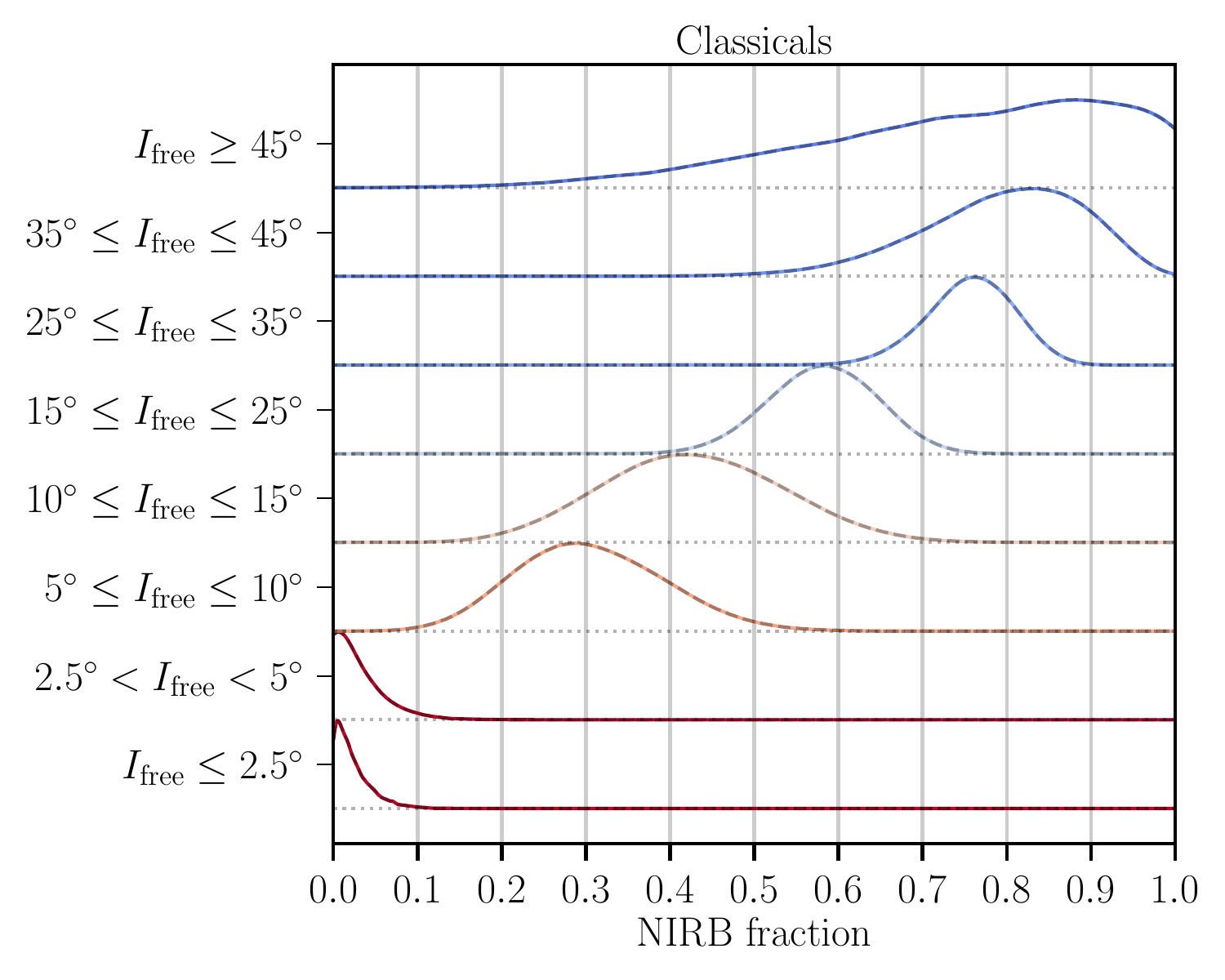}
	\includegraphics[width=0.32\textwidth]{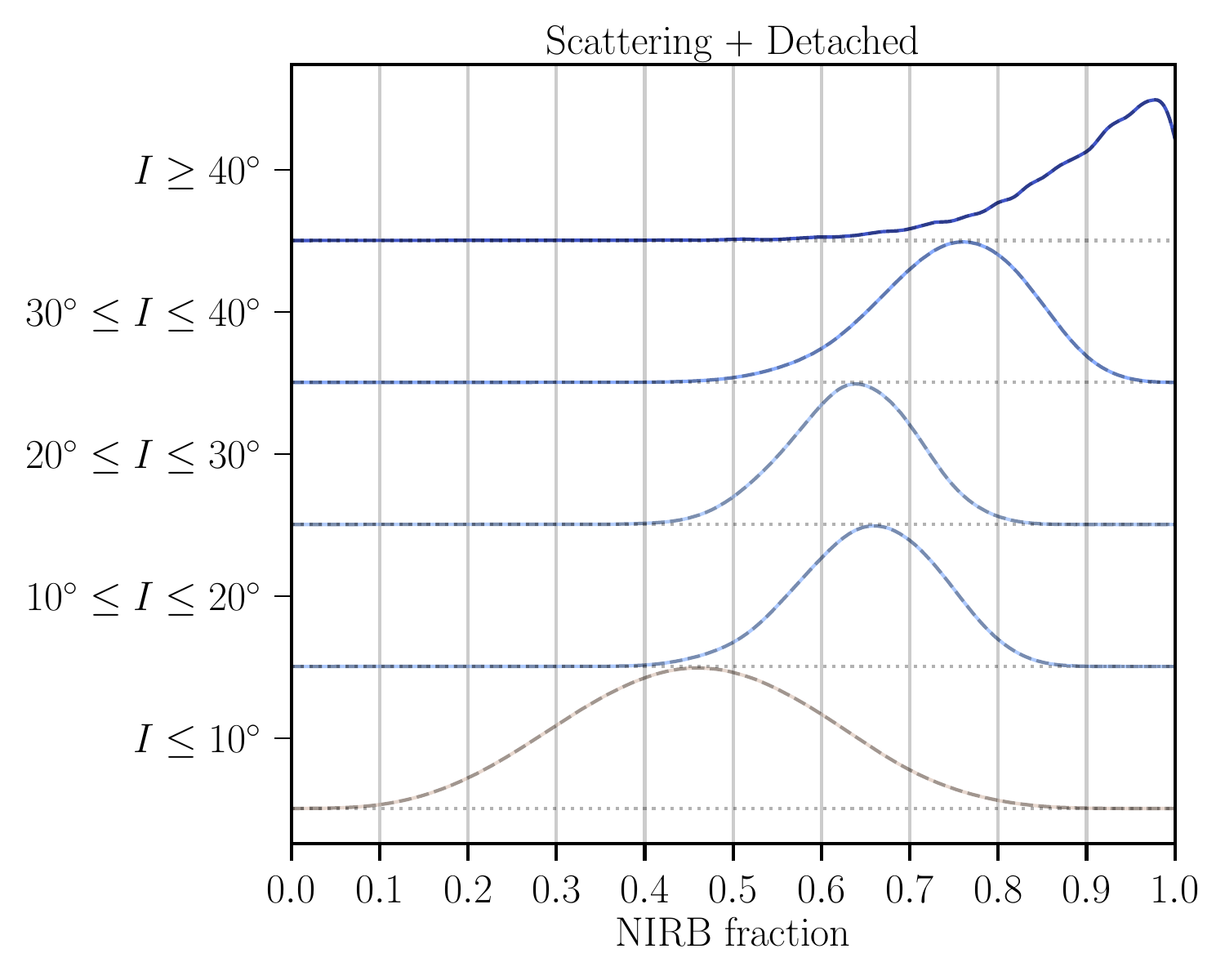}
	\includegraphics[width=0.32\textwidth]{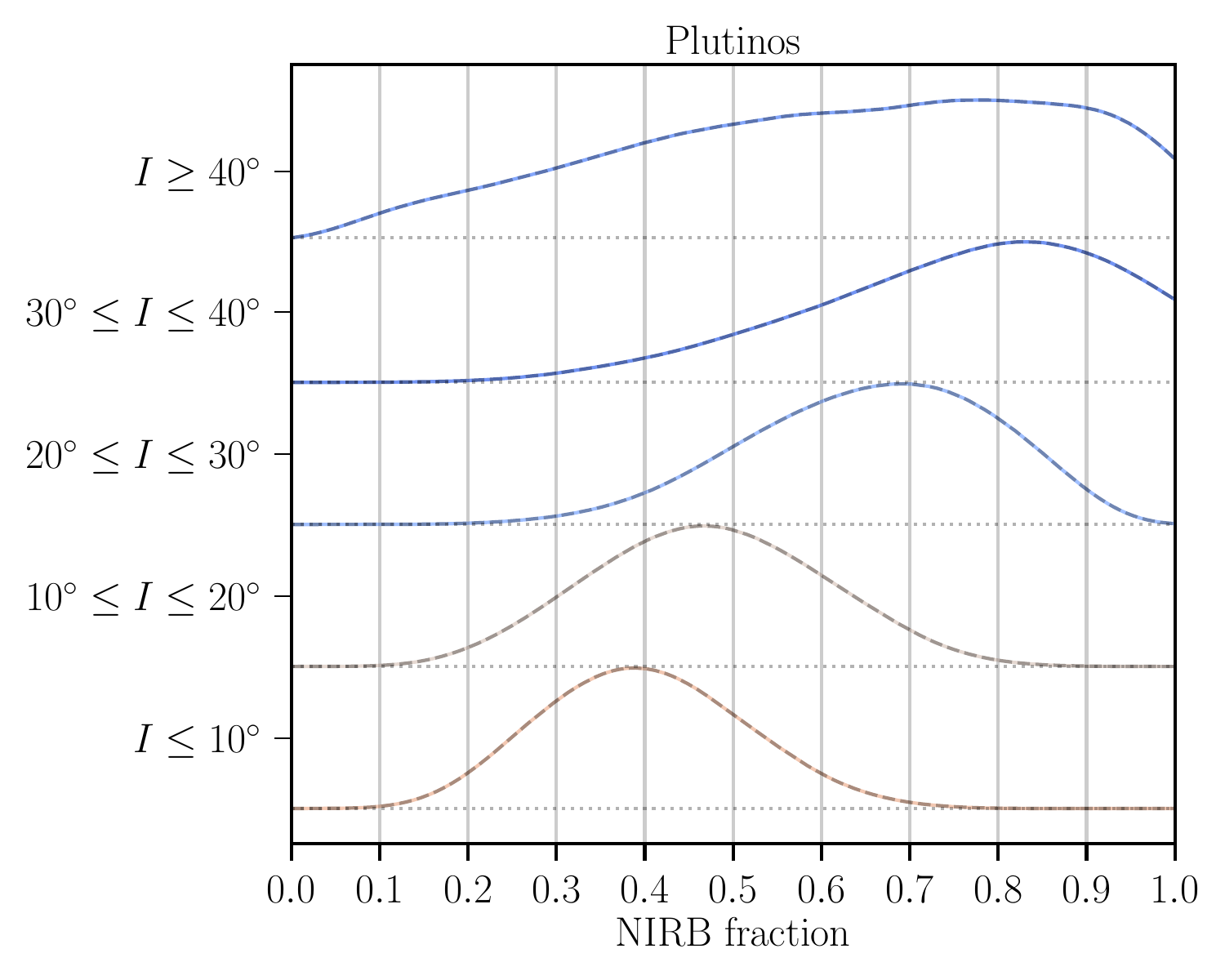}\\
	\caption{Similar to Figure \ref{im:fclass}, where the TNOs are now split into bins of inclination. The left panel shows the Classical population (HC+CC), the center panel shows the joint set of Scattering and Detached objects, and the right panel shows the Plutinos.}
\label{im:finc}
\end{figure}

Figure \ref{im:finc} shows the evolution of $f_\nirb$ among subsets of the TNOs defined by inclination angle (for the Classical population, we use free inclinations).  At left, the merged CC+HC sample is divided into 8 bins of increasing $I$ and the model with 2 physical components is fit to the ensemble with free $f_{\nirb,d}$ for each inclination bin $d$. By definition, the 2 lowest-$I$ bins are the CC's, where a pure-NIRF population is favored.  As expected, higher-$I$ bins (the HC's) show higher $f_\nirb,$ but it also seems that this fraction is strongly inclination-dependent within the HC population.  This is quantitatively confirmed by a decisive Bayesian evidence ratio, $\mathcal{R} = 6.588$, favoring distinct $f_\nirb$ values for each HC inclination bin over a common $f_\nirb$ for all the HC bins. We note that this gradient in NIRF/NIRB population could underlie much of the correlation between inclination and visible spectral slope reviewed by \citet{marsset2019}.  But analysis of the ColOSSOS data by \citet{marsset2023}, separating TNOs into their BrightIR and FaintIR color groups analgous to our own NIRB/NIRF, suggest that an anticorrelation between visible spectral slope and inclination exists even within the BrightIR class.  This suggests that we investigate potential differences between the red and blue ends of the NIRB sequence, which we will do in the Section~\ref{sec:split}.

The center and right panels of  Figure~\ref{im:finc} slice the Plutino and the combined scattering+detached classes into inclination bins.
The visually apparent trend toward higher $f_\nirb$ at higher inclinations---with the $I>40\arcdeg$ bin consistent with 100\% NIRB---is not, however, quantitatively significant.  Indeed the evidence ratio for distinct vs common $f_{\rm NIRB}$ across inclination bins is$\mathcal{R} = -0.27$ for the Plutinos, and $\mathcal{R} = 0.743$ for the joint scattering+detached classes, offering no strong preference for either scenario. We revisit and quantify the $f_\nirb$ dependency with inclination in Section \ref{sec:inclination}.

\subsection{Goodness of fit}
We conclude this section by testing whether the 2-component model is a good description of the underlying data. To do so, we first evaluate whether the GMM color model derived in Section \ref{sec:colormodel} is a good fit to the observed color distribution (with the outliers excluded). Our goodness-of-fit statistic will be the $\log\likeli$ in Equation \ref{eq:loglike} (including only the color terms).  We produce a distribution of expected $\log\likeli$ by drawing $10^5$ sets of 696 TNOs from the best-fit GMM to the data.  For each sampled TNO, we create a color ``swarm'' by shifting the measurement distribution of a real TNO to the newly sampled colors as in Section~\ref{sec:outliers}.
For each set $k$ of 696 simulated measurements, we repeat the GMM training procedure of  Section \ref{sec:gmm} to find the best GMM for this realization of the full population, and then retain the resultant $\log\likeli_k$ of the realization.  A $p$-value for the real data is defined as the fraction of the realizations for which $\log\likeli_k < \log\likeli_{\rm data}.$ We arrive at $p = 0.918$, indicating that the real data (with outliers excluded) are a plausible realization of the GMM.


The same goodness-of-fit process is repeated for the rolling-power-law distributions of $H_r$.  
We create a realization of the 696 truth values of $H_r$ for TNOs by sampling from the model with the likelihood-maximizing values of
$\{\theta,\theta^\prime\}$ and $\{ f_\alpha\}$, preserving the number of objects per dynamical class.  To place measurement errors to these synthetic $H_r$ measurements, we assign the measurement errors in $H$ of the $n^\mathrm{th}$-brightest real TNO in each dynamical class to the $n^\mathrm{th}$-brightest TNO of the dynamical class in the realization. This gives the smallest objects the largest uncertainties in $H,$ as in the real data.
Then, we re-minimize the likelihood in Equation \ref{eq:loglike}. Repeating this procedure $5 \times 10^4$ times, we find that the resultant $(\theta,\theta^\prime)$ values are shifted by $(0.039\pm0.022, -0.024\pm0.019)$ from the input values, suggesting immaterially small biases in the process.
We can also use this procedure to derive a $p$-value from $\log\likeli$ as above, obtaining $p = 0.860$, once again indicating a good fit. 

\begin{figure}[ht!]
	\centering
	\includegraphics[width=0.49\textwidth]{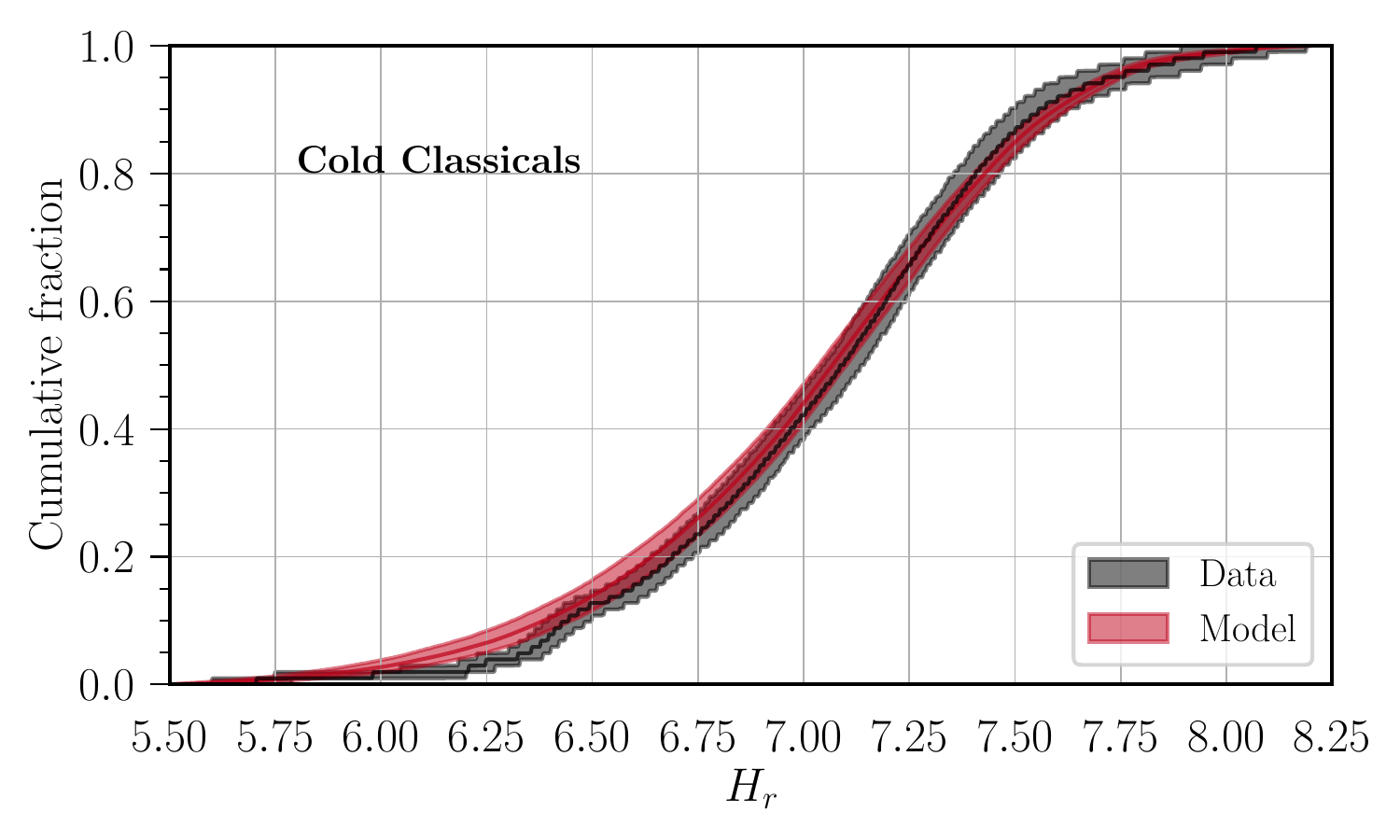}
	\includegraphics[width=0.49\textwidth]{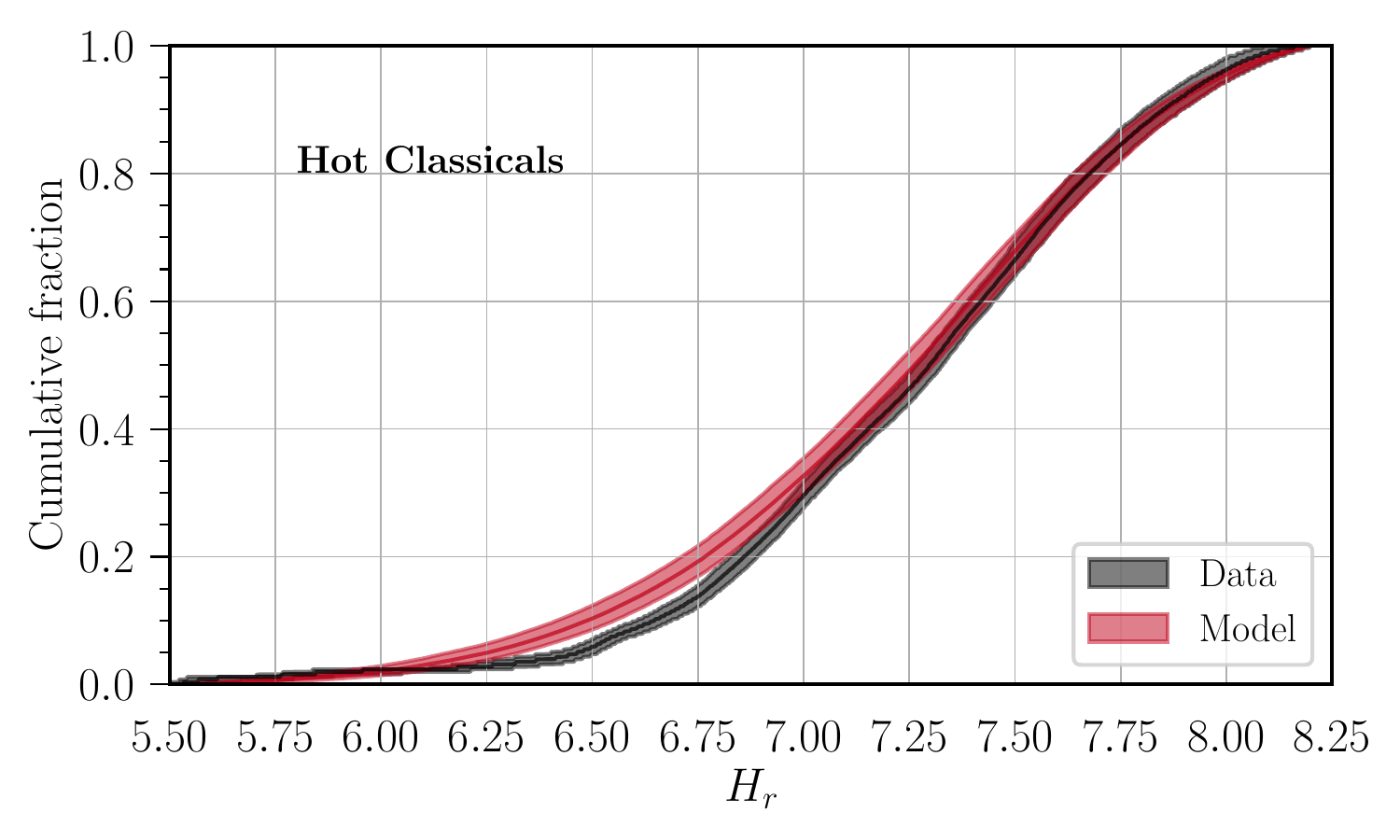}
	\includegraphics[width=0.49\textwidth]{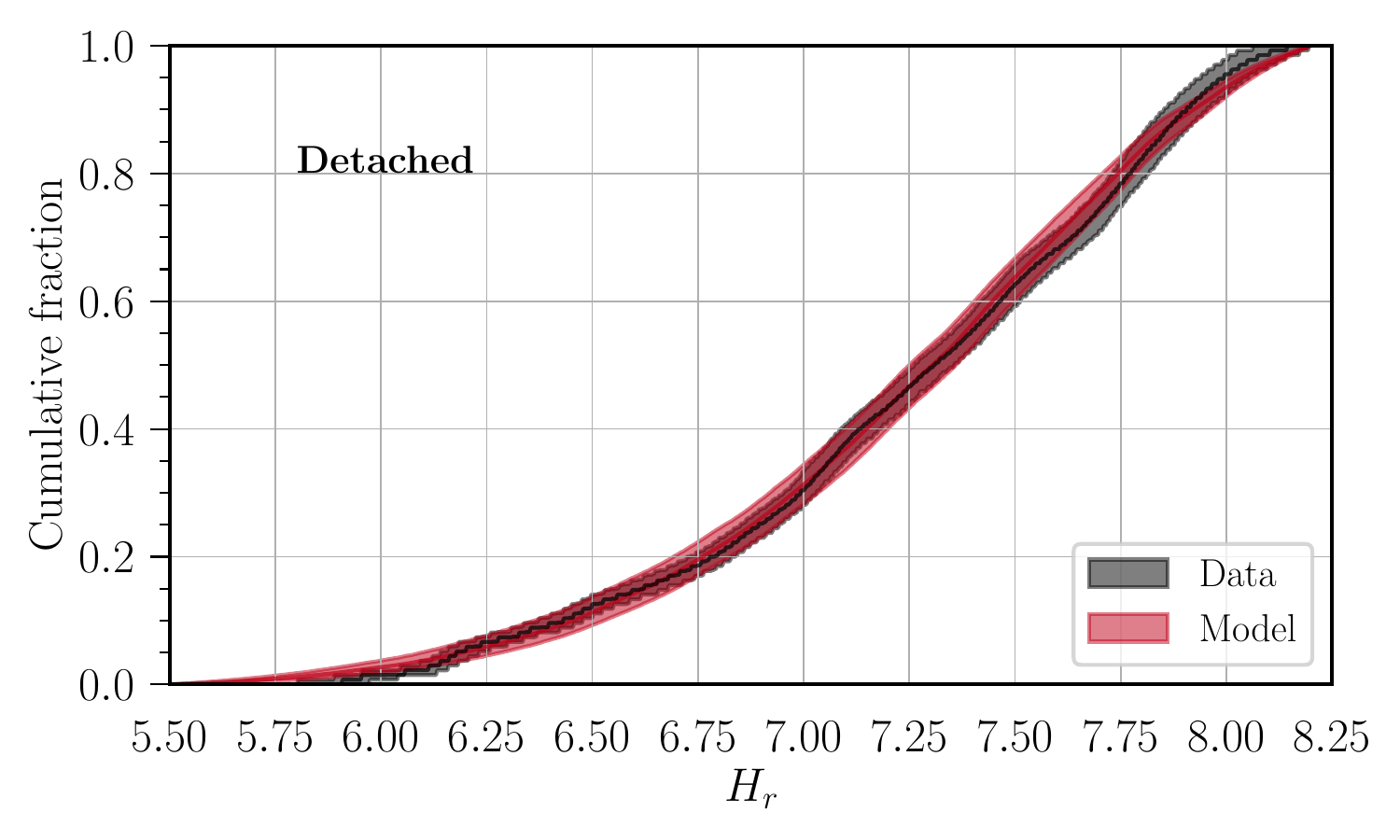}
	\includegraphics[width=0.49\textwidth]{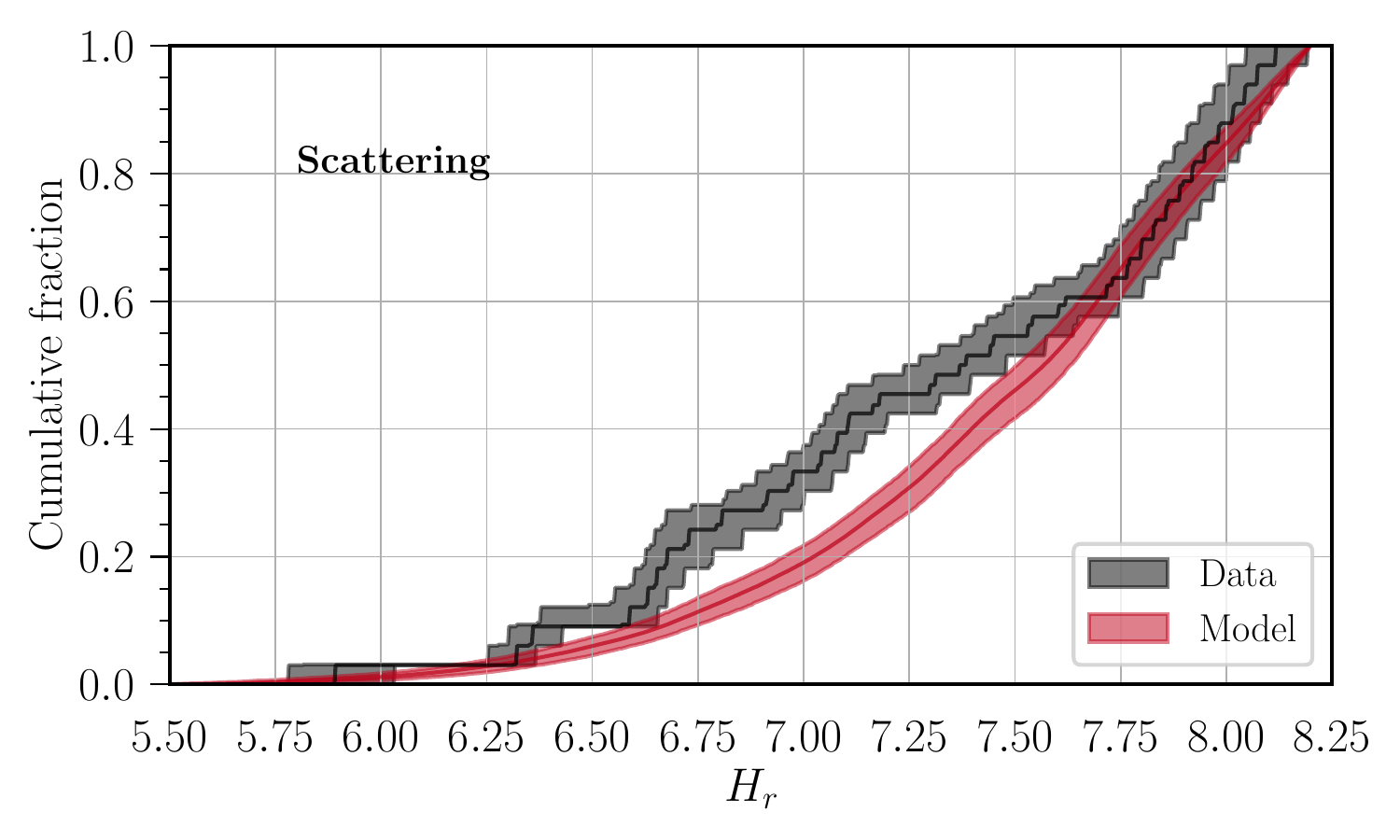}
	\includegraphics[width=0.49\textwidth]{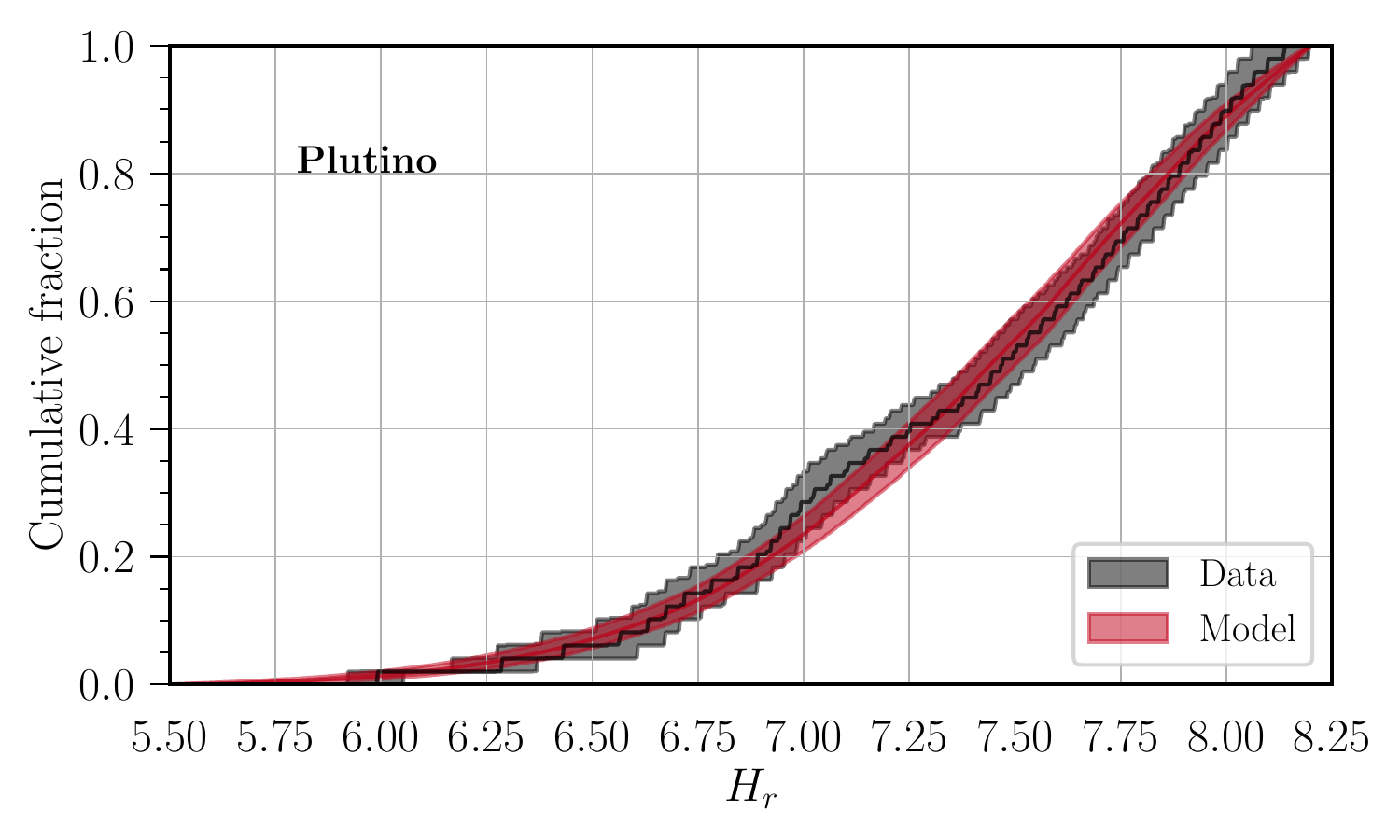}
	\includegraphics[width=0.49\textwidth]{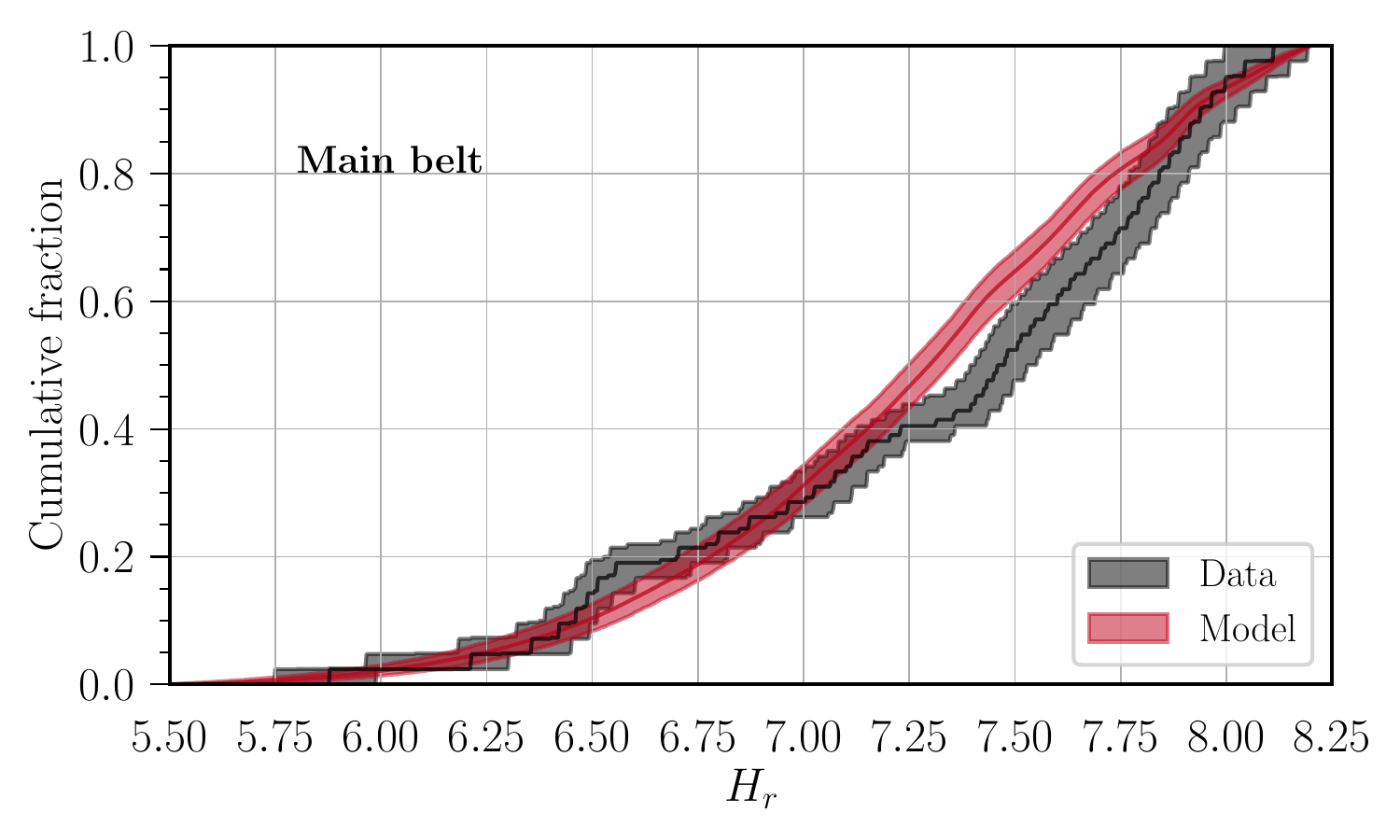}
	\caption{Each panel shows the 68\% limits of the data distributions per dynamical class (black) as well as the best-fit $p(H_r)$ distribution multiplied by the selection function for the corresponding dynamical subset (red). In all cases, the data distribution has been constructed by sampling independent realizations of the $H_r$ for each object, while the model distribution samples from the posterior of the likelihood.\label{im:absmagdist}}
\end{figure}

As a further test of the rolling-power-law parameterization of $p(H_r),$ we compare the observed cumulative distributions of $H_r$ in each dynamical class to the cumulative distributions obtained from draws from the rolling power law that have been subjected to our selection criteria and measurement errors.
Figure \ref{im:absmagdist} shows the resulting data and model curves, and Table~\ref{tb:ksad} presents the $p$-values obtained by comparison of the data to the model distributions with the
Kolmogorov-Smirnov (KS) and Anderson-Darling statistics \citep{press2007numerical}. Both tests show that for all dynamical subsets
are consistent with the null hypothesis that the data are sampled from the best-fit models.

\begin{deluxetable}{lcc|lcc|lcc}[ht!]
	\tabletypesize{\footnotesize}
	\tablecaption{$p$-values of the KS and AD statistics for the $H_r$ distribution\label{tb:ksad}}
	\tablehead{\colhead{Subset} &  \colhead{KS} &  \colhead{AD} & \colhead{Subset} &  \colhead{KS} &  \colhead{AD} & \colhead{Subset} &  \colhead{KS} &  \colhead{AD}}
	\startdata
	CC & 0.878 & 0.697 & HC & 0.748 & 0.241 & Detached & 0.989 & 0.916 \\
	Scattering & 0.255 & 0.301 & Trojan & 0.622 & 0.787 & Inner belt & 0.051 & 0.213 \\ 
	Plutino & 0.982 & 0.999 & Main Belt & 0.231 & 0.466 & 2:1 & 0.221 & 0.408 \\ 
	5:2 &  0.677 & 0.574 & Distant & 0.621 & 0.747 & & &
	\enddata
\end{deluxetable}

\section{Are two components enough?}
\label{sec:split}
In the previous section we constructed physical families based on 2 components of color distributions, and showed that the populations of the current dynamical families within a given color distributions have statistically indistinguishable $H$ and $A$ distributions, supporting the hypothesis that each dynamical family is a mixture of two common origin populations.
Here we ask: Is there any evidence for the current dynamical classes to consist of \emph{more} than 2 distinct physical families?

We approach this question by repeating the procedures of Section~\ref{sec:nirbnirf} but beginning with 3 color distributions instead of 2.  Each of the GMM components listed in Table~\ref{tb:gmm} is nearly a one-dimensional sequence through the 3-dimensional color space, as can be seen if one takes the eigenvalues $\lambda_1>\lambda_2>\lambda_3$ of the covariance matrices: for both NIRB and NIRF, $\lambda_1\approx 100 \lambda_2\approx1000 \lambda_3.$ We split the NIRB component in half along its major axis, leaving us with three color components indexed by $\alpha \in \{\nirb+, \nirb-, \nirf\},$ where the $+$ and $-$ denote the redder and bluer halves of the sequence, respectively.  We can then model each DES TNO dynamical population $d$ as a mixture of components $\beta=(\alpha,d).$ If the $H$ and $A$ distributions of a given color component are consistent across dynamical classes, then these 3 color components are consistent with being the underlying constituents of all dynamical distributions.  If we find evidence that $f_{\nirb+ | d} \ne f_{\nirb-,d}$ for some class $d$, this would show that there is more differentiation of physical properties among the dynamical classes than can be fully described by our 2-component model.  We can likewise ask these questions when splitting the NIRF component into redder and bluer halves to look for differential contributions to the dynamical classes.

Mathematically, this construction can be defined by changing the vector basis of the vector $ \mathbf{c} - \boldsymbol{\mu}_\nirb$ in Equation \ref{eq:gmm} to the basis formed by the eigenvectors of $\Sigma_\nirb$. Calling this new vector $\mathbf{c}' = (c'_1, c'_2, c'_3)$, where the $c'_i$ is the component corresponding to the eigenvalue $\lambda_i$, we can then define the probability distribution
\begin{equation}
	p_{\nirb\pm}(\mathbf{c} | \boldsymbol{\mu}_\nirb, \Sigma_\nirb) \equiv 
	\begin{cases}
		2 \mathcal{N}(\mathbf{c}|\boldsymbol{\mu}_\nirb, \Sigma_\nirb), & \text{if } c'_1 \gtrless 0\\
		0, & \text{otherwise}  
	\end{cases}.
\end{equation}
These $p(\vecc)$ distributions are properly normalized to $\int \mathrm{d} \mathbf{c} \, p_{\nirb\pm} (\mathbf{c}) = 1$, although they are no longer Gaussian, and their  mean colors are no longer $\boldsymbol{\mu}_\nirb$. The original 2-component GMM is nested within this model, as the case $f_{\nirb+,d} = f_{\nirb-,d}.$

\subsection{Splitting the NIRB component}
We begin by evaluating differentiation within the more populous NIRB component. The model now has 3 physical components for each dynamical subpopulation $d$, with the fractions constrained to the triangular face of a simplex defined by $f_{\nirb+|d}+f_{\nirb-|d}+f_{\nirf|d}=1.$ The natural prior for the $f$'s is uniformity across this triangle, which is described by the Dirichlet distribution (a.k.a. multivariate $\beta$ distribution) with parameters $\alpha_i=-1.$  

As in Section~\ref{sec:plaw}, we test whether the lightcurve parameters ($\bar{A}$ and $s$) vary across the two halves of the NIRB component by finding a log evidence ratio of $\mathcal{R} = 1.771,$ slightly favoring a shared distribution but not reaching Jeffrey's ``strong'' level.  Similarly, the evidence that the two absolute magnitude distribution parameters $\theta,\theta^\prime$ are shared is $\mathcal{R} = -0.475$.
The evidence that all four parameters physical parameters are shared is $\mathcal{R} = 1.180$. The DES data thus neither favor nor disfavor a shared distribution strongly. For simplicity, we proceed with the model where the NIRB+ and NIRB- populations do share $H_r$ and $A$ distributions.

\begin{figure}[ht!]
	\centering
	\includegraphics[width=\textwidth]{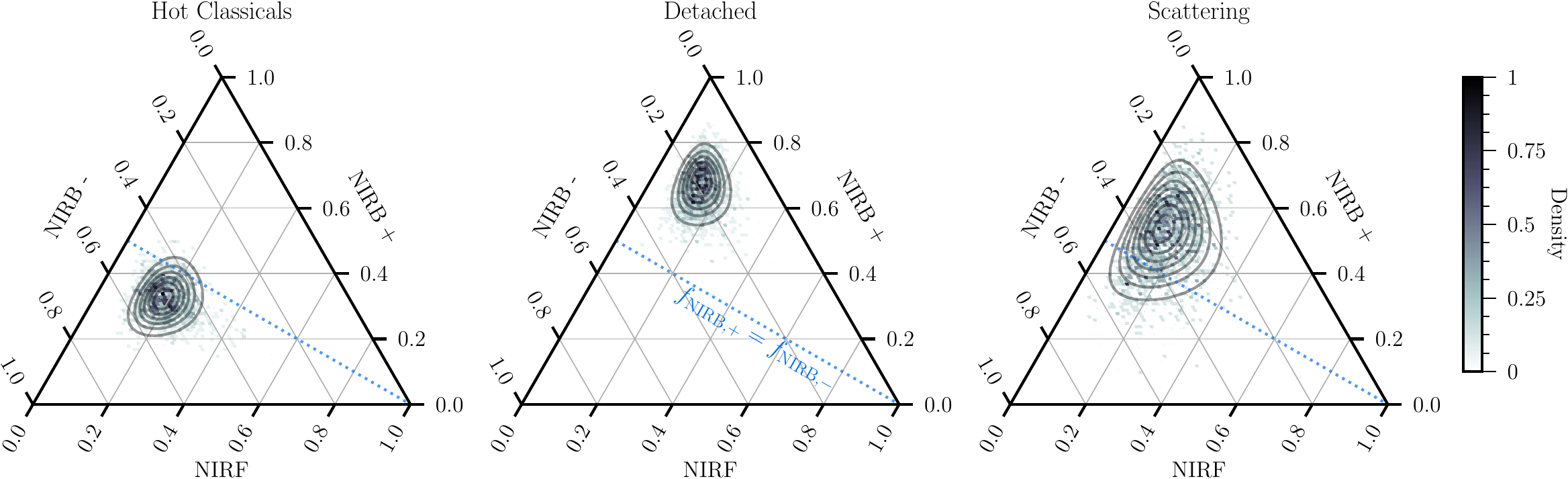}
	\caption{Ternary histograms of $(f_\nirf, f_{\nirb+}, f_{\nirb-})$ posterior samples HC, detached, and scattering populations are plotted as gray points in the three panels.  Also plotted are contours of the 3-dimensional Dirichlet distribution that best fits the sampled $\{f_\alpha \}$.
The blue dotted line represents the locus where $f_{\nirb+} = f_{\nirb-}$, indicating equal occupancy of the two sides of the NIRB Gaussian distribution.\label{im:ternary}}
\end{figure}

After marginalizing over the $p_\beta(H)$ and $p_\beta(A)$ parameters, we test for differences in the ratio of NIRB+ to NIRB- occupation in the $f_{\nirb, \pm}$ posterior distributions for the NIRB-dominated classes. The posterior samples of the set of $\{f_\alpha\}$ for the HC, detached and scattering populations are presented in Figure \ref{im:ternary}. In each panel, we also show the contours of the Dirichlet distribution that best fits the sampled $\{ f_\alpha \}$ values, using the maximum likelihood method of \citep{Minka2000} to derive Dirichlet parameters.

The visual impression is that the detached TNOs (and probably the scattering objects) preferentially like on the red side of the NIRB locus, with the HCs favoring the blue side.  This difference is confirmed by an evidence ratio of $\mathcal{R} = -4.043$ showing very strong evidence that the HC and detached populations do \emph{not} share a common ratio of $f_{\nirb+}$ to $f_{\nirb-}.$  A shared ratio for the  scattering and detached objects is strongly favored by $\mathcal{R} = 2.673$, while the evidence for shared HC and scattering parameters is inconclusive at  $\mathcal{R} = 0.224$.

While the fraction $f_\nirb$ is nearly identical for the HC, detached, and scattering populations, their distributions \emph{within} the NIRB color sequence do not agree, indicating a deviation from the two-component model.   This would require the existence of a spatial gradient of color within the NIRB birth population, with subsequent differential migration into the HC vs the detached (and possibly scattering) population.  Alternatively, post-migration surface color changes were different for the HCs than the detached objects.

\subsection{Splitting the NIRF component}

Next we test for differences in the occupation of the NIRF component. This time, we can include constraints for the primarily-NIRF CC population. We begin by allowing distinct values of $(\bar{A}, s)$ on the red and blue sides of the NIRF component. We see that the two values for $\bar{A}$ are nearly identical in the two populations. The sharpness parameters disagree, however: $s_{\nirf,+} = 1.12 \pm 0.10$, $s_{\nirf,-} = 0.66 \pm 0.10$. This $3.1\sigma$ difference in $s$ leads to an evidence ratio $\mathcal{R} = - 4.552$ against a common distribution. This implies that the distribution of lightcurve amplitudes is broader for redder NIRF TNOs. We do not see a significant difference in the ($\theta,\theta^\prime$) parameters between the NIRF+ and NIRF- populations, with evidence ratio $\mathcal{R} = -1.398$. Thus, we proceed with two NIRF components that differ only in their LCA distributions, but share their $H_r$ distributions.

\begin{figure}[ht!]
	\centering
	\includegraphics[width=\textwidth]{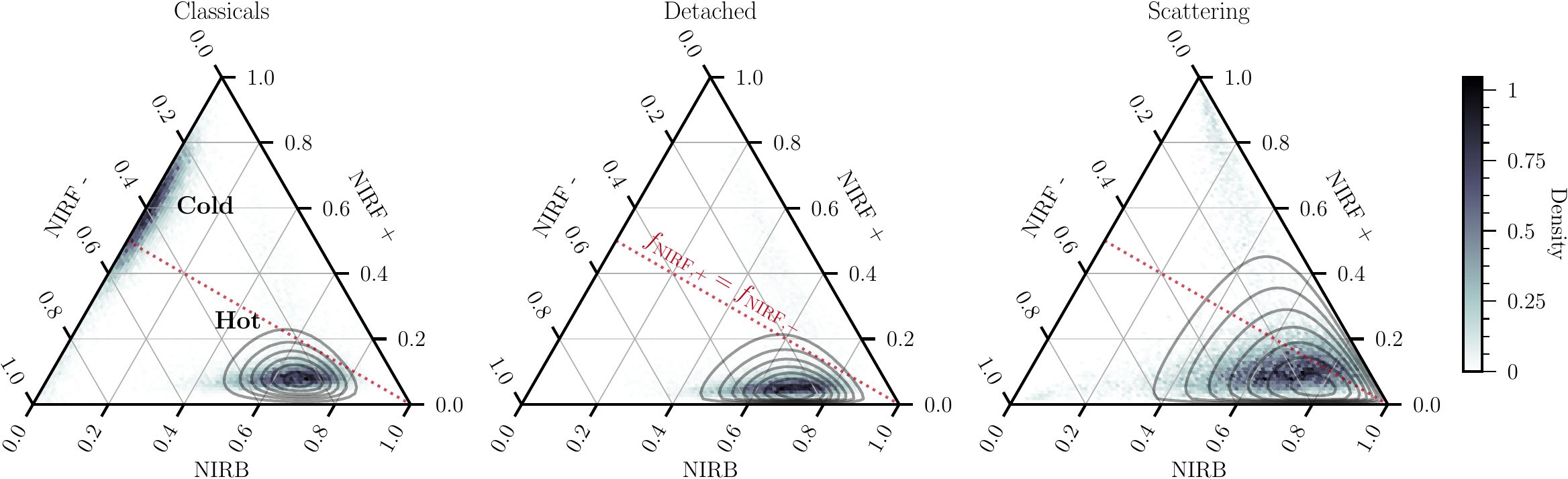}
	\caption{Similar to Figure \ref{im:ternary}, but showing $(f_\nirb, f_{\nirf+}, f_{\nirf-})$ posterior samples. The red dotted line represents the locus where $f_{\nirf+} = f_{\nirf-}$. The first panel shows both the Hot and Cold Classical posterior samples.}
\label{im:ternary_nirf}
\end{figure}

We can also evaluate the distribution of the $f_{\nirf,\pm}$ for each of the dynamical classes. Figure \ref{im:ternary_nirf} shows the distribution of $\{f_\alpha\}$ for the CC, HC, detached and scattering objects. Defining $f_{-|\nirf} = f_{\nirf-}/(f_{\nirf+}+f_{\nirf-})$ to denote the fraction of NIRFs on the blue side of the sequence, we have that the CC population has $f_{-|\nirf} = 0.57 \pm 0.07$, consistent with equal membership in the two components. The evidence ratio for the hot Classicals, detached and scattering objects is $\mathcal{R} = 5.504$, showing decisive evidence for a shared distribution among the three populations, where we find $f_{-|\nirf} = 0.247 \pm 0.039$, $4\sigma$ below the CC value.

\section{The inclination distribution for the NIRB and NIRB populations}\label{sec:inclination}

\subsection{Measuring the inclination distribution}
As demonstrated in Section \ref{sec:nirbnirf} and Figure \ref{im:finc}, $f_\nirb$ seems to depend on inclination within the HC population. To quantify this trend, we measure the inclination distribution of the NIRB and NIRF components of the trans-Neptunian populations. Since \des\ has substantial coverage to high ecliptic latitudes, this test will also provide strong constraints to the shape of the inclination distribution for each dynamical subpopulation. In addition, \cite{Huang2022} compute ``free'' inclinations ($I_\mathrm{free}$) for all \des\ main belt TNOs. This quantity is better conserved over the age of the Solar System for members of the classical population, and so is preferred to the ecliptic inclination when measuring the inclination distribution of this population. We also use this methodology to estimate $I_\mathrm{free}$ for the simulated objects used to determine the completeness functions, so that we can measure $p(\mathcal{S}|I_\mathrm{free})$.

The framework presented in Section \ref{sec:framework} still applies for this test, as we can safely assume that the measurements of inclination for each object are essentially noiseless. 

We use the von Mises-Fischer distribution \citep{matheson2023,Malhotra2023} as a convenient and statistically well-motivated functional family for the inclination distribution,
\begin{equation}
	p(I|\kappa) = \frac{\kappa}{e^\kappa - e^{-\kappa}} \exp(\kappa \cos I) \sin I,
\end{equation}
where $\kappa$ is the concentration parameter. For large $\kappa$, the distribution peaks at inclinations $I \approx \kappa^{-1/2}$ radians (see \citealt{matheson2023} for a full derivation and discussion). This means that larger $\kappa$ imply a more concentrated distribution, that is, a distribution with lower inclination objects. To improve the numerical stability of the HMC procedure and impose the (necessary) constraint that $\kappa > 0$, we will use $\log\kappa$ as the free parameter in our likelihoods.

\subsection{Inclination distribution of the joint physical + dynamical subpopulations}
\label{sec:vMF}

\begin{figure}[ht!]
	\centering
	\includegraphics[width=0.495\textwidth]{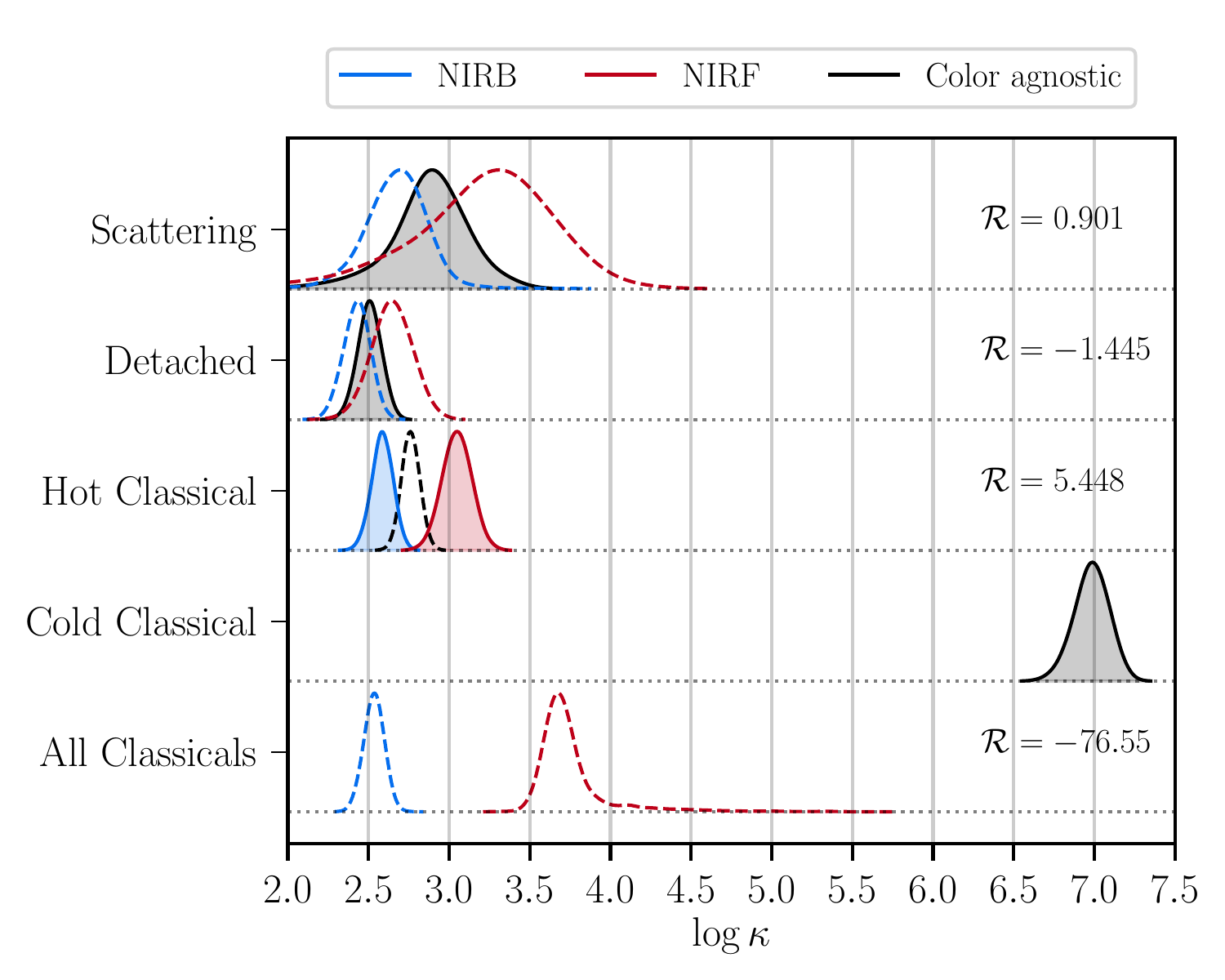}
	\includegraphics[width=0.495\textwidth]{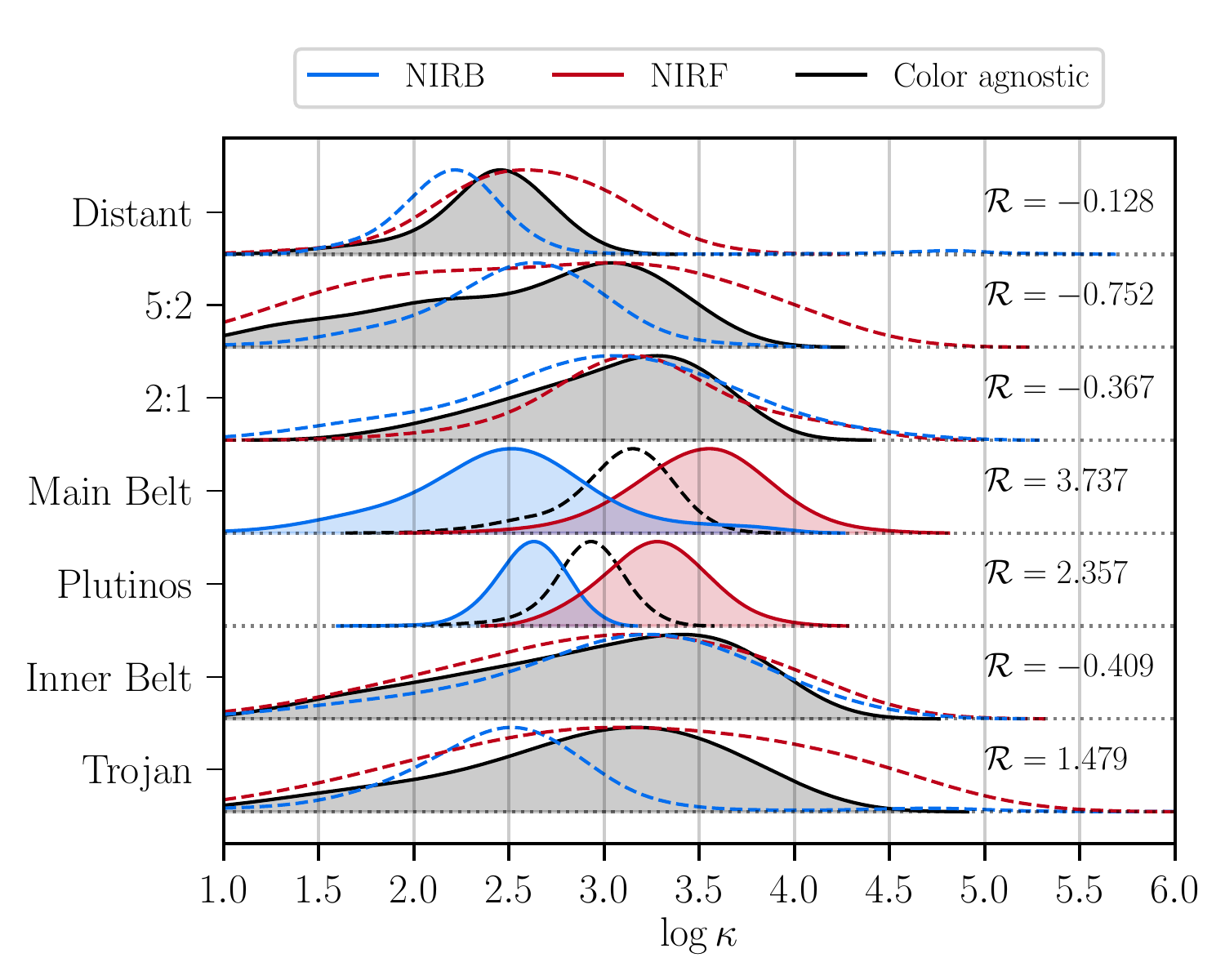}

	\caption{Histograms of the posterior samples for the inclination distribution concentration $\log\kappa$, with each row representing one dynamical subpopulation. The blue (red) curves represent the NIRB (NIRF) component, while the black curves represents the case where the two color components share their inclination distribution. In all cases, the shaded curve with the solid line represents the statistically preferred model, with the alternative option having a dashed curve. The evidence ratio $\mathcal{R}$ for each subset is also shown, and a positive value shows a preference for the two component model. The left panel shows the non-resonant populations, with the last set of distributions representing the posteriors for the set of all Classical objects, while the two rows above it represent the Cold and Hot Classical subsets. The right panel shows the resonant populations. The Cold Classicals are dominated by the NIRF component, so the NIRB distribution is unconstrained. In this case, we present only the color agnostic model. Similarly, the Trojan and 5:2 resonators are primarily dominated by the NIRB component, and so the NIRF component is poorly constrained.\label{im:kappa}}
\end{figure}

Figure \ref{im:kappa} shows the $\log\kappa$ posterior distributions within each dynamical subset, in models where there are distinct $\kappa_\nirb$ and $\kappa_\nirf$ (in blue and red, respectively), or when the whole dynamical subpopulation shares a common $\kappa$ (in gray). Among the non-resonant classes, only the HCs decisively favor a
``color-inclination bimodality,'' with a more concentrated NIRF component (larger $\kappa$). This does not mean, however, that there are no NIRF objects with high inclinations: as can be seen from Figure \ref{im:aeicolor}, the most inclined NIRF HC has $I = 42.3\degr$.
It does mean, however, that the current dynamical states of the HCs retain some memory of whether they originated in the birth region of the NIRF colors vs the NIRB colors.  For the scattering and detached classes, the NIRB and NIRF classes have been mixed by migration processes so that they are dynamically indistinguishable in the DES catalogs.

A similarly decisive result is also seen for the resonant  TNOs within the main belt range of $a.$ The Plutinos yield an $\mathcal{R}$ that is slightly above Jeffreys's threshold for ``strong'' evidence favoring inclination segregation of the NIRB and NIRF members.


In the previous sections, we have treated the classical component of the Kuiper belt as two distinct subpopulations, HC and CC, split by $I_{\rm free}.$ By dividing the classicals strictly by NIRB or NIRF color and introducing distinct inclination parameters $\kappa_\nirb$ and $\kappa_\nirf,$ we can ask whether the HC NIRFs are consistent with being the high-$I$ tail of the CC population, as opposed to having the HC and CC NIRFs be distinct.
Our data decisively rule out this scenario of a single NIRF component describing the (free) inclination distribution of both the cold and hot classical populations.  The very negative $\mathcal{R}$ value in the lower row of the left panel of Figure~\ref{im:kappa} rejects this case in favor of a trimodal classical distribution composed of a CC population heavily concentrated on the invariable plane, a less concentrated group of NIRF HCs, and an even more inclined group of NIRB HCs.

We also test whether there is any detectable distinction between the inclination distributions of the blue and red halves of the NIRB distribution, as defined in Section~\ref{sec:split}.  This is done by comparing the evidence ratio between a hypothesis with a single $\kappa_\nirb$ vs a model with distinct $\kappa_{\nirb+}$ and $\kappa_{\nirb-}.$ Neither the HCs ($\mathcal{R} = 0.678$), nor scattering ($\mathcal{R} = -0.805$), nor detached ($\mathcal{R} = -0.438$) populations exhibit any strong preference between a joint vs common inclination width of the red and blue ends of the NIRB sequence.

\begin{figure}[ht!]
	\centering
	\includegraphics[width=0.32\textwidth]{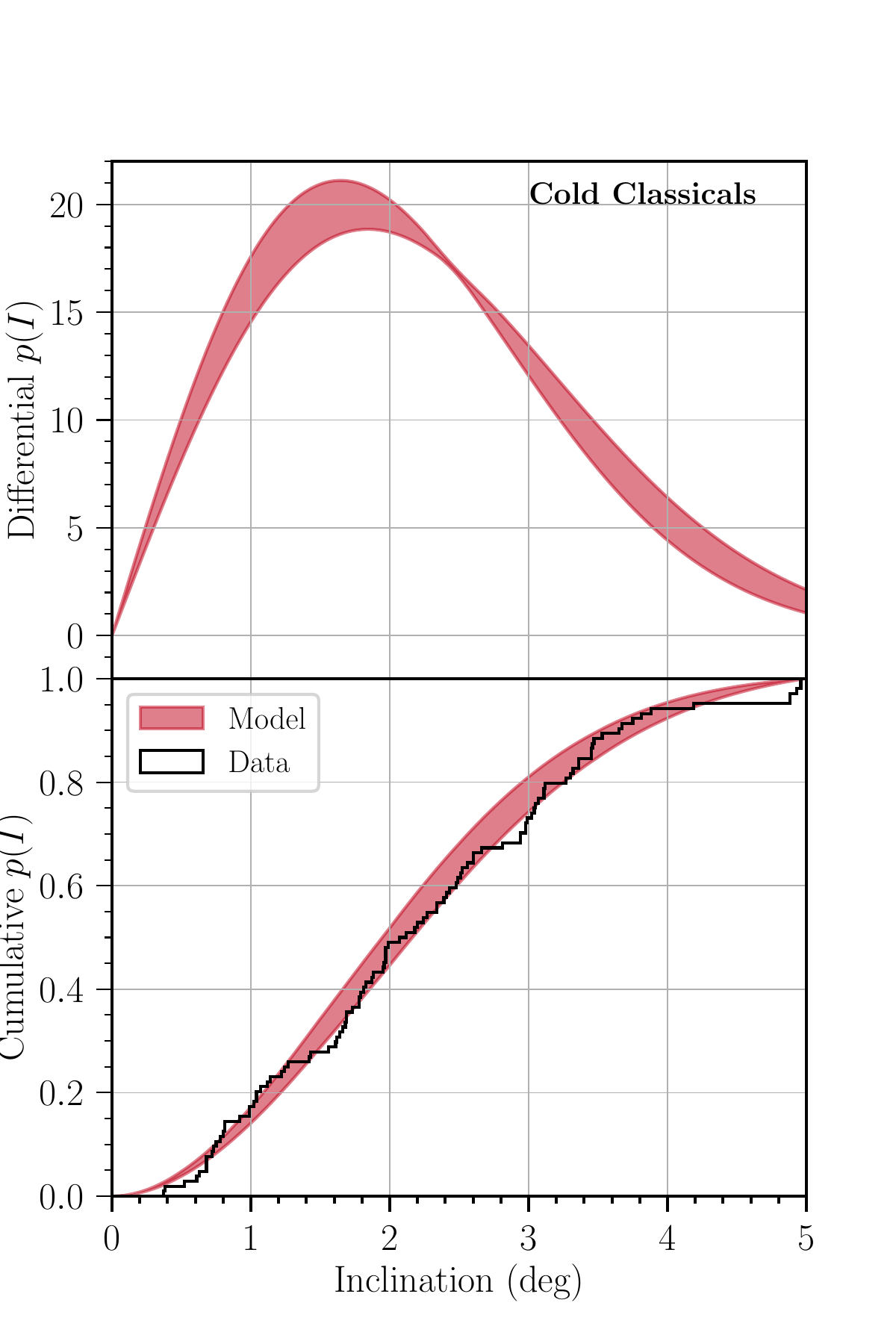}
	\includegraphics[width=0.32\textwidth]{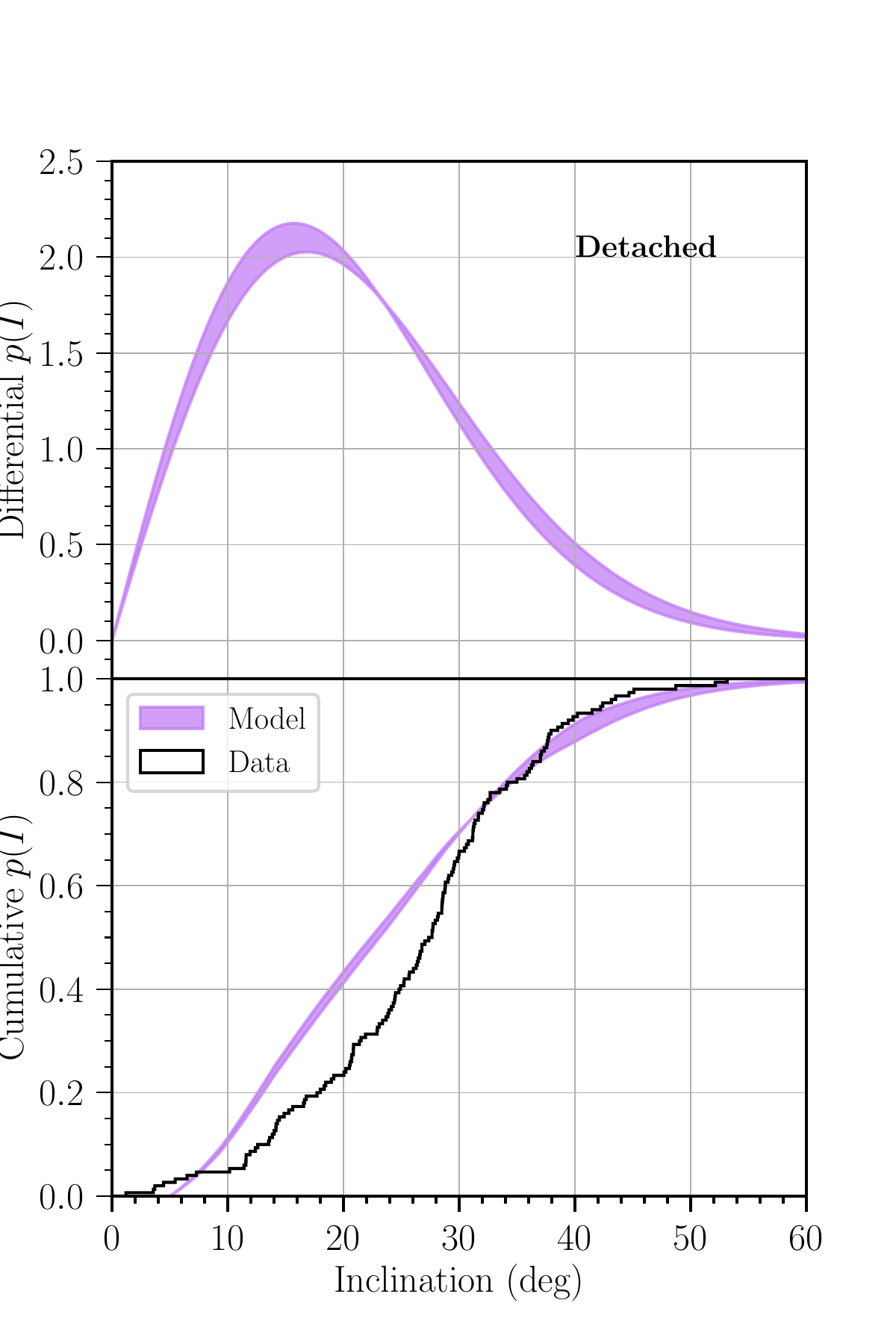}
	\includegraphics[width=0.32\textwidth]{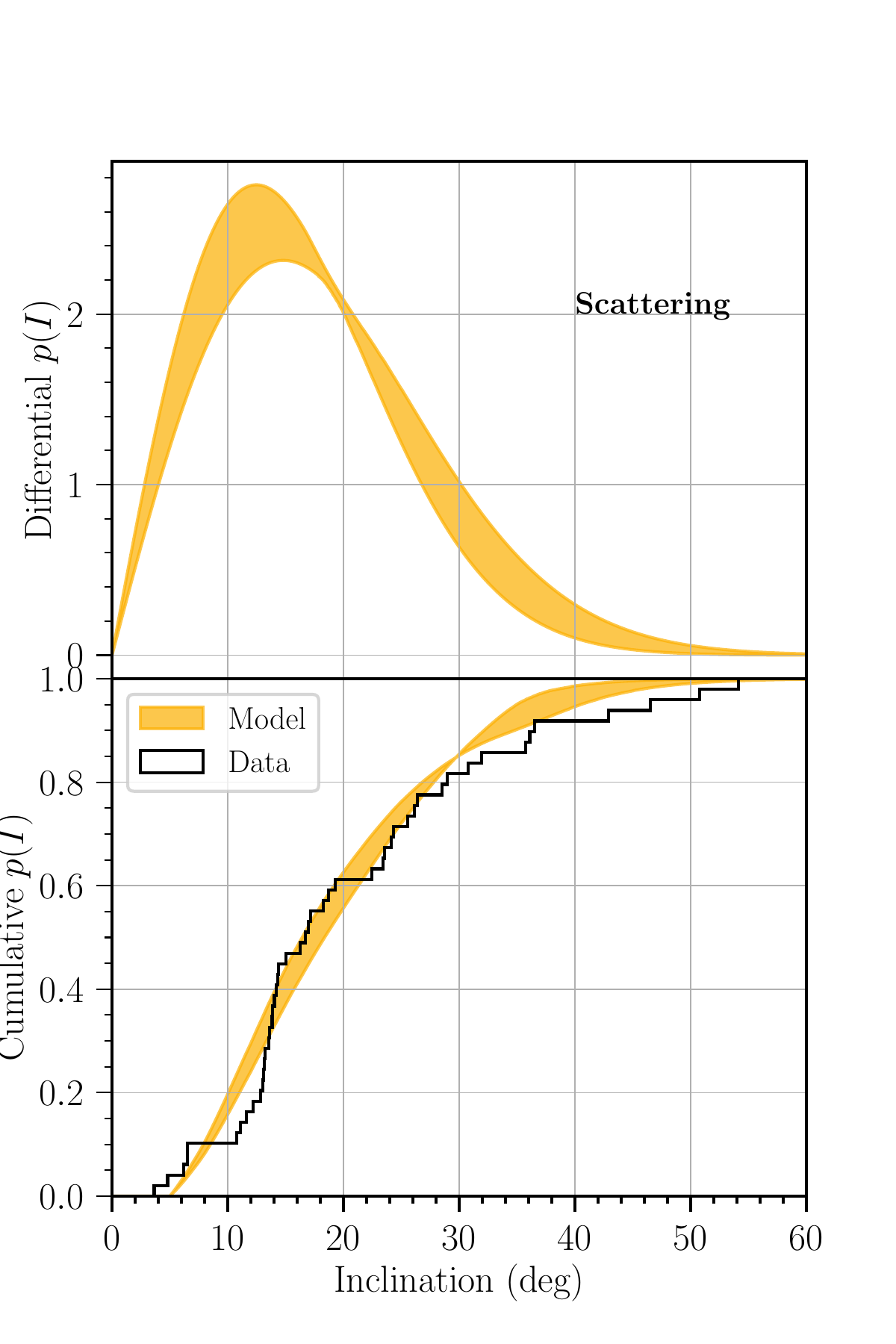}
	\includegraphics[width=0.32\textwidth]{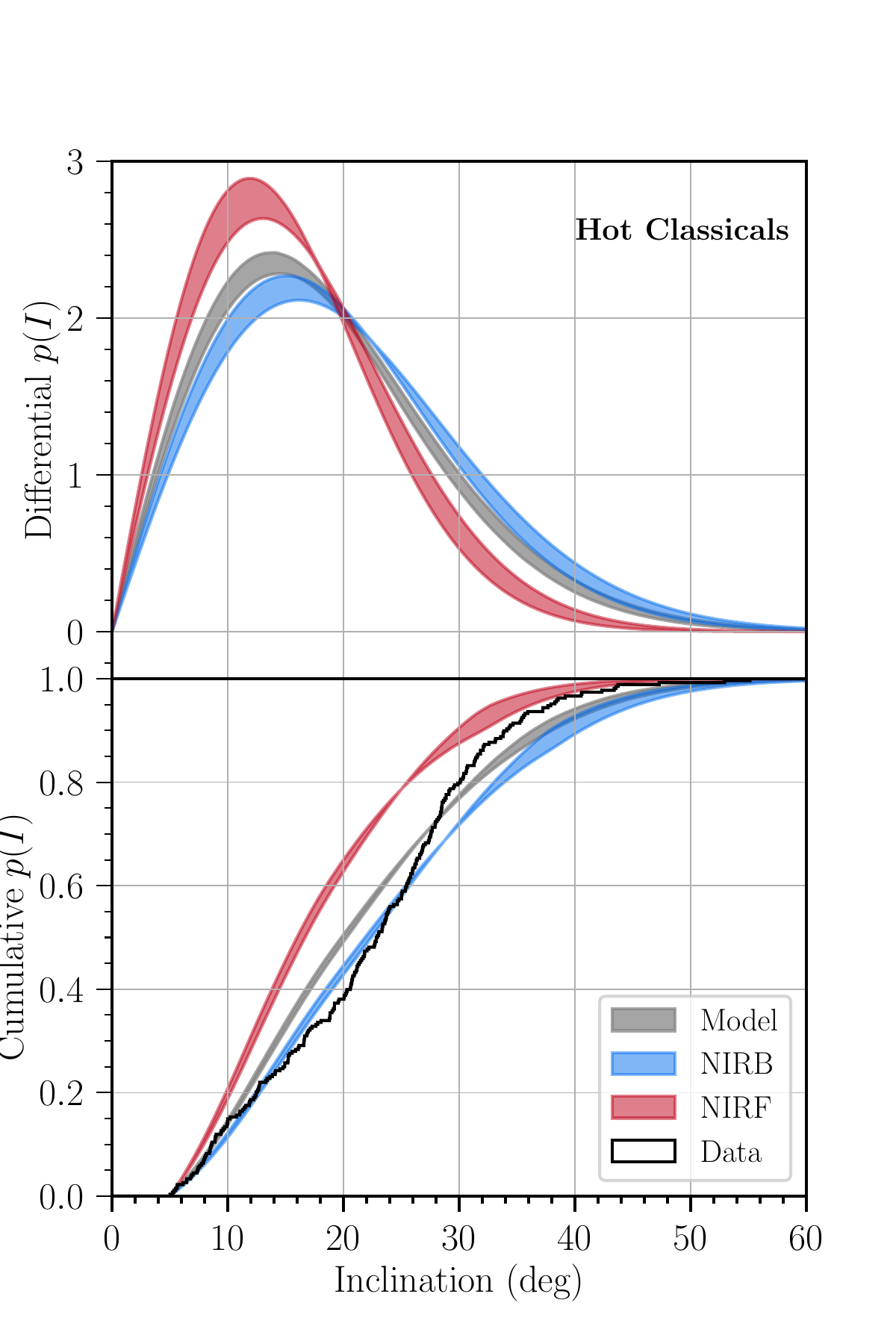}
	\includegraphics[width=0.32\textwidth]{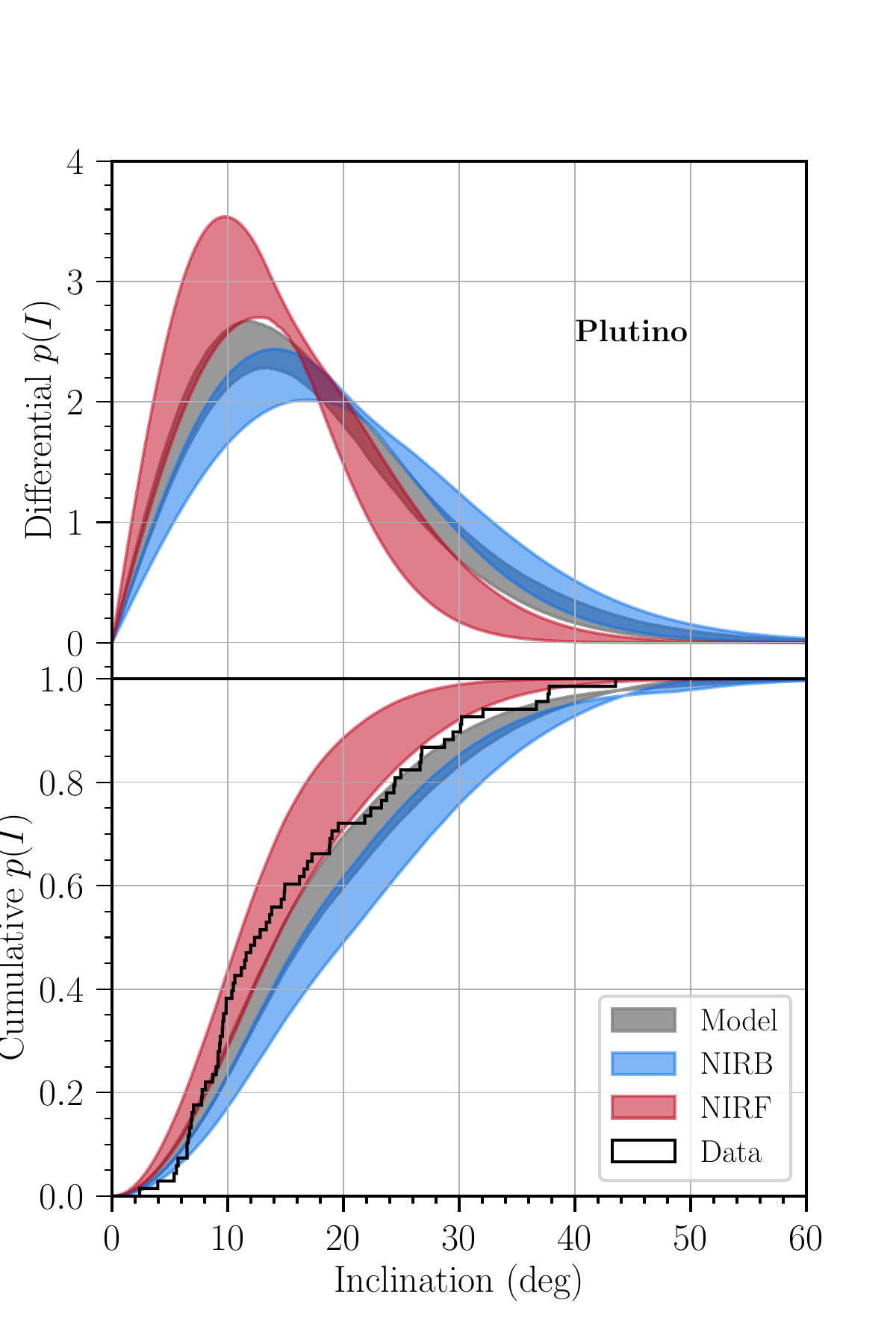}
	\includegraphics[width=0.32\textwidth]{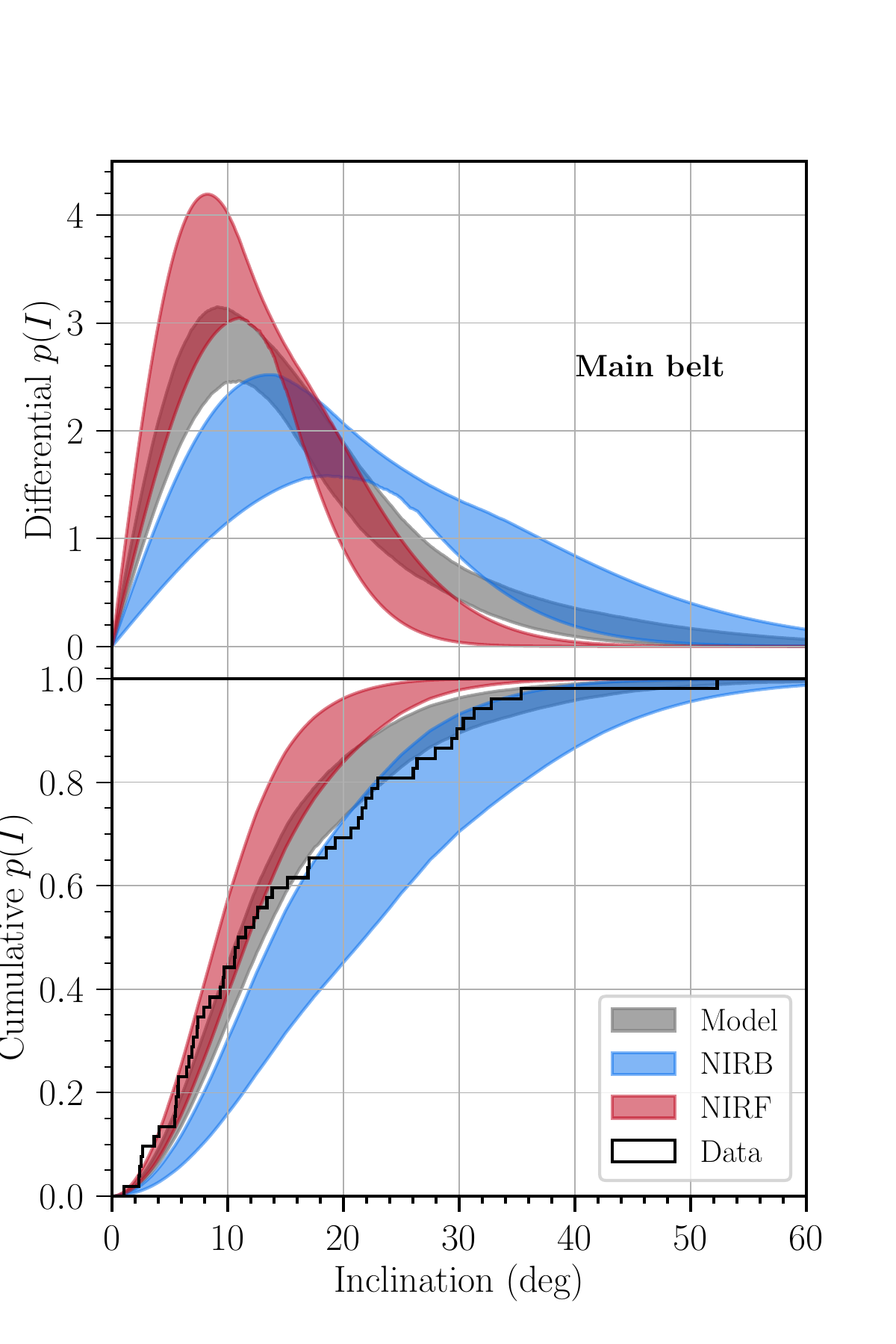}
	\caption{Each panel has differential (top) and cumulative (bottom) histograms of the 68\% confidence range of the model inclination distributions fit to the labelled dynamical class, where the model either a single von Mises-Fischer distribution or the sum of two for the NIRF and NIRB components (whichever is preferred by the evidence ratio).  The model curves include the observational selection function.
The cumulative distribution of the detected TNOs is shown in black.  The cold and hot Classicals show the free inclination (instead of the ecliptic inclination), and had their cumulative histograms truncated to $I_\mathrm{free} < 5\deg$ and $I_\mathrm{free} > 5\deg$, respectively. In addition to the best-fit model, the hot Classical, Plutino and main belt resonant population also shows the confidence limits for the NIRB (blue) and NIRF (red) model components.  The detached and HC inclination distributions are not as well fit by the models.}
\label{im:incdist}
\end{figure}

Figure \ref{im:incdist} allows visual examination of whether the von Mises-Fischer distributions are a good model for the inclination distributions of each dynamical classes.  For the CC, scattering, Plutino and main belt resonant populations, the best-fit von Mises-Fischer distributions for the NIRF and NIRB components sum to a cumulative $I$ distribution that is a good visual match to the data.
The HC and detached populations however, indicate poor fits, as has also been noticed by \cite{Malhotra2023} for the HC case---the actual distributions are more ``hollowed out'' at low inclinations than the models. Further investigation of alternative functional forms, then, is needed to find a more statistically appropriate description of these populations. Despite these differences, the HC NIRF/NIRB bimodality is still statistically robust. Furthermore, it explains the trend seen in Figure \ref{im:finc} (which does not impose any model to the inclination distribution). 

\subsection{Population estimates}
\label{sec:counts}
With the outputs of all the Markov chains in hand, we can use the Poisson distribution for $N$ derived in Equation~\ref{eq:Npoisson}
to infer the number of objects in a range of $H_r$ for each trans-Neptunian subpopulation. 
For dynamical subset $d$ with $N_{\mathrm{sample},d}$ objects in our sample, maximum-posterior estimate of the total number of objects $N_d$ in the $H_{r,\mathrm{min}} \leq H_r \leq H_{r,\mathrm{max}}$ range (from Equation \ref{eq:numbers}) is
\begin{equation}
	N_d =  \frac{N_{\mathrm{sample},d}}{\sum_{\beta} f_{\beta} S_{\beta}} \int_{H_{r,\mathrm{min}}}^{H_{r,\mathrm{max}}} \mathrm{d}H_r\, \sum_{\beta}  f_\beta p_\beta(H_r).
\end{equation}

In principle, the total number of objects over ranges of other characteristics can be similarly computed. The uncertainty in these numbers can be directly estimated by sampling from the chains, and accounting for Poisson fluctuations in the number of detections by replacing $N_{\mathrm{sample},d}$ by a random draw from a Poisson distribution with this mean value. We present these estimates for all of our subsets in Table \ref{tb:pop}.  For each population $d$, we use either a single inclination distribution or distinct NIRB and NIRF inclination distributions, whichever is preferred by the Bayesian evidence ratios in  Section \ref{sec:nirbnirf}. 
Our uncertainty estimates do \emph{not} include uncertainties in the completeness corrections derived in Appendix~\ref{sec:completeness}.  The $S_\beta$ have the most noise for types of orbits and $H$ values that are likely to generate $\lesssim 1$ detection in the DES observations, and the number estimates will not include any portions of a dynamical class whose $H$ and orbits bring them into observability by DES.  One must be careful in comparing population estimates of the detached objects, in particular, to others results, to insure that the selection regions and/or extrapolation into unselectable regions are comparable.

We can also use these results to derive mass estimates, for a given choice of albedo and density. To relate our absolute magnitude to a TNO volume, following \cite{1916ApJ....43..173R}, we have for a spherical body with uniform surface:
\begin{align}
	H_r = m_{\sun,r} - 2.5 \log_{10} \frac{\nu_r R^2}{(1 \, \mathrm{au})^2} 
  \\ \implies  R(H_r) = \frac{1.496 \times 10^8 \, \mathrm{km}}{\nu_r^{1/2}} 10^{-0.2 (H_r - m_{\sun,r})}
  \label{eq:RHr}
\end{align}
where $m_{\sun,r} =  -26.96$ is the magnitude of the Sun \citep{Willmer2018}, $\nu_r$ is the geometric albedo in the $r$ band and $R$ is the radius.   For a spherical object, $V = 4 \pi R^3/3,$ but approximately zero of our TNOs are likely to be spherical.   A crude estimate of the effect of asphericity on the volume estimate of a TNO with given lightcurve-averaged apparent magnitude $H_r$ comes from considering an oblate spheroid rotating around a short axis normal to the line of sight.  If the short and long semi-major axes of the body are $a$ and $c$, respectively, then the minimum apparent flux is $f_{\rm mean}(1-A) \propto a^2$, and the lightcurve maximum is $f_{\rm mean}(1+A) \propto ac$, hence $c/a=(1+A)/(1-A).$ The volume is $V =  4\pi a^2c/3 =  4\pi R^3 (1+A)\sqrt{1-A}/3$ if $R$ is derived from Equation~\ref{eq:RHr} using the mean flux to estimate $H.$ Integrating the $(1+A)\sqrt{1-A}$ factor over the $A$ distributions that best fit the NIRB and NIRF populations leads to $\lesssim5\%$ changes in the estimated mean volume from the spherical ($A=0$) case.  This is much smaller than the uncertainties in albedo and density, so we will ignore shape effects.

The total mass of the population, then, becomes:
\begin{subequations}
\begin{align}
	M  =  &  \int \mathrm{d}H_r\, n(H_r) M(H_r) =  N \sum_\beta f_\beta \int  \mathrm{d}H_r  \, p_\beta(H_r) M(H_r) \\ & = N \times K \times \sum_\beta f_\beta \frac{\langle \rho_\beta \rangle}{\langle \nu_{r,\beta} \rangle^{3/2}} \int \mathrm{d}H_r \, p_\beta(H_r) 10^{-0.6 H_r} , 
\end{align}
\end{subequations}
where
\begin{equation}
	K = \frac{4\pi}{3} \times 10^{0.6 m_{\sun,r}} \times \left(1.496 \times 10^ 8\, \mathrm{km} \right)^3
\end{equation}
is a constant. Note that we have taken the average albedo $\nu_r$ and density $\rho$ for each family $\beta$. As we can adopt $p_\nirf(H_r) = p_\nirb(H_r)$ within the accuracy of the DES survey, the sum over the species $\beta$ then only depends on potential differences in albedos and densities.

\citet{lacerda2014a} use thermal-IR measurements to estimate albedos, finding that the bluer objects have a mean albedo $\nu_r = 0.05$, while redder objects (such as CCs) are more reflective with $\nu_r = 0.15$. We adopt these values to be representative of our NIRB and NIRF classes, respectively. Trans-Neptunian densities are less constrained for objects in our size range, requiring a combination of a well measured binary system and either occultation even or thermal measurements are needed to obtain the size of each object. Following \cite{2020tnss.book..153M}, we adopt a reference value of $\rho = 1\, \mathrm{g}/\mathrm{cm}^3$ for both classes. 

We report our mass estimates in our reference size range, as well as extrapolations to $5 \leq H_r \leq 12$, in Table \ref{tb:pop}. \change{We note that while DES can only observe the L4 Trojan cloud \citep{Lin2019}, the simulation of Appendix \ref{sec:completeness} implies that the figures given for total Trojan counts inherit an assumption of equal occupation of the L4 and L5 clouds.}

\begin{deluxetable}{l||ccc||ccc|c}[ht!]
	\tabletypesize{\footnotesize}
	\tablecaption{Population estimates\label{tb:pop}}
	\tablehead{ \colhead{} &  \multicolumn{3}{c}{Number of objects $5.5 \leq H_r \leq 8.2$}  &  \multicolumn{4}{c}{Mass estimates ($\times 10^{-4} M_\Earth$)}  \\  \colhead{Subset} & \colhead{Joint}  & \colhead{NIRB} & \colhead{NIRF} & \colhead{$5.5 \leq H_r \leq 8.2$}  & \colhead{NIRB} & \colhead{NIRF} & \colhead{$5 \leq H_r \leq 12$} }
	\startdata
	CKBO & $5556^{+606}_{-595}$ & $64^{+72}_{-40}$ & $5477^{+607}_{-583}$ & $2.69^{+0.41}_{-0.36}$& $0.15^{+0.17}_{-0.09}$ & $2.51^{+0.35}_{-0.33}$ & $7.39^{+1.12}_{-0.93}$  \\
	HKBO & $11487^{+816}_{-815}$ & $6599^{+636}_{-612}$ & $4866^{+612}_{-566}$ & $18.0^{+1.4}_{-1.3}$ & $15.8_{-2.2}^{+2.2}$ & $2.23^{+0.36}_{-0.31}$ & $49.4^{+5.7}_{-4.9}$ \\
	Detached & $17878_{-1627}^{+1775}$ & $12147^{+1524}_{-1373}$ & $5654^{+1078}_{-948}$ & $31.7^{+4.9}_{-4.7}$ & $29.0^{+4.8}_{-4.3}$ & $2.59^{+0.58}_{-0.48}$ & $87.2^{+11.6}_{-10.3}$  \\
	Scattering & $2298^{+448}_{-419}$ & $1525^{+380}_{-341}$ & $738^{+284}_{-228}$ & $3.97^{+1.06}_{-0.86}$ & $3.62^{+1.04}_{-0.85}$ & $0.34^{+0.14}_{-0.11}$ & $10.9^{+2.6}_{-2.3}$  \\ \hline
	Trojan & $139^{+78}_{-56}$ & $93_{-41}^{+57}$ & $43_{-24}^{+38}$ & $0.25_{-0.10}^{+0.15}$ & $0.22_{-0.11}^{+0.15}$ & $0.02_{-0.01}^{+0.01}$ & $0.68^{+0.39}_{-0.28}$\\
	Inner belt & $621_{-173}^{+173}$ & $342_{-119}^{+145}$ & $253_{-96}^{+130}$ & $0.94_{-0.30}^{+0.37}$ & $0.81_{-0.29}^{+0.37}$ & $0.12_{-0.05}^{+0.06}$ & $2.59_{-0.80}^{+0.96}$\\
	Plutinos & $2453_{-343}^{+372}$ & $1312_{-248}^{+289}$ & $1312_{-236}^{+274}$ & $3.66_{-0.371}^{+0.81}$ & $3.13_{-0.68}^{+0.78}$ & $0.51_{-0.12}^{+0.14}$ & $10.1_{-1.8}^{+2.1}$\\	
	Main belt & $6042_{-921}^{+1026}$ & $1682_{-709}^{+726}$ & $4292_{-888}^{+1100}$ & $5.9_{-1.5}^{+1.8}$ & $4.0_{-1.7}^{+1.8}$ & $1.9_{-0.4}^{+0.5}$ & $16.5_{-4.3}^{+4.8}$ \\	
	2:1 & $3358^{+866}_{-769} $ & $1186_{-470}^{+569}$ & $2092^{+736}_{-604}$ & $3.8^{+1.5}_{-1.1}$ & $2.8_{-1.1}^{+1.4}$ & $0.96_{-0.29}^{+0.35}$ & $10.6_{-3.2}^{+3.8}$ \\ 
	5:2 & $2818_{-771}^{+894}$ & $2189_{-803}^{+695}$ & $553_{-294}^{+444}$ & $5.5_{-1.7}^{+2.0}$ & $5.2_{-1.6}^{+2.0}$ & $0.25_{-0.13}^{+0.20}$ & $15.1_{-4.6}^{+5.5}$ \\ 
	Distant & $15146_{-2800}^{+2610}$ & $8969_{-2029}^{+2318}$ & $5913_{-1625}^{+2024}$ & $24.2_{-5.3}^{+6.2}$ & $21.3_{-5.1}^{+6.2}$  & $2.7_{-0.8}^{+1.0}$ & $67_{-14}^{+17}$  \\ \hline
	Total & $77155_{-6242}^{+11795}$ & $36250_{-6190}^{+4488}$ & $38678_{-5531}^{+18649}$ & $104_{-14}^{+14}$ & $85_{-17}^{+15}$ & $18_{-3}^{+8}$ & $288_{-30}^{+34}$ \\
	\enddata
	\tablecomments{All values correspond to the 68\% confidence limits derived from the HMC chains.}

\end{deluxetable}

\section{Discussion and implications}
\label{sec:discussion}
The analysis presented herein is purely data-driven: no pre-existing population models of the TNO population were assumed, aside from the separability of the ${\bf c}, H_r,$ and $A$ distributions within a physical family, and rotational symmetry round the ecliptic plane for the populations of non-resonant TNOs.  This leads to statistically robust constraints---our measurements are ``observable'' properties of the TNO sample that are valid for any population model.  We have not coupled the color distributions with specific models of planetesimal formation and migration models, as done \eg\ by \citet{buchanan2022}, instead discussing herein the general properties that any model must have to be consistent with our data, and supplying enough information that any investigator can check quantitative compatibility of their model with our inferences on the true population.

\subsection{Framework}
\label{sec:framework2}
The analysis of the full DES population requires novel techniques for statistical inference of physical and dynamical properties from the largest catalog to date of trans-Neptunian objects orbital elements, absolute magnitudes, colors, and lightcurve amplitudes.  The DES catalog differs from the ColOSSOS data used by \citet{fraser2023} to investigate color distributions in that ColOSSOS obtains followup multiband observations, usually 1 object at a time, of a subset of OSSOS discoveries, aiming to integrate long enough to place each source unambigously in \eg\ one of the modes of the $g-r$ color distribution, which is a substantial investment in telescope time, and requires careful treatment of the selection process and success rate for followup.  DES, on the other hand, has color data over the complete sample without the need for expensive followup data or additional selection effects---but the per-object color uncertainties for many members of the sample are too large to resolve the color differences between families.  Our methods must therefore accomodate probabilistic assignment of individual objects to test population models.  The reliability of such results depends critically on having accurate estimates of the flux uncertainties and selection criteria of the survey.  Fortunately, for a multi-purpose sky survey such as DES or the upcoming LSST, there are many science pursuits that require, and will share the effort to obtain, careful point-source characterization.

Characterization of the variability distribution of TNOs has also generally been limited to sources with dedicated followup observations, which are expensive and involve some bias in the choice of followup targets.  Studies such as \citet{Showalter2021} that collect data from the literature are additionally subject to a strong unknown ``publication bias,'' in that many observers will only publish light-curve data if variability is strongly detected, with poor or absent characterization of upper limits for other sources.  The many observations per source in DES allow us to produce a well-characterized $p(A)$ for every source, even if there is no definitive detection of variability, and combine these into constraints on the underlying distributions.

We have adapted the Gaussian mixture model technique to accomodate the inference of underlying color distributions from data with varying levels of measurement error and unknown number of clusters.  The method allows us to infer, for example, a bimodality in the intrinsic $r-i$ color distribution (Figure~\ref{im:model}), which is not visible in the ColOSSOS data of \citet{fraser2023} and would not be apparent in a histogram of our point-estimate $r-i$ colors either.
While tailored for our analysis of the color distribution of TNOs, the algorithm can be applied to any other similar clustering problems. The mathematical framework established in Section \ref{sec:framework} substantially extends previous Bayesian analyses of trans-Neptunian size distributions, and enables a joint characterization of TNO colors, sizes, and lightcurve properties, as a function of families of objects and with rigorous treatment of measurement uncertainties, as well as perform robust hypothesis tests between competing representations of the data. Our methodology is will scale gracefully to future data sets, particularly from the LSST \citep{Ivezic2019}, that will produces catalogs of 10--100$\times$ more TNOs than DES with multi-year astrometric and photometric data.

\subsection{Two physical families across dynamical groups}
\label{sec:2fam}
Our unsupervised Gaussian mixture analysis independently arrives at a similar conclusion to \cite{fraser2023}: there are (at least) two color families in the TNO region, with the main color sequences ``NIRB'' and ``NIRF'' seen in Figure~\ref{im:model}.
The HC, detached and scattering objects, and most of the mean-motion resonances, are all NIRB-dominated, while the CCs are exclusively NIRF to within our measurement accuracy.  With the fundamental assumption that surface colors in our $H$ regime were established at birth, and the current paradigm that CCs were born \textit{in situ} but the excited populations have been scattered outwards to their current locations, our data very strongly supports the conclusion that the NIRF population formed external to the region that sourced the NIRB component.

The interpretation of the color families as birth populations is bolstered by our finding that all members of the NIRB color sequence are consistent with a common absolute magnitude distribution, and a common variability distribution, independent of which dynamical class they currently occupy, and likewise for the NIRF TNOs.

Assuming that NIRF and NIRB colors are indeed established at birth, the NIRB/NIRF fractions of current dynamical populations inform their dynamical histories.  We find that the populations of HC, scattering, and detached TNOs, plus the Neptune Trojans and most resonances, are consistent with $f_\nirb=0.6$ to $0.7.$ A common $f_\nirb$ would mean that, to first order, these dynamical classes' members are drawn from a common range of heliocentric birth distance, as would occur if their dynamical pathways all involved the same first step with a randomized final placement.  An important exception is that the 5:3, 7:4, and possible 2:1 resonances each contain significantly larger fraction of NIRF TNOs than the other dynamically excited classes.  This would be the case if many of the 5:3 and 7:4 resonators were formed as CCs close to their current semi-major axes, fell into their resonances after or in the late stages of Neptune's migration, then had their inclinations excited by, for example, Kozai cycles \citep{Lykawka2005}.  They would be mixed in these resonances with NIRB objects transported during migration.

With the larger samples from LSST, it should be possible to measure $f_\nirb$ more precisely in each class and resonance, and see if any further differentiation of dynamical pathways is discernible.

\subsection{Comparing NIRF and NIRB}
\label{sec:compare}
The DES data are consistent with \emph{all} TNOs sharing a common $H$ (and presumably size) distribution in the range $5.5<H_r<8.2$ (Figure~\ref{im:rolling}) This pushes the agreement in $p(H)$ noted among dynamical classes by  \citet{Adams2014} to a magnitude deeper, and extends the common $p(H)$ noted between the two modes of $g-i$ color by \citet{Wong2017}, and between the dynamically cold and hot populations by \citet{petit2023}.
The preference for a rolling power-law distribution for all dynamical subpopulations also agrees with the deviation from a single power law seen in the Plutino population by \cite{Alexandersen2016} and in the scattering population by \cite{Lawler2018a}.  Our result, then, brings together several hints of a universal $p(H)$ in this size range seen in other datasets. This is a strong argument that the physical process of planetesimal formation was coherent across the entire span of TNO's birth populations, despite the change in local environments implied by the NIRF/NIRB color transition and some heliocentric radius. Any subsequent collisional evolution of the size distribution would need to be small and/or similar enough in the dynamically hot and cold populations to maintain the observed $p(H)$ agreement.

The light-curve amplitude distributions $p(A)$ are discernibly different between the NIRF and NIRB populations, in contrast to the common $p(H),$  with the NIRF TNOs being more photometrically variable than the NIRBs (Figures \ref{im:lcadynamics} and \ref{im:lcadist}).  This difference can be induced by the NIRF objects being less homogeneous in surface albedo; comprising higher-ellipticity bodies; having a higher fraction of contact binaries; having a stronger tendency of rotation poles to be normal to the ecliptic; or any combination of these effects \citep[see][for a review]{Showalter2021}.
One economical hypothesis would be that all TNOs were created with spin poles close to normal to the ecliptic plane, which the NIRFs maintain, but having spin poles randomized during migration to make the NIRF objects more isotropic and less apparently variable. A weak argument against this scenario is that the only small CC with a known spin pole, Arrokoth, has its pole close to the ecliptic plane \citep{Stern2019}. \change{This scenario also seems to be ruled out by our very strong determination that the dynamically excited NIRF population is more variable than the NIRB TNOs even when removing the CCs. As such, our analysis points towards surface features and shapes as the likely scenario, motivating further investigations of these trends using the upcoming LSST data.}  


\subsection{Inclination information}
\label{sec:inc}
The inclination-distribution measurements of the DES TNO populations in Section~\ref{sec:vMF} improve on previous efforts by having larger numbers of sources, and by the substantial high-ecliptic-latitude coverage of the DES imaging. TNO emplacement models first must be able to reproduce the observed widths of each dynamical group's inclination distribution (as measured by $\log\kappa$ in the von Mises-Fisher distribution), but also must reproduce any correlations with color.

\subsubsection{Population widths}
Figure~\ref{im:kappa} shows that the detached population has the broadest inclination distribution of all dynamical subpopulations (lowest $\log\kappa$), with the NIRB HCs and scattering populations similar or wider, as expected due to dynamical effects related to Neptune's migration \citep{Gomes2003}. Comparisons between our results and the inclination distributions produced by different migration scenarios, then, are of interest \citep[see, \eg][]{Nesvorny2015,volk2019}. Similarly, our analysis also shows a rich structure in the inclination distribution of the resonant subpopulations, being an additional constraint for migration scenarios.

Our measurement of the cold Classical inclination distribution is in direct agreement with previous measurements \citep{VanLaerhoven2019,Malhotra2023}, indicating the undisturbed nature of this population \citep{Dawson2012,Nesvorny2015b,Batygin2020}. Our data also conclusively exclude any hope that the 
the NIRF HCs are the high-$I$ tail of the \textit{in-situ} CC population, despite their similar physical properties.

The inclination distribution of our NIRF component of the HCs,  $\kappa_\nirf = 20.9 \pm 1.9,$ agrees with that measured for full HC population by \citet{Malhotra2023}, who report $\kappa = 19.7 \pm 1.9$, ($2\sigma$ limits). Our analysis, however, demonstrates the need for two distributions, with a more inclined NIRB component with $\kappa_\nirb = 13.25 \pm 0.88$.  As \cite{Malhotra2023} used the entire Minor Planet Center (MPC) sample for their analysis, the agreement between their measurements and our NIRF components indicates that high-$I$ HCs are relatively poorly represented in the MPC, reflecting the tendency for TNO surveys to target fields near the ecliptic plane. Furthermore, without the color information as a discriminant, there is no other simple way to decouple these two distributions, which explains why, since \cite{Brown2001}, the HC inclination distribution has been treated as a single component.

\subsubsection{Inclination-color relations}

Several of the tests in this paper search for additional information on the birth-to-present dynamical pathways beyond the basic NIRB/NIRF split.  Essentially, we ask whether the current dynamical groups ``remember'' anything further about their birth circumstances beyond their $f_\nirb$ value.  Section~\ref{sec:vMF} tests whether the current dynamical state, namely inclination, differs for the NIRF vs NIRB components of a given dynamical family.  For the HC, detached, and scattering TNOs, we can consider the NIRF and NIRB members of the dynamical family to represent TNOs born in the outer and inner regions of their source population, respectively.   As seen in Figure~\ref{im:kappa}, the HCs show conclusive differences between the inclination distributions of their NIRF and NIRB populations.  In each case, the NIRB population has lower $\kappa$ ($\Rightarrow$ higher inclinations) than the NIRF.  For the HCs, this means that those that have migrated further in $a$ (the NIRBs) have ended up with larger present-day inclinations.

The difference in inclination widths of the HCs' NIRF and NIRB populations seen in Figure~\ref{im:kappa} also manifests as the inclination dependence of $f_\nirb$ visible in Figure~\ref{im:finc}, and seen in previous works \cite[\eg][]{marsset2019}.  Such a correlation has been posited in the dynamical models of \cite{Nesvorny2020}, but we note that our sample span a broader range of inclinations than those produced by their model. \cite{marsset2023} claim that, beyond the color-inclination correlation, colors and inclinations are also correlated \emph{within} their ``brightIR'' class for the HCs---similar to our NIRB population. Despite a $\gtrsim 3\times$ larger sample of HCs that span a broader range of inclinations than those used by \cite{marsset2023}, we were not able to detect differences in the inclination distributions for the redder and bluer halves of the HC NIRB population. 

The right panel of Figure~\ref{im:kappa} shows that statistically significant differences in inclination widths are present between the NIRF and NIRB components of the combined 7:4 and 5:3 resonances lying within the classical belt; likewise for the 3:2 Plutinos just interior to the classical belt.  These are the two resonance samples with the largest number of DES detections, hence similar patterns could be present but undetected within other resonances.
If the Main Belt resonants are a mix of migrated objects (mostly NIRB) with former CCs (which are NIRF) captured into the resonance near the end of (or after) Neptune's migration, then the implication of our observations is that the migrated objects have moved into higher-inclination orbits than the former CC's.  The Plutinos also show evidence that higher inclinations are currently held by objects that have migrated longer distances within the resonance.

We also note an apparent disagreement between our $f_\nirb$ for the Plutinos and the equivalent quantity measured by \cite{pike2023}: in our units, they report $f_\nirb \approx 0.11$, while we see $f_\nirb \approx 0.56$. This discrepancy could result from an underestimate of the high-inclination population in their sample, because we find that the higher-$I$ Plutinos are more likely to be NIRB, leading to an underestimate of $f_\nirb.$ 
The most inclined Plutino in the \cite{pike2023} sample has $I = 24.9\degr$, while our Plutinos extend to $I = 43.0\degr.$ This has also been noted by \cite{cameron}. 
Another point of comparison for the Plutinos is the measurement by \cite{matheson2023} of $\kappa \in [27.3, 36.3]$ ($3\sigma$ bounds) for the full sample of 431 reliably known Plutinos. Our result for the full DES Plutino population is  $\kappa = 17.9 \pm 4.0$, which $2 \sigma$ below the lower $3\sigma$ bound of \cite{matheson2023}.  We measure $\kappa_\nirf = 25.6 \pm 7.5$ and $\kappa_\nirb = 13.7 \pm 2.6$---the former is consistent with their result, and it is possible that the methods for correcting for completeness of the full MPC sample are under-representing the higher-$I$, NIRB-dominated Plutinos.

\subsection{Departures from a two-component model}
\label{sec:departures}
A simple scenario has all CCs born exterior to some heliocentric radius while all dynamically excited TNOs form interior to this radius, with the boundary lying somewhat exterior to the primordial transition from NIRB to NIRF surfaces.
The inclination studies provide strong evidence this scenario is inadequate, since the HCs and Main Belt resonators show different inclination distributions for NIRF and NIRB members, a ``memory'' of location within their birth regions.  There are other statistically significant trends in the DES catalog that are in conflict with the simple two-zone scenario.

First, there are objects in the selected $H_r$ range whose colors deviate significantly from the NIRB/NIRF model. While some of these can be assigned to the Haumea collisional family, whose unusual colors have been attributed to a relatively recent exposure of the interior of a large parent body, the other color outliers we identify require a different mechanism. One simple explanation is that these objects have poor photometric measurements, and high-$S/N$ follow-up observations would show that these belong to the NIRB or NIRF classes. However, if our color measurements are accurate, another explanation will be needed: have these objects formed in different regions of the Solar System, then transported as interlopers into the trans-Neptunian region? Have these objects experienced unusual surface evolution, or, less likely, are these objects part of some other (unidentified) collisional family?   These questions motivate followup observations of the outliers.

Second, we find in Section~\ref{sec:split}, and illustrated in Figures~\ref{im:ternary} and \ref{im:ternary_nirf}, that not all the dynamical classes sample equally from the red and blue ends of the NIRB and NIRF color sequences.  Let us postulate \citep[as per][]{Nesvorny2020, buchanan2022} that color distribution was determined by radius in the Solar System's natal protoplanetary disk, with each color family being formed in a different stratum of the protoplanetary disk, and that migration processes are color-independent.  Then
if each of these regions (namely NIRF and NIRB) had a homogeneous color distribution, then any sampling of objects formed inside them would also produce a homogeneous distribution, and the NIRF (or NIRB) members of all dynamical classes must share a color distribution within the NIRF (NIRB) locus. It is clear from our analysis, however, that this is not the case: The HC NIRBs are bluer, on average, than detached$+$scattering NIRBs.  HC and detached NIRFs are significantly bluer, on average, than CC NIRFs.  Two possibilities exist: one is that post-natal processing changes surface colors, \eg\ some collisional evolution or radiolysis that depends on late-time location and/or dynamical state.  A more interesting scenario, though, might be that there is some further radial compositional stratification or gradient within both the NIRB and NIRF formation regions, coupled with radius-dependent transport into the current HC and detached dynamical states.
With the larger TNO samples and more definitive color assignments to be available from LSST, we might be able to better localize (in the current dynamical space) the regions of gradients in occupancy of the color loci, and use this to unwind the originating natal gradients and the differential transport mechanisms.

\subsection{Population estimates}
\label{sec:pops}
Comparison of our total population estimates (Table~\ref{tb:pop}) with previous estimates reveals agreements and some puzzling discrepancies. Our estimates for the Plutino, 2:1 and 5:2 resonant populations \change{agree with (at similar or better precision than) the OSSOS estimates \citep{Volk2016,Chen2019};} and our main belt populations (5:3 and 7:4) agree with the estimates given by \cite{Gladman2012}, once we convert their $H_g$ using the nominal $g-r$ for the NIRF population.

Our population estimate for the detached population is consistent with (and less uncertain than) that of \cite{beaudoin2023}, and shows that the size of the detached population is comparable to that of the main Kuiper belt. This large population size and the broad inclination distribution of the detached populations imply that a substantial fraction of TNOs that the upcoming LSST will find are from this population, although this population will be less represented in surveys of the ecliptic plane such as DEEP \citep{Trilling2024}. We should reiterate that detached TNOs with high perihelia that have zero selection probabilities at DES depth regardless of orbital phase will not be counted in our population estimates---direct comparison of different surveys' estimates for this population should probably include careful matching of selection criteria.

\cite{crompvoets2022} estimated that a large population of objects in distant resonances is needed to explain the number of such objects in the OSSOS (and accompanying) surveys. Here, we directly confirm their measurement: we see that the population of objects in resonances beyond 60 au is comparable to the detached population. These objects also have a wide inclination distribution and have a similar color occupation distribution to the detached population. It has been posited that a portion of the detached population is composed of ``resonant dropouts,', objects that were transported by Neptune's migration and were dropped off sunward of the resonance (see \citealt{Nesvorny2016,Kaib2016}, see also \citealt{lawler2019} and \citealt{Bernardinelli2022} for observational tests of this hypothesis). Such similarities, then, favor this common origin scenario for these two populations.

One discrepancy is that our total CC and HC population inferences are $1.5-1.8\times$ smaller than those of \citet{kavelaars2021} and \citet{petit2023} for the same $H_r$ range.  The best-fit absolute magnitude distribution from \cite{kavelaars2021} with the normalization in \cite{petit2023} predicts that DES would have discovered $\approx 150$ CCs with the CFEPS-L7 orbital distribution model \citep{Petit2011}, whereas we find 102 (Table~\ref{tb:sample}). The underlying surveys are fairly complementary: OSSOS is deeper, while DES is wider (with an effective CC search area of $\approx 200 \, \deg^2$). A careful joint analysis of the surveys would resolve this inconsistency. Despite such differences, \citet{petit2023} report their ratio of HCs to CCs as 2.2, consistent with our $1.9 \pm 0.2.$

Our total size for the scattering population is also inconsistent with that of \cite{Lawler2018a}: after scaling their absolute magnitude distribution to our limits: their reported population size is $\approx 2.3\times$ larger than ours. This discrepancy can be partially attributed to differences in the dynamical classification, that is, the distinction between scattering and detached objects, but also derived from our shallower $H_r$ distribution. We note, however, that our population estimate for this population agrees with (and is less uncertain than) that of \cite{Petit2011}.

Our population estimates show that the number of NIRB and NIRF objects are comparable in the entire TNO region: while the NIRB population is clearly dominant in the dynamically excited populations, these numbers are balanced by the large number of CKBOs and the abundance of NIRF in the main belt resonances and the 2:1 population. However, even with the comparable numbers, the majority of the trans-Neptunian mass comes from the NIRB population. This is primarily due to the smaller albedos, therefore demanding larger objects at the same $H_r$. This is compatible with scenarios where there is a transition in surface density in the protoplanetary disk \citep[see, for example,][]{Nesvorny2020}. We note that, in the same size range, our mass estimates for the cold Classical belt are substantially smaller than those of \cite{petit2023}, who reports a $\approx 2.8\times$ larger mass for the entire cold belt when compared to ours, and \cite{napier2024}, who reports a $2.3\times$ larger mass. This difference can partially be traced to our inconsistent population sizes compared to \cite{petit2023} (note that \citealt{napier2024}'s small area leads to a poor constraint of the normalization or the bright end of the $H_r$ distribution), and the less steep behavior of our best-fit $p(H)$ when extrapolated to $H_r > 8.2$.

\section{Summary}
\label{sec:summary}
We examine the relations between the physical properties (color, size, shape)  and the dynamical properties of the sample of 696 trans-Neptunian objects with $5.5<H_r<8.2$ detected in the Dark Energy Survey, which yield constraints on the ``birth'' locations and emplacement processes of TNOs currently in different dynamical states.  All objects have, in addition to multiple years of astrometric observations, measurements of absolute magnitudes, colors, and lightcurve amplitudes, as well as well-characterized discoverability conditions.  While the measurements of individual TNOs' colors and lightcurve amplitudes are not all sufficiently precise to assign definitive color classifications or variability states, the uncertainties on all measurements are carefully characterized so that we can infer accurate population-level statistics using probabilistic assignments. Our main findings are as follows:
\begin{itemize}
\item We confirm and strengthen the observation by \citet{fraser2023} that the colors of the great majority of TNOs lie within two narrow loci in $griz$ color space, which we call near-infrared bright or faint (NIRB or NIRF), according to their relative $i$ and $z$-band magnitudes at a fixed visible $g-r$ color. The parameters of the modelled Gaussian NIRB/NIRF color distributions are in Table~\ref{tb:gmm}. There are statistically significant outliers, including but not limited to Haumea-family members, which could be interlopers from the inner solar system. (Section~\ref{sec:colormodel}, Figure~\ref{im:model}, discussion in Section~\ref{sec:2fam})
\item The NIRB-colored members of all dynamical classes are consistent with sharing a common $H_r$ distribution and common variability distribution.  Likewise, the physical properties of NIRF-colored objects are independent of dynamical state, to the accuracy attained by the DES catalog.  This is consistent with a scenario in which NIRF and NIRB color families are in fact birth populations that have been dispersed into present-day dynamical states. (Sections~\ref{sec:plaw}, \ref{sec:2fam})  
\item The NIRB and NIRF populations share a common absolute-magnitude distribution over our $H_r$ range, broadening the similarities in $H$ distributions noted by \citet{Adams2014}, \citet{Wong2017}, and \citet{petit2023} among more limited TNO subsets.  This implies a similarity in the planetesimal formation process in this size range across the heliocentric distance range that spans both the NIRF and NIRB birth regions. We fit a rolling power law to $p(H_r)$ and decisively exclude a simple power-law (no rolling) case, but we cannot distinguish between a rolling power law and other parameterizations of the curvature. (Section~\ref{sec:physical}, Figure~\ref{im:rolling})
\item On the other hand,  objects from the NIRF component have signficantly larger lightcurve amplitudes than those from the NIRB family, suggesting either some difference in birth shape/spin distributions, or some alteration during the migration of NIRB objects.  The DES data do not have the power to discriminate these scenarios, but LSST data should. (Section~\ref{sec:physical}, Figure~\ref{im:lcadist}, Section~\ref{sec:compare})
\item We determine the fraction $f_\nirb$ of each dynamical population, finding that the CCs are consistent with $f_\nirb=0,$ in agreement with many previous color-dynamic studies.  Assuming the CCs were formed \textit{in situ}, the NIRB population was formed interior to the NIRF.  The HC, scattered, and detached TNOs are, at present precision, consistent with a common $f_\nirb\approx0.7,$ suggesting they share a common origin ``footprint'' (but see below).  The mean motion resonances do, however, show significant variation in $f_\nirb,$ with resonances lying within the classical belt being relatively rich in NIRF sources, consistent with some of these being CCs captured after the main Neptune migration was complete. (Section~\ref{sec:fractions}, Figure~\ref{im:fclass}, Section~\ref{sec:2fam})
	\item The inclination distribution of NIRF HCs is narrower than the inclination distribution of NIRB HCs, implying that the current HCs ``remember'' to some extent whether they were born in the inner NIRB vs outer NIRF part of the HC birth regions.  Post-formation dynamics have \emph{not} mixed all birth orbits into a common present-day phase space.  A similar pattern of narrower inclination widths for NIRF colors is detected among Plutinos (3:2) and the Main Belt (7:3, 5:4) resonators. (Section~\ref{sec:vMF}, Figure~\ref{im:kappa}, Section~\ref{sec:inc}.)

        \item Features in the joint color-dynamics distribution of the DES TNOs require a more complex history than having just 2 physically homogeneous birth populations.  The NIRB members of the HC population tend to be bluer than the NIRB members of the detached and scattering populations, and the NIRF members of the CC population favor the redder end of the NIRF sequence relative to the HC, detached, and (possibly) scattering populations.  Unless there are post-migration surface color shifts, these observations require the occupation of each sequence to have differed as a function of heliocentric brith distance, along with radius-dependent selection into the current dynamical classes.  (Section~\ref{sec:split}, Figures~\ref{im:ternary} and \ref{im:ternary_nirf}, Section~\ref{sec:departures})
        \item The combination of characterization in absolute magnitude and inclination allows us to derive population and mass estimates for all trans-Neptunian subpopulations represented in the DES dataset. We find that there are similar numbers of NIRB and NIRF objects in the trans-Neptunian region, but the NIRB population accounts for most of the trans-Neptunian mass if the albedo trends described by \citet{lacerda2014a} are correct.  The population number estimates are in apparent disagreement with other recent estimates, motivating some future joint analyses across surveys. (Section~\ref{sec:counts}, Table~\ref{tb:pop}, Section~\ref{sec:pops})
\end{itemize}

Our analysis is the first to our knowledge to try to simultaneously characterize multiple physical properties (color, size, lightcurve amplitudes) and dynamical information.  While our own discussion offers general comments on the interpretation of the patterns that we have detected in the joint dynamical-physical space, we have not tested any specific scenarios against the data.
We hope these results will serve as strong quantitative constraints for future models of the planetesimal formation and migration in the outer solar system, and motivate the creation of models that track both dynamical state and color/size of TNOs.    The DES results show that the dynamical evolution from birth to present has \emph{not} completely mixed the different birth regions into indistinguishable current states; there does remain information beyond the 2-population model in the current physical-dynamical distribution which can be used to discriminate dynamical histories.  The $>10$-fold increase in sample size from LSST should sharpen these measures of differential birth properties and migration.
 
The  HMC and likelihood computation software, as well as all data produced (completeness estimates and HMC chains) will be available for open use at \url{https://github.com/bernardinelli/des_tno_likelihood} after acceptance of this contribution to the journal. The GMM package is provided as a standalone package at \url{https://github.com/bernardinelli/gmm_anyk}.

\acknowledgments

\emph{Contribution Statements:} P. H. B. led the analysis, developed the GMM algorithm, implemented the numerical calculations and produced the figures and results. G. M. B. derived the likelihood form and produced the initial HMC code. P. H. B. and G. M. B. contributed equally to the text. All other authors have contributed to the construction and operation of the Dark Energy Survey. 

\emph{Software}: This work made use of the following public libraries: \textsc{Numpy} \citep{Numpy}, \textsc{SciPy} \citep{SciPy}, \textsc{Astropy} \citep{Astropy2013,Astropy2018}, \textsc{Matplotlib} \citep{Matplotlib}, \textsc{IPython} \citep{iPython}, \textsc{Numba} \citep{numba}, \textsc{Jax} \citep{jax2018github}, \textsc{SymPy} \citep{sympy}, \textsc{Rebound} and \texttt{WHFast} \citep{rein2012,Rein2015}, \textsc{GetDist} \citep{Lewis2019}, \textsc{mpltern} \citep{yuji_ikeda_2023_8025024}

We thank B. Proudfoot and D. Ragozzine for discussions related to the Haumea family, J. J. Kavelaars for discussions about the OSSOS absolute magnitude distribution, M. Belyakov for discussions about the color distribution results, and H. W. Lin for a thorough review of this manuscript prior to submission.

P.H.B. acknowledges support from the DIRAC Institute in the Department of Astronomy at the University of Washington. The DIRAC Institute is supported through generous gifts from the Charles and Lisa Simonyi Fund for Arts and Sciences, and the Washington Research Foundation.
G.M.B. acknowledges support from NSF grants AST-1515804 and AST-2205808 during the course of this work.

This work used Anvil at Purdue University \citep{Anvil} through allocation PHY240190 from the Advanced Cyberinfrastructure Coordination Ecosystem: Services \& Support \citep[ACCESS,][]{ACCESS} program, which is supported by National Science Foundation grants \#2138259, \#2138286, \#2138307, \#2137603, and \#2138296

Funding for the DES Projects has been provided by the U.S. Department of Energy, the U.S. National Science Foundation, the Ministry of Science and Education of Spain,
the Science and Technology Facilities Council of the United Kingdom, the Higher Education Funding Council for England, the National Center for Supercomputing
Applications at the University of Illinois at Urbana-Champaign, the Kavli Institute of Cosmological Physics at the University of Chicago,
the Center for Cosmology and Astro-Particle Physics at the Ohio State University,
the Mitchell Institute for Fundamental Physics and Astronomy at Texas A\&M University, Financiadora de Estudos e Projetos,
Funda{\c c}{\~a}o Carlos Chagas Filho de Amparo {\`a} Pesquisa do Estado do Rio de Janeiro, Conselho Nacional de Desenvolvimento Cient{\'i}fico e Tecnol{\'o}gico and
the Minist{\'e}rio da Ci{\^e}ncia, Tecnologia e Inova{\c c}{\~a}o, the Deutsche Forschungsgemeinschaft and the Collaborating Institutions in the Dark Energy Survey.

The Collaborating Institutions are Argonne National Laboratory, the University of California at Santa Cruz, the University of Cambridge, Centro de Investigaciones Energ{\'e}ticas,
Medioambientales y Tecnol{\'o}gicas-Madrid, the University of Chicago, University College London, the DES-Brazil Consortium, the University of Edinburgh,
the Eidgen{\"o}ssische Technische Hochschule (ETH) Z{\"u}rich,
Fermi National Accelerator Laboratory, the University of Illinois at Urbana-Champaign, the Institut de Ci{\`e}ncies de l'Espai (IEEC/CSIC),
the Institut de F{\'i}sica d'Altes Energies, Lawrence Berkeley National Laboratory, the Ludwig-Maximilians Universit{\"a}t M{\"u}nchen and the associated Excellence Cluster Universe,
the University of Michigan, the National Optical Astronomy Observatory, the University of Nottingham, The Ohio State University, the University of Pennsylvania, the University of Portsmouth,
SLAC National Accelerator Laboratory, Stanford University, the University of Sussex, Texas A\&M University, and the OzDES Membership Consortium.

Based in part on observations at Cerro Tololo Inter-American Observatory, National Optical-Infrared Astronomy Observatory, which is operated by the Association of
Universities for Research in Astronomy (AURA) under a cooperative agreement with the National Science Foundation.

The DES data management system is supported by the National Science Foundation under Grant Numbers AST-1138766 and AST-1536171.
The DES participants from Spanish institutions are partially supported by MINECO under grants AYA2015-71825, ESP2015-66861, FPA2015-68048, SEV-2016-0588, SEV-2016-0597, and MDM-2015-0509,
some of which include ERDF funds from the European Union. IFAE is partially funded by the CERCA program of the Generalitat de Catalunya.
Research leading to these results has received funding from the European Research
Council under the European Union's Seventh Framework Program (FP7/2007-2013) including ERC grant agreements 240672, 291329, and 306478.
We acknowledge support from the Australian Research Council Centre of Excellence for All-sky Astrophysics (CAASTRO), through project number CE110001020, and the Brazilian Instituto Nacional de Ci\^encia
e Tecnologia (INCT) e-Universe (CNPq grant 465376/2014-2).

This manuscript has been authored by Fermi Research Alliance, LLC under Contract No. DE-AC02-07CH11359 with the U.S. Department of Energy, Office of Science, Office of High Energy Physics. The United States Government retains and the publisher, by accepting the article for publication, acknowledges that the United States Government retains a non-exclusive, paid-up, irrevocable, world-wide license to publish or reproduce the published form of this manuscript, or allow others to do so, for United States Government purposes.

\bibliography{references}
\bibliographystyle{aasjournal}

\newpage
\appendix
\section{Estimating distributions with Gaussian Mixture Models}
\label{sec:gmm}

\subsection{Gaussian mixture model review}
We wish to determine the optimal underlying color distribution implied by our TNO sample. To this end, we will posit that the underlying distribution for the colors $\mathbf{c}$ is a mixture model given by a combination of $K$ components. For a general measurement $\mathbf{x}$ in $d$ dimensions, we have that the mixture model is
\begin{equation}
	p(\mathbf{x}|\boldsymbol\theta) = \sum_{\alpha=1}^K f_\alpha p(\mathbf{x}|\boldsymbol\theta_\alpha), \label{eq:gmmmodel}
\end{equation}
where the $\alpha^\text{th}$ component is described by a probability distribution $p(\mathbf{x}|\boldsymbol\theta_\alpha)$, and the parameters $\boldsymbol\theta = \bigcup_\alpha \{\boldsymbol\theta_\alpha, f_\alpha \} $ are subject to the constraint $\sum_\alpha f_\alpha = 1$ to ensure that $p(\mathbf{x}|\boldsymbol\theta)$ is normalized. In a Gaussian mixture model (GMM), the probability distribution $p(\mathbf{x}|\boldsymbol\theta_\alpha)$ is a multivariate Gaussian with mean $\boldsymbol\mu_\alpha$ and covariance matrix $\boldsymbol\Sigma_\alpha$, that is,
\begin{equation}
	p(\mathbf{x} | \boldsymbol\theta_\alpha) = \mathcal{N}(\mathbf{x}|\boldsymbol\mu_\alpha, \boldsymbol\Sigma_\alpha) \equiv \frac{1}{\sqrt{(2\pi)^d \det(\Sigma_\alpha)}} \exp\left[-\frac{1}{2}(\mathbf{x} - \boldsymbol\mu_\alpha)^\top \boldsymbol\Sigma_\alpha^{-1} (\mathbf{x} - \boldsymbol\mu_\alpha)\right].
\end{equation}

We define the likelihood of the $N$ data points $\mathcal{X} \equiv \bigcup_{i=1}^N\{\mathbf{x}_i \}$ as
\begin{equation}
	\mathcal{L} (\mathcal{X}|\boldsymbol\theta) \equiv \prod_{i=1}^N \sum_{\alpha=1}^K f_\alpha p(\mathbf{x}_i | \boldsymbol\theta_\alpha).\label{eq:likegmm}
\end{equation}
This is an \emph{uncategorized} likelihood, as we do not have the information of which component $k$ is the parent for sample $i$. We define the indicator $\mathcal{Z} = \bigcup_i \{\mathbf{z}^{(i)} \}$, such that the $\alpha^\text{th}$ entry of $\mathbf{z}^{(i)}$ is $z_\alpha^{(i)} = 1$ if $i$ was generated by component $\alpha$, and $0$ otherwise. We define the likelihood of the complete data $\mathcal{D} = \{ \mathcal{X}, \mathcal{Z} \}$ as
\begin{equation}
	\mathcal{L}(\mathcal{D} | \boldsymbol\theta) \equiv \prod_{i=1}^N \sum_{\alpha=1}^K \left[f_\alpha p(\mathbf{x}_i | \boldsymbol\theta_\alpha) p(z_{\alpha}^{(i)} | \mathbf{x}_i,\boldsymbol\theta_\alpha) \right]^{z_\alpha^{(i)}}.
\end{equation}
From Bayes' theorem,
\begin{equation}
	q_{i\alpha} \equiv p(z_\alpha^{(i)} | \mathbf{x}_i,\boldsymbol\theta_\alpha) = \frac{p(\mathbf{x}_i|z_\alpha^{(i)},\boldsymbol\theta_\alpha) p(z_\alpha^{(i)})}{\sum_\beta p(\mathbf{x}_i|z_\beta^{(i)},\boldsymbol\theta_\beta) p(z_\beta^{(i)})}. \label{eq:probgmmbasic}
\end{equation}
We set $p(z_\alpha^{(i)}) \propto 1$, that is, all components are equally likely.

The expectation-maximization (EM) algorithm for GMMs works by alternating the expectation (E) step by computing Equation \ref{eq:probgmmbasic} and using its value to update the model parameters by taking weighted moments of $\mathbf{x}$:
\begin{subequations}
	\begin{align}
		f_\alpha           & =  \frac{1}{N} \sum_i q_{i\alpha}                                                                                     \\
		\boldsymbol\mu_\alpha    & =  \frac{1}{\sum_i q_{i\alpha}} \sum_i q_{i\alpha} \mathbf{x}_i                                                            \\
		\boldsymbol\Sigma_\alpha & = \frac{1}{\sum_i q_{i\alpha}} \sum_i q_{i\alpha} (\mathbf{x}_i - \boldsymbol\mu_\alpha) (\mathbf{x}_i - \boldsymbol\mu_\alpha)^\top
	\end{align}
\end{subequations}
These steps, applied successfully, converge to a local maximum of $\mathcal{L}(\mathcal{D}|\boldsymbol\theta)$ \citep{dempster1977,wu1983}.

The standard GMM framework requires that the number of components $K$ is known (or posited) a priori - something that we do not necessarily have in our case. The final values for $\boldsymbol\mu_\alpha$ are also dependent on the initialization algorithm, as the EM steps can converge to a local, but not necessarily global, maximum of the likelihood. We will modify the likelihood in Equation \ref{eq:likegmm} to account for the unknown $K$, as well as make the algorithm more robust to the initial values and incorporate the generalizations by \cite{bovy2011,melchior2018} to account for measurement errors and selection effects.

\label{sec:ugmm}
\subsection{Unsupervised Gaussian mixture model with noisy and incomplete data}
\cite{figueiredo2002} proposed a modification of the general mixture model framework so that the number of components $K$ does not need to be defined a priori. Rather, a range $K\in[K_\mathrm{min},K_\mathrm{max}]$ is specified, and the modified E and M steps are such that components whose amplitudes $\alpha_k$ become too small are annihilated. Their algorithm is also more robust to the initialization as, by starting with a number of components $K_\mathrm{max} \gg K_\mathrm{true}$, these are more likely to cover the entire parameter space, and redundant components compete with each other until all but one are annihilated. This means that the number of components $K$ is a free parameter in the model, and the optimal number is found in an unsupervised manner.

Here, we combine the \cite{figueiredo2002} approach with the extreme de-convolution of \cite{bovy2011} and its modifications for missing data by \cite{melchior2018}. We do not know, for any given sample $i$, the true value $\mathbf{x}_i$. What is known is the measured value $\mathbf{y}_i = \mathbf{x}_i + \mathbf{e}_i$, where $\mathbf{e}_i$ is a (Gaussian) noise drawn from $\mathcal{N}(\mathbf{e}_i|\mathbf{0}, \mathbf{S}_i)$, where $\mathbf{S}_i$ is the covariance matrix of the measurement. Following \cite{bovy2011}, we write the likelihood of $\mathbf{y}_i$ by marginalizing over $\mathbf{x}_i$,
\begin{equation}
	p(\mathbf{y}_i |\mathbf{S}_i, \boldsymbol\theta) =  \sum_\alpha f_\alpha \int \mathrm{d}\mathbf{x}_i \, \mathcal{N}(\mathbf{y}_i|\mathbf{x}_i, \mathbf{S}_i) \mathcal{N}(\mathbf{x}_i|\boldsymbol\mu_\alpha, \boldsymbol\Sigma_\alpha) = \sum_\alpha f_\alpha \mathcal{N}(\mathbf{y}_i|\boldsymbol\mu_\alpha, \boldsymbol\Sigma_\alpha + \mathbf{S}_i).
\end{equation}
We can modify our likelihood (Equation \ref{eq:likegmm}) for the noisy data $\mathcal{Y} = \bigcup_i \{ \mathbf{y}_i, \mathbf{S}_i \}$ and also penalize it due to a selection process $p(\mathcal{S}|\mathbf{x})$
\begin{equation}
	\mathcal{L}(\mathcal{Y} | \boldsymbol\theta) = \frac{\prod_i \sum_\alpha f_\alpha \mathcal{N}(\mathbf{y}_i| \boldsymbol\mu_\alpha, \boldsymbol\Sigma_\alpha + \mathbf{S}_i) }{\left[\int\mathrm{d} \mathbf{x}' \, p(\mathbf{x}'|\boldsymbol\theta) p(\mathcal{S}|\mathbf{x}')\right]^N}  .
\end{equation}
Note that, implicitly, we are assuming that the data is ``missing at random'' \citep{rubin1976}, that is, $p(\mathcal{S}|\mathbf{x})$ is independent of $p(\mathbf{x}|\boldsymbol\theta)$, so the likelihood is also independent of the missing data. The corrections in the GMM due to missing samples are found by a rejection sampling procedure between the EM steps. We draw $N'$ samples $\mathbf{x}_i'$ from the current model, apply noise to these samples so that $\mathbf{y}'_i = \mathbf{x}'_i + \mathbf{e}_i'$, where $\mathbf{e}_i'$ is obtained by sampling a Gaussian distribution with a covariance matrix $\mathbf{S}'_i$ drawn from the distribution of $\{\mathbf{S}_i\}$ and applying the selection function: a given sample $i$ is rejected if $p(\mathcal{S}|\mathbf{y}_i') \leq u$, where $u\sim \mathrm{Unif}(0,1)$. This process is repeated until the number of accepted samples $N'_\mathrm{acc}$ is consistent with the $68\%$ limits of a Poisson distribution given by $N$ \citep{melchior2018}. We define the subset of the rejected data with $N'_\mathrm{rej}$ samples as $\mathcal{M} = \bigcup_{i=1}^{N'_\mathrm{rej}} \{ \mathbf{y}'_i, \mathbf{S}'_i\}$.

Finally, we need to account for the unknown number of mixture components $K$. To do so, we apply the improper Dirichlet prior for the amplitudes $\{ f_\alpha \}$ from \cite{figueiredo2002},
\begin{equation}
	p(\{f_\alpha\}) =N^{-K(N_\mathrm{par}+1)/2}\exp\left[-\frac{N_\mathrm{par}}{2} \sum_{\alpha, f_\alpha > 0} \log f_\alpha \right], \label{eq:dirichlet}
\end{equation}
where $N_\mathrm{par}$ is the number of free parameters in each component $\alpha$. For a Gaussian mixture, $N_\mathrm{par} = d + d(d+1)/2$. The inclusion of this prior translates to a modification in the amplitude calculation in the M step, so that the amplitudes $\{ f_\alpha \}$ are penalized by the number of parameters each component needs to estimate, which allows for a component to have $f_\alpha = 0$ and, if this happens, the component is discarded by the algorithm. We refer the reader to \cite{figueiredo2002} for a full derivation of the procedure.

We now can write the full EM algorithm for this unsupervised mixture model. The expectation step is as in \cite{melchior2018}:
\begin{subequations}
	\label{eq:Estepcomplete}
	\begin{align}
		q_{i\alpha}          & = \frac{f_\alpha \mathcal{N}(\mathbf{y}_i |\boldsymbol\mu_\alpha, \boldsymbol\Sigma_\alpha + \mathbf{S}_i)}{\sum_\beta f_\beta \mathcal{N}(\mathbf{y}_i |\boldsymbol\mu_\beta, \boldsymbol\Sigma_\beta + \mathbf{S}_i)}, \\
		\mathbf{b}_{i\alpha} & = \boldsymbol\mu_\alpha + \boldsymbol\Sigma_\alpha (\boldsymbol\Sigma_\alpha + \mathbf{S}_i)^{-1} (\mathbf{y}_i - \boldsymbol\mu_\alpha),                                                                               \\
		\mathbf{B}_{i\alpha} & = \boldsymbol\Sigma_\alpha - \boldsymbol\Sigma_\alpha(\boldsymbol\Sigma_\alpha + \mathbf{S}_i)^{-1} \boldsymbol\Sigma_\alpha.
	\end{align}
\end{subequations}
The two additional parameters, $\mathbf{b}_{i\alpha}$ and $\mathbf{B}_{i\alpha}$, are, respectively, the conditional estimates of $\mathbf{y}_i$ and $\mathbf{S}_i$, given each component $\alpha$ \citep{bovy2011}. An important detail here is that this is computed $\forall i \in \{\mathcal{Y}, \mathcal{M}\}$, that is, this is computed both for the observed data $\mathcal{Y}$ and the realizations of the missing data $\mathcal{M}$. If there is no selection effect being considered, only the measurements in set $\mathcal{Y}$ are required.

The maximization step includes the corrections for the noisy data as well as the prior in Equation \ref{eq:dirichlet}:
\begin{subequations}
	\label{eq:MstepAdaptive}
	\begin{align}
		f_\alpha            & =  \frac{\max\left(0, \sum_{i\in \{\mathcal{Y}, \mathcal{M}\}} q_{i\alpha} - \frac{N_\mathrm{par}}{2}\right)}{\sum_\beta \max\left(0, \sum_{i\in \{\mathcal{Y}, \mathcal{M}\}} q_{i\beta} - \frac{N_\mathrm{par}}{2}\right)},             \\
		\boldsymbol\mu_\alpha    & =  \frac{1}{\sum_{i\in \{\mathcal{Y}, \mathcal{M}\}} q_{i\alpha}} \sum_{i\in \{\mathcal{Y}, \mathcal{M}\}} q_{i\alpha} \mathbf{b}_{i\alpha},                                                                                                \\
		\boldsymbol\Sigma_\alpha & = \frac{1}{\sum_{i\in \{\mathcal{Y}, \mathcal{M}\}} q_{i\alpha}} \sum_{i\in \{\mathcal{Y}, \mathcal{M}\}} \left[ q_{i\alpha} (\mathbf{b}_{i\alpha} - \boldsymbol\mu_\alpha) (\mathbf{b}_{i\alpha} - \boldsymbol\mu_\alpha)^\top + \mathbf{B}_{i\alpha} \right].
	\end{align}
\end{subequations}
If a component is annihilated (that is, $f_\alpha = 0$), the E-step is recomputed so that its probability mass is redistributed for the other components. The procedure is repeated until $K = K_\mathrm{min}$. There is the possibility that, for a given $K$, no component is annihilated and the likelihood converges to a (potentially local) maximum. If that happens, the component with the smallest $f_\alpha$ is discarded, so that the entire range of $K$ can be explored. The optimal model, as expected, maximizes $p(\{ f_\alpha \} )\mathcal{L}(\mathcal{Y}|\boldsymbol\theta)$, which properly penalizes the likelihood with respect to number of components. The inclusion of the missing samples $\mathcal{M}$ means that the converge is stochastic \citep{nielsen2000}, and either more computational steps or oversampling of the missing data $\mathcal{M}$ are required to ensure convergence.

This software is available for open use at \url{https://github.com/bernardinelli/gmm_anyk}.

\section{The Bayesian evidence ratio}
\label{sec:evidence}
\change{New appendix}

Given two competing hypotheses $\mathcal{H}_1$ and $\mathcal{H}_2$ to explain observational data $D$, the (log) Bayesiean evidence ratio is defined as
\begin{equation}
  \mathcal{R} = \log \frac{p(D|\mathcal{H}_2)}{p(D|\mathcal{H}_1)} = \log \frac{ \int d\boldsymbol{q}_1\, \likeli(D | \boldsymbol{q}_1) p(\boldsymbol{q}_1)}
  { \int d\boldsymbol{q}_2\, \likeli(D | \boldsymbol{q}_2) p(\boldsymbol{q}_2)},
\end{equation}
where $\boldsymbol{q}_{1,2}$ are the free parameters of the two competing hypotheses for the likelihood $\likeli$ of obtaining data $D,$ and $p(\boldsymbol{q}_{1,2})$ are priors on these parameters.
In our  typical application, our two hypotheses are that (1) each dynamical subset $d$ has a distinct set of parameters $\boldsymbol{q}_d$  to generate its data $D_d,$ vs (2) all subsets share a common set of parameters $\boldsymbol{q}$ for the distribution of physical parameters.  In each case, the fractions $f$ of mixtures are free to vary over the simplex to best fit the data.  The evidence ratio is in this case
\begin{equation}
  \mathcal{R}
= \log \frac{\prod_d \int \mathrm{d}\boldsymbol{q}_d \, p(\boldsymbol{q}_d) p(\boldsymbol{q}_d | D_d)}{\int \mathrm{d}\boldsymbol{q} \, p(\boldsymbol{q})\,\prod_d  p(\boldsymbol{q} | D_d)}. \label{eq:evidence}
\end{equation}
where
\begin{equation}
	p(\boldsymbol{q}| D) \equiv \int \mathrm{d}\!f\, \likeli(\{\boldsymbol{q}, f\} | D)
\end{equation}
is the likelihood of the parameters of interest after marginalizing over the other parameters of the overall model, including the occupation fractions.

In order to avoid performing the integrals in Equation \ref{eq:evidence} over the high-dimensional parameter spaces directly, we can use our HMC outputs to approximate this integral. We use the Laplace-Metropolis estimator of \cite{Lewis1997}. We have that
\begin{equation}
	p(D|\mathcal{H}) = \int \mathrm{d}\boldsymbol{q} \, p(\boldsymbol{q}) \mathcal{L}(\boldsymbol{q} | D) \approx (2\pi)^{P/2} |\mathbf{H}^*|^{1/2} p(\boldsymbol{q}^*) \mathcal{L}(\boldsymbol{q}^*|D).
\end{equation}
Here, $P$ is the number of parameters, $\boldsymbol{q}^*$ is the value that maximizes $p(\boldsymbol{q}) \likeli (\boldsymbol{q}|D)$, and $\mathbf{H}^*$ is the inverse Hessian of this function evaluated at $\boldsymbol{q}^*$. In practice, these can be estimated by the mean and covariance of the HMC samples, respectively. Note that, while the HMC procedure also produces estimates for the $f_\beta$, these, by construction, cannot be used by the Laplace-Metropolis estimator: the simplex constraint imposes a Dirichlet prior on these variables, and Dirichlet-distributed variables have singular covariance matrices. To avoid this issue, we compute the integral over the $f_\beta$ directly, using the fixed $\boldsymbol{q}^*$. This procedure is justified, as it is equivalent to changing the order of integration among two sets of variables. 

\section{Hamiltonian Monte Carlo}
\label{sec:hmc}
We present here a simple introduction to the Hamiltonian Monte Carlo (HMC) algorithm---we refer the reader to \cite{neal2011}, \cite{gelman2013bayesian} and \cite{betancourt2017} for more complete presentations of this technique. As a variant of Markov Chain Monte Carlo, the goal of HMC is to produce samples $\mathbf{q}_i$ from some probability distribution $p(\mathbf{q}).$ In our case the parameters $\mathbf{q}$ include both the population fractions $f_\beta$ and any of the parameters $\theta,\theta^\prime,\bar A, s$ of our physical distributions.  The probability distribution is the product of the likelihood in Equation~\ref{eq:loglike} multiplied by any priors assigned to the parameters.

Following classical mechanics, we assume a Hamiltonian physical system with canonical positions $\mathbf{q}$ and momenta $\mathbf{p}$. We define the kinetic energy of the system as $K(\mathbf{p})$ and the potential energy as $U(\mathbf{q})$. The Hamiltonian, a quantity analogous to the total energy of the system, is
\begin{equation}
	\mathcal{H}(\mathbf{q},\mathbf{p}) = K(\mathbf{p}) + U(\mathbf{q}).
\end{equation}
For a HMC system, the target parameters are the canonical positions, and we assign the potential energy to be the log of the target probability:
\begin{equation}
	U(\mathbf{q}) = -\log\left[ p(\mathbf{q}) \mathcal{L}(\mathbf{q})\right].
      \end{equation}
We posit a kinetic energy term that is quadratic in the canonical momenta, with some chosen mass matrix $\mathbf{M}$:
\begin{equation}
	K(\mathbf{p}) = \frac{1}{2} \mathbf{p}^\top \mathbf{M}^{-1} \mathbf{p}.
\end{equation}
The key to HMC is to sample from a joint distribution $p(\mathbf{p},\mathbf{q}) = \exp(-\mathcal{H}).$ 
The first step of the HMC is simply the generation of new momenta $\mathbf{p}$ from the conditional distribution $\sim p(\mathbf{p}|\mathbf{q})$; with the quadratic kinetic energy, this is a draw from a multivariate Gaussian, which is fast and exact.  The second step updates the positions and the momenta by using the (Hamiltonian) dynamics of the system, following the standard equations of motion
\begin{equation}
	\frac{\mathrm{d} \mathbf{q}}{\mathrm{d} t} =\frac{\partial \mathcal{H}}{\partial \mathbf{p}}; \quad \frac{\mathrm{d}\mathbf{p}}{\mathrm{d}t} = -\frac{\partial\mathcal{H}}{\partial \mathbf{q}}.
\end{equation}
For a perfect numerical integrator, $\mathcal{H}$ and hence $p(\mathbf{p},\mathbf{q})$ are conserved during this integration, and integration over some time period $t$ leads to a new sample of $(\mathbf{p}^\star,\mathbf{q}^\star)$ for the HMC chain.  A numerical integration is necessarily inexact, and we retain or discard this sample using the Metropolis-Hastings probability of acceptance
\begin{equation}
	\min\left[1, \exp(-\mathcal{H}(\mathbf{q}^\star,\mathbf{p}^\star) + \mathcal{H}(\mathbf{q},\mathbf{p}))\right].
\end{equation}
The main implementation challenge in HMC is in the calculation of the gradient of the log likelihood with respect to the model parameters, so that the equations of motion can be integrated. Despite this complexity compared to simpler Metropolis-Hastings samplers, the addition of the Hamiltonian structure to the problem increases the chance of an accepted state during sampling in a high-dimensional space.

The only complication in the application to our problem is implementation of the system of constraints on the fractional parameters $f_\beta,$ which must each be $\ge 0$ and must satisfy $\sum f_\beta=1.$  The former constraint can be automatically satisfied by a change of variable from $f_\beta$ to $q_\beta \equiv \log f_\beta.$  The simplex constraint is implemented by using the \texttt{RATTLE} algorithm for numerical integration of a constrained Hamiltonian system, as described by \citet{brubaker}.  All of the derivatives of our likelihood, priors, and constraints are straightforward, and in fact 
amenable to the use of the automatic differentiation package \texttt{jax} \citep{jax2018github}.



\section{Completeness estimates}\label{sec:completeness}
To calculate the terms $S_\beta$ in the likelihood in Equation~\ref{eq:loglike} for the DES data given a population model, we need to execute the integrals over the selection probability $p(\mathcal{S} | \mathbf{c},H,A,\mathbf{P})$ given in Equation~\ref{eq:Sbeta}.
Object discovery in \des\ uses detections on images taken in the $griz$ bands \citep{Bernardinelli2019,Bernardinelli2022}, and each dynamical class has its own observational biases, dominated, primarily, by $a$, $q$ and $I$: larger-$a$ objects can only be discovered when they are near perihelion \citep[see, \eg][for a thorough discussion]{Bernardinelli2020}; objects at higher perihelia can only be detected with brighter $H$; and populations with a higher density of high-inclination objects also have a larger area of the sky where objects can be discovered, which inherently changes their discoverability due to the large high ecliptic latitude coverage of the \des\ footprint. 

The value of $p(\mathcal{S} | \mathbf{c},H,A,\mathbf{P})$ for any specified values of the conditional variables can be estimated
using the \des\ survey simulator\footnote{\texttt{DESTNOSIM}, \url{https://github.com/bernardinelli/DESTNOSIM}}.  To conduct the specified integration of this, we need to posit a model for the distribution $p_\beta(\mathbf{c},H,A,\mathbf{P})$ across the dynamical class under study.  Our hypotheses provide models the distributions of $\mathbf{c}, H_r,$ and $A,$ but do not provide the full distribution of a dynamical class over orbital parameters.
As a proxy model, we assume that the true distribution of non-resonant dynamical classes has azimuthal and temporal symmetry, \ie\ the distributions over orbital parameters $\Omega$ and $\mathcal{M}$ are uniform between 0 and $2\pi.$   We make the further assumption that argument of perihelion $\omega$ is uniform:
\begin{equation}
	p(\Omega,\omega,\mathcal{M}) = u(\Omega) u(\omega) u(\mathcal{M}),\label{eq:uniform}
\end{equation}
where $u$ is uniform between 0 and $2\pi.$   Our choice of $I$ and $H_r$ distributions are described below, and depend on whether we are modeling the underlying distributions of $I$ and $H_r$ or not.  This leaves the $a$ and $e$ distribution of the true population unspecified.  We will approximate this as the sum of a series of $\delta$ functions such that the dynamical distribution of our \emph{observed} population would be exactly reproduced after applying the survey selection function, namely:
\begin{align}
  p(a,e,I,\Omega,\omega,\mathcal{M}, H_r, \mathbf{c}) & = \sum_j \frac{\delta(a - a_j) \delta (e - e_j)}{p(\mathcal{S}|a_j,e_j, I, H_r, \mathbf{c})}   p(\Omega,\omega,\mathcal{M}), \\
  p(\mathcal{S}|a_j,e_j,I,H_r, \mathbf{c}) & \equiv \int dI\,d\Omega\,d\omega\,d\mathcal{M}\, p(\mathcal{S} | a_j, e_j, I, \Omega, \omega, \mathcal{M}, H_r, \mathbf{c}) p(\Omega,\omega,\mathcal{M}).
                                             \label{eq:Spop}
\end{align}
In essence we bootstrap the observed population into a full population (for the purposes of conducting the integral in Equation~\ref{eq:Sbeta}) but weighting each TNO orbit by the inverse of its selection rate for a chosen $I, H_r,$ and color.  When using the simulator to calculate each selection probability in Equation~\ref{eq:Spop}, we can give the simulated object a light curve amplitude $A$ if desired.



For a resonant object in mean motion resonance $p$:$q$, we also need to find the subset of angles $\{\Omega,\omega,\mathcal{M}\}_{p:q} \subset [0,2\pi)^3$ that produce a librating angle. To do so, we follow the procedure outlined in \cite{Bernardinelli2022} to derive dynamical classifications for our real objects: we integrate the orbits of each synthetic object for 10 Myr and determine the fraction of time that the resonant angle 
\begin{equation}
	\sigma = p (\mathcal{M} + \omega + \Omega) - q (\mathcal{M}_\mathrm{N} + \omega_\mathrm{N} + \Omega_\mathrm{N}) + (q-p) (\Omega + \omega)
\end{equation} 
librates. If the object librates for more than 50\% of the time, we include it in the resonant subset; this is consistent with the criterion for a ``resonant candidate'' object in \cite{Bernardinelli2022}. The subscript $\mathrm{N}$ refers to the relevant angles for Neptune. To reduce the computation complexity of the problem, only Neptune is treated as an active particle in the simulation. 

Finally, we treat the inclinations $I$ and absolute magnitudes $H_r$ in three different ways, depending on the necessity of the model being evaluated:
\begin{itemize}
	\item For the color distribution analysis of Section \ref{sec:colormodel}, we have that $ p(I) p(H_r) \propto \sum_j \delta(H_r - H_{r,j}) \delta(I - I_j)$;
	\item For the absolute magnitude distribution analysis, we maintain the inclination distribution as a $\delta$ function, but have a $p(H_r)$ uniform in the $3 \leq H_r \leq 10$ range;
	\item When simultaneously modelling the $H_r$ and $I$ distributions, we also consider an inclination distribution sinusoidally distributed in the $0\degr \leq I \leq 90\degr$ range. This choice leads to fewer clones at inclinations $I \geq 60\degr$, where the \des\ distribution is approximately constant at a fixed magnitude (see Figure 5 of \citealt{Bernardinelli2022}) and no objects were discovered.
\end{itemize}

\begin{figure}[ht!]
	\centering
	\includegraphics[width=0.495\textwidth]{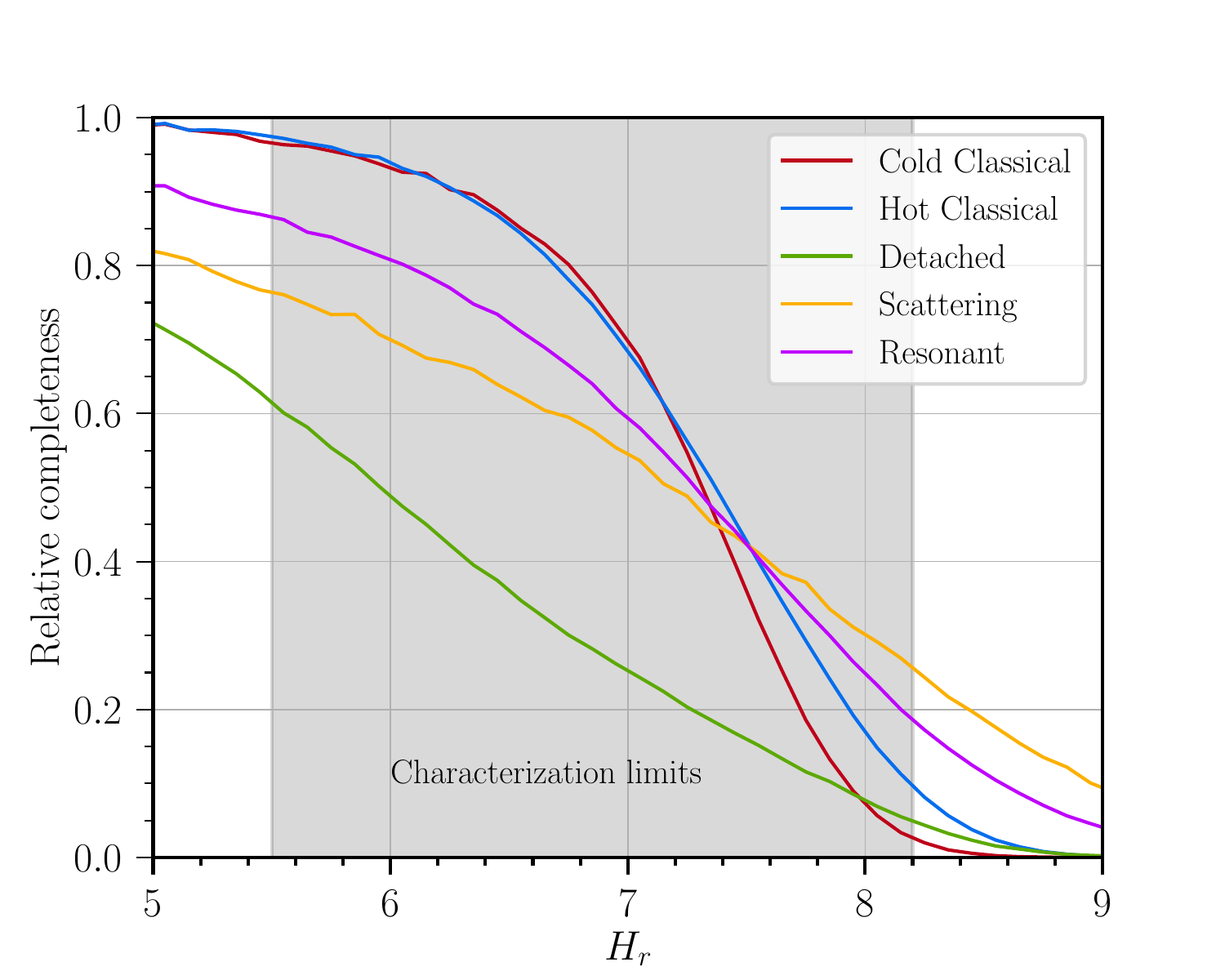}
	\includegraphics[width=0.495\textwidth]{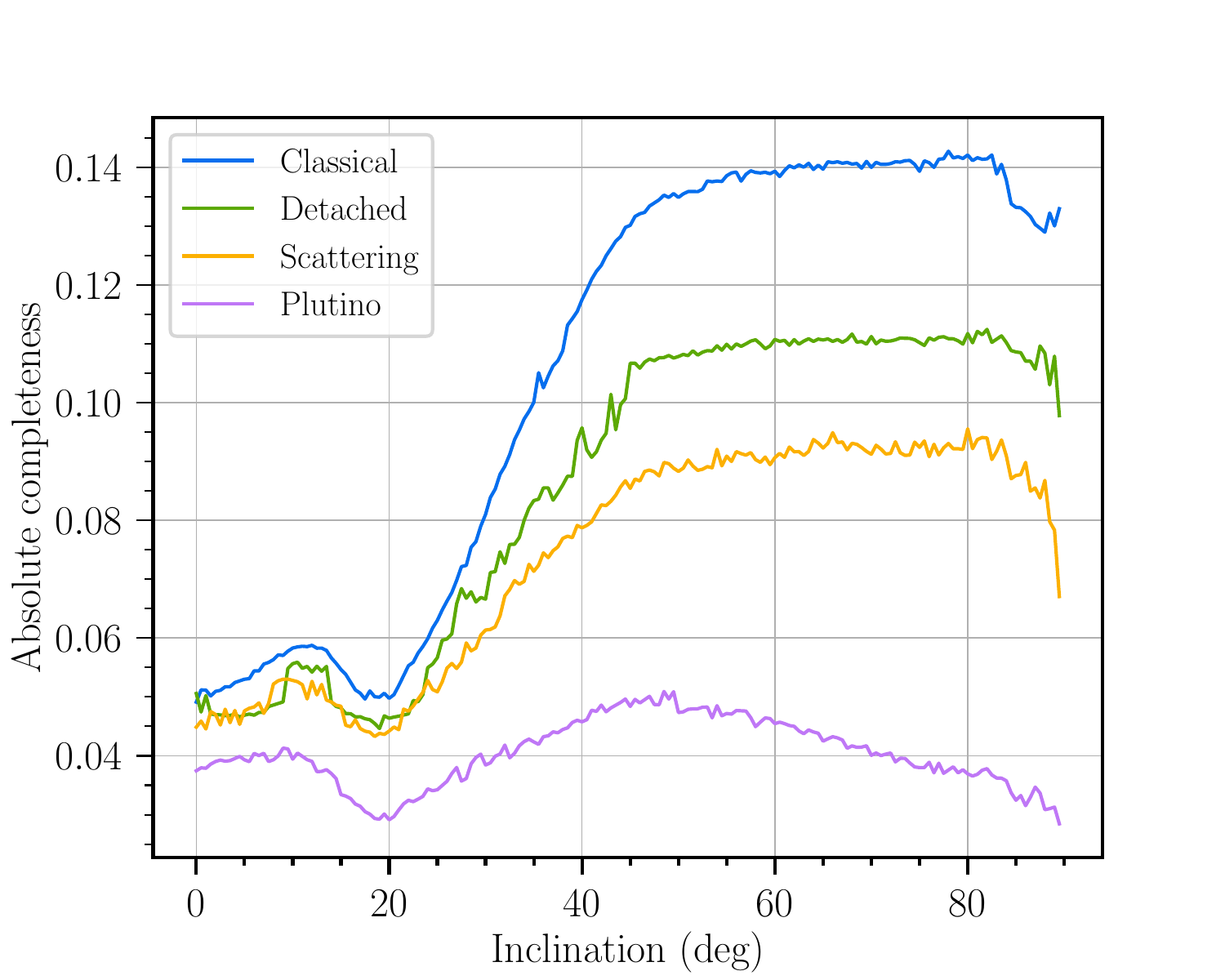}\\
	\includegraphics[width=1\textwidth]{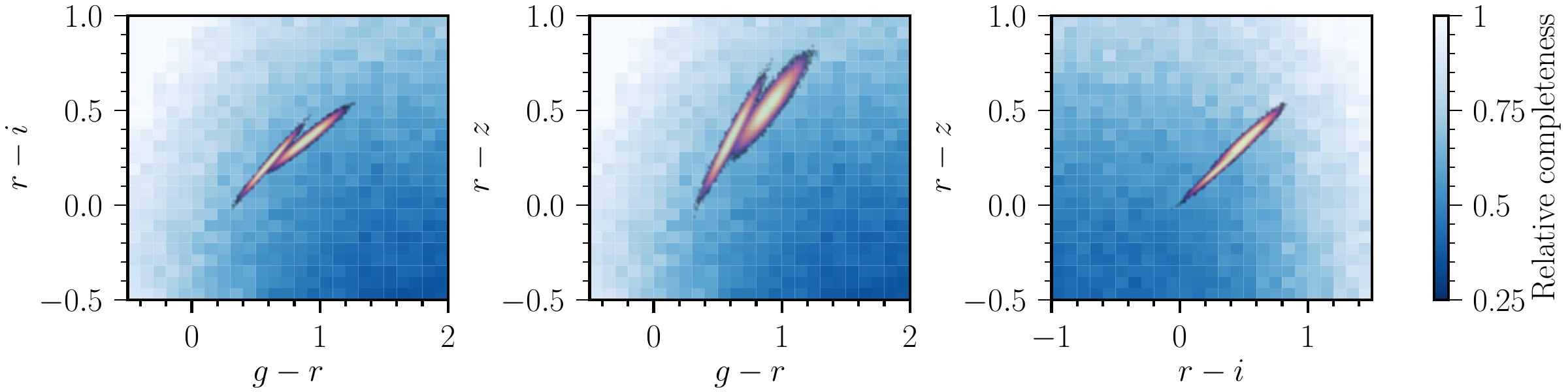}

	\caption{Completeness as a function of $H_r$ (top left), inclination (top right) and the three color pairs (bottom) after marginalizing over the other relevant parameters. While the color and $H_r$ figures show a relative completeness, so that the maximum value of the selection function is 1, the inclination completeness represents an absolute number that is indicative of the size of the \des\ footprint: \des\ covered $\approx 12\%$ of the sky, while the larger relative completeness accounts for object drifting into the survey footprint over the six years of the survey. The $H_r$ and inclination completeness are subdivided as a function of orbital class and, for simplicity, we only show the Plutino case for the inclination distribution. In the color completeness panel, we also show the resulting color model as in Figure \ref{im:model}.}
\label{im:completeness}
\end{figure}

In the last case, we expect that the marginalized distribution $p(\mathcal{S}|I, H_r=3)$ will represent the fraction of the area of the sky available to each population that is covered by the \des\ footprint at inclination $I$. In total, the estimates for this completeness functions required over 2 billion synthetic TNOs. We illustrate our completeness in Figure \ref{im:completeness}.

\allauthors


\end{document}